\documentclass[review]{elsarticle}

\usepackage{color} 
\usepackage{amsmath} 

\usepackage{lineno,hyperref}
\modulolinenumbers[5]

\journal{Elsevier}

\begin{document}

\begin{frontmatter}

\title{Particulate immersed boundary method for complex fluid-particle interaction problems with heat transfer}

\author[myspecialaddress]{Hao Zhang}
\ead{hao@cttc.upc.edu}
\author[mymainaddress]{Haizhuan Yuan}
\author[myspecialaddress]{F.Xavier Trias}
\author[myespecialaddress]{Aibing Yu}
\author[mysecondaryaddress]{Yuanqiang Tan}
\author[myspecialaddress]{Assensi Oliva\corref{mycorrespondingauthor}}
\cortext[mycorrespondingauthor]{Corresponding author}
\ead{cttc@cttc.upc.edu}
\address[myspecialaddress]{Heat and Mass Transfer Technological Center, Technical University of Catalonia,
 Terrassa, Barcelona 08222, Spain}
 \address[mymainaddress]{ School of Mathematics and Computational Science, Xiangtan University, Hunan 411105, China}
\address[myespecialaddress]{Department of Chemical Engineering, Monash University, Vic 3800 Australia}
 \address[mysecondaryaddress]{ School of Mechanical Engineering, Xiangtan University, Hunan 411105, China}

\begin{abstract}

In our recent work [H. Zhang, F.X. Trias, A. Oliva, D. Yang, Y. Tan, Y. Sheng. PIBM: Particulate immersed boundary method 
for fluid-particle interaction problems. Powder Technology. 272(2015), 1-13.], a particulate immersed boundary method (PIBM) 
for simulating fluid-particle multiphase flow was proposed and assessed in both two- and three-dimensional applications. 
In this study, the PIBM was extended to solve thermal interaction problems between spherical particles and fluid. The Lattice 
Boltzmann Method (LBM) was adopted to solve the fluid flow and temperature fields, the PIBM was responsible for the no-slip 
velocity and temperature boundary conditions at the particle surface, and the kinematics and trajectory of the solid particles 
were evaluated by the Discrete Element Method (DEM). Four case studies were implemented to demonstrate the capability of the 
current coupling scheme. Firstly, numerical simulation of natural convection in a two-dimensional square cavity with 
an isothermal concentric annulus was carried out for verification purpose. The current results were found to have good 
agreements with previous references. Then, sedimentation of two- 
and three-dimensional isothermal particles in fluid was numerically studied, respectively. The instantaneous 
temperature distribution in the cavity was captured. The effect of the thermal buoyancy on particle behaviors was discussed.
Finally, sedimentation of three-dimensional thermosensitive particles in fluid was numerically investigated.
Our results revealed that the LBM-PIBM-DEM is a promising scheme for the solution of complex fluid-particle interaction 
problems with heat transfer.

\end{abstract}

\begin{keyword}
 Lattice Boltzmann Method \sep Particulate Immersed Boundary Method \sep Discrete Element Method 
 \sep Fluid-particle interaction  \sep Heat transfer   
\end{keyword}

\end{frontmatter}


\section{Introduction}

In recent years, numerical simulations based on combined Lattice Boltzmann Method (LBM) ~\cite{qian1992lattice} and 
Discrete Element Method (DEM)~\cite{hao8} have gained popularity in the fluid-particle interaction problems. 
The coupling scheme is quite attractive because the calculation of both sides is fairly local at the particle or even 
sub-particle scale. 
This feature provides more freedom on the particle geometry and the choice of interaction laws. The commonest way 
for the coupling is to solve the fluid field at the particle scale while the solid  
particles are treated as moving boundaries~\cite{hu1996direct}. The macro fluid quantities such as velocity and 
temperature are enforced to be coincident with the solid boundaries. This target is not easy to be achieved numerically since 
the solid boundary may not be at the same places with the lattice nodes. Therefore, the Immersed Boundary Method 
 (IBM)~\cite{Peskin1977220} is adopted to polish the stepwise representation and overcome the numerical oscillation 
 when the moving boundary crosses the LBM grids. Meanwhile, the hydrodynamic force and torque exerted on the particle 
 can be evaluated through the numerical correction. Finally, a proper particle tracking technique like the DEM~\cite{hao8} 
 is required to make the calculation cycle closed. 
 
For calculating the fluid-solid interaction force, there are several available schemes such as the penalty 
method~\cite{Feng2004602}, the direct forcing method~\cite{Feng200520} and the modified bounce-back rule proposed by 
Niu et al.~\cite{Niu2006173}. Comparing with the former two, the third scheme is more efficient due to the 
fact that the forcing term is simply evaluated based on the momentum exchange method. This scheme was thereafter tested in 
the Drafting-Kissing-Tumbling (DKT) problem~\cite{Niu2006173} and simulation of several particulate 
systems~\cite{wu2010}. However, 
neither of these two studies have considered more than two solid particles because the inter-particle collisions were roughly 
treated by the Lennard-Jones potential equation. In our previous work~\cite{ZHANGCAF2014}, we reported a combined 
LBM-IBM-DEM method based on the scheme of Niu et al.~\cite{Niu2006173}. The new combined strategy was employed to simulate 
the dynamic process of sedimentation of 504 particles in a two-dimensional square cavity. The advantage of this strategy is 
that the particle-particle interaction rules are governed by theoretical contact mechanics. Therefore, it has great potential 
to be a promising method since no artificial parameters are required during the calculation of both fluid-particle and 
particle-particle interaction forces. However, the major weakness of that is the low computational efficiency. In the 
combined LBM-IBM-DEM method, one solid particle is represented by a set of small Lagrangian points. The interaction between 
the Lagrangian points and fluid lattice nodes provides detailed information of hydrodynamic behaviors. On the other hand, 
the force and torque exerted on the particles is a summation action over all the relevant lattice nodes nearby. It is 
pointed out by Yu and Xu~\cite{yu2003particle} that the difficulty in particle-fluid flow modeling is mainly related to solid 
phase rather than fluid phase. A proper simplification on the fluid phase can be therefore 
tolerated especially when treating a system where inter-particle collisions  
dominate~\cite{Tan2012137,Zhang2012467,zhang2015powderEffect}. For this reason, we further proposed a Particulate 
Immersed Boundary Method (PIBM)~\cite{Zhang20151PIBM} to improve the computational efficiency of the original LBM-IBM-DEM 
scheme. The basic idea of the PIBM is to remove the constraints between the Lagrangian points and thus each Lagrangian point 
is treated as one single solid particle. Different to the conventional 
LBM-DEM based simulations~\cite{Han20071080,cui2014coupled}, the size of solid particles is allowed to be less than one fluid 
lattice in the PIBM. By doing so, detailed geometry of the solid particles is absent in the coupling whereas much more superior 
computational convenience is brought. The novel LBM-PIBM-DEM scheme has been successfully applied to three-dimensional 
simulation of sedimentation process involving $8125$ solid particles on a single CPU. In this study, we adopt the 
LBM-PIBM-DEM scheme to solve the thermal interactions between spherical particles and fluid. The LBM is adopted to solve 
the fluid flow and temperature fields, the PIBM is responsible for the no-slip velocity and temperature boundary conditions 
at the particle surface and the kinematics and trajectory of the solid particles are evaluated by the DEM. To the best 
knowledge of the authors, no relevant research has been reported before. 

However, this is not the first attempt of using LBM or DEM to analyze the heat transfer phenomenon in a fluid-solid 
interaction system. Since He et al.~\cite{he1998novel} proposed a dual-LBM approach which uses a density distribution 
function to simulate hydrodynamics meanwhile another temperature distribution function to simulate thermodynamics for 
heat transfer. The methodology has been widely adopted by various researchers 
(For example:~\cite{shu2005lattice,han2008modelling,hu2013natural,hu2015study} and others). It is worthwhile mentioning that 
Han et al.~\cite{han2008modelling} proposed a numerical approach to account for the
thermal contact resistance between contacting surfaces of fluid and solid. They introduced a numerical case of a 
two-dimensional thermal cavity filled with solid particles to investigate the heat convection and conduction in the 
particulate system. The solid particles in the work of Han et al.~\cite{han2008modelling} were keeping stagnant and 
thus briefly played a role to construct complex solid boundary structures to test their model. 
Feng et al.~\cite{feng2008discrete,feng2009discrete} proposed a Discrete Thermal Element Method (DTEM) to model the heat 
transfer in the systems involving a large number of circular particles. Again, these work mainly focused on the heat 
conduction between solid-solid or solid-fluid whereas the dynamic behavior of the solid particles were ignored. It is 
interesting to study the thermal convection of particulate flow in a fluid with intense inter-particle 
collisions
~\cite{gan2003direct,feng2009heat,zhou2009particle, zhou2010new,hou2012computational1,hou2012computational2,Feng201462}. 
However, none of these work was 
conducted based on LBM nor in three dimensional due to the enormous computational cost. In this study, 
we use the PIBM to conquer the limit.

The remainder of the paper is organized as follows. To make this paper self-contained, the mathematics of the 
three-dimensional LBM, PIBM and DEM were briefly introduced in Section~\ref{LBMPIBMDEM}. In Section~\ref{Numericalresults}, 
case studies of (1) Natural convection in a two-dimensional square cavity with a concentric annulus, (2) Sedimentation of 
two-dimensional isothermal particles in fluid, (3) Sedimentation of three-dimensional isothermal particles in 
fluid and (4) Sedimentation of three-dimensional thermosensitive particles in fluid were presented. The 
numerical results were discussed. Finally, conclusions were given in Section~\ref{conclusion}.

\section{Governing equations}\label{LBMPIBMDEM}

\subsection{Lattice Boltzmann model with single-relaxation time collision}

 \begin{figure}
 \centering
 \includegraphics[width=0.5\textwidth]{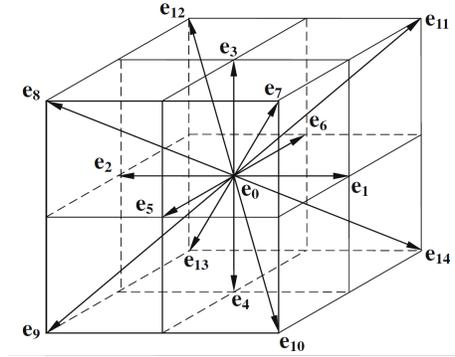}
 \vskip-0.2cm
 \caption{Schematic diagram of the D3Q15 model\cite{Wu20105022}.} \label{PIBMsch}
 \end{figure}

We consider the simulation of the incompressible Newtonian fluids where the LBM-D2Q9 and D3Q15 models~\cite{qian1992lattice} 
are adopted for the two- and three-dimensional calculations, respectively. In this section, the equation systems of the 
D3Q15 model are presented. For the two-dimensional ones, the readers are referred to our previous study~\cite{ZHANGCAF2014}. 
The three-dimensional spatial distribution of the fluid velocities is shown in Figure~\ref{PIBMsch} which can be expressed by 

\begin{eqnarray}
e_{\alpha} = 
\left\{
\begin{array}{ll}
   (0,0,0)c & \alpha=0 \\
   (\pm 1,0,0)c, (0,\pm 1,0)c, (0,0,\pm 1)c & \alpha=1-6  \\
   (\pm 1,\pm 1,\pm 1)c & \alpha = 7-14 
\end{array}
\right. 
\end{eqnarray}

\noindent where $c$ is termed by the lattice speed. The formulation of the lattice Bhatnagar-Gross-Krook model is

\begin{equation}\label{lbm2}
f_{\alpha}(r+e_{\alpha}\delta_{t},t+\delta_{t})=f_{\alpha}(r,t)-\frac{f_{\alpha}(r,t)-f_{\alpha}^{eq}(r,t)}{\tau_{f}}+F_{\alpha}\delta_{t}
\end{equation}
\begin{equation}\label{lbm2.5}
g_{\alpha}(r+e_{\alpha}\delta_{t},t+\delta_{t})=g_{\alpha}(r,t)-\frac{g_{\alpha}(r,t)-g_{\alpha}^{eq}(r,t)}{\tau_{g}}+G_{\alpha}\delta_{t}
\end{equation}

\noindent where $f_{\alpha}(r,t)$ and $g_{\alpha}(r,t)$ represent the fluid density and temperature distribution functions, 
respectively, $r=(x,y,z)$ stands for the space position vector, $t$ denotes time, and $\tau_{f}$ and $\tau_{g}$ denote the 
non-dimensional relaxation times which can be calculated by~\cite{he1998novel}

\begin{equation}\label{lbm2.6}
\tau_{f}=  \dfrac{\sqrt{Pr}L_{c}u_{c}}{\sqrt{Ra}c_{s}^{2}\delta_{t}}+0.5
\end{equation}
\begin{equation}\label{lbm2.7}
\tau_{g}=  \dfrac{L_{c}u_{c}}{\sqrt{PrRa}c_{s}^{2}\delta_{t}}+0.5
\end{equation}

\noindent where $c_{s}=c/\sqrt{3}$ is the lattice speed of sound, $L_{c}$ and $u_{c}=\sqrt{g \beta L_{c} \Delta T}$ are the 
characteristic length and velocity, respectively, and $Pr$ and $Ra$ are the Prandtl and Rayleigh numbers, respectively. 

\begin{eqnarray}
Pr=\frac{\nu_{f}}{\alpha_{f}}
\end{eqnarray}

\begin{eqnarray}
Ra=\frac{g \beta \Delta T L_{c}^{3}}{\alpha_{f} \nu_{f}}
\end{eqnarray}

\noindent where $\nu_{f}$ is the kinematic viscosity of fluid, $\alpha_{f}$ is the thermal diffusivity, 
 $g$ is the gravity, and $\beta$ is the thermal expansion coefficient.
 The source terms, $F_{\alpha}\delta_{t}$ and $G_{\alpha}\delta_{t}$, in Equations~\ref{lbm2} and ~\ref{lbm2.5} are given 
 in Section~\ref{GOPIBM}. The equilibrium density and temperature distribution functions, $f_{\alpha}^{eq}(r,t)$ 
 and $g_{\alpha}^{eq}(r,t)$, can be written as

\begin{equation}\label{lbm3}
f_{\alpha}^{eq}(r,t)= \rho_{f} \omega_{\alpha} [1+3(e_{\alpha}\cdot u)+\frac{9}{2}(e_{\alpha}\cdot u)^{2}-\frac{3}{2}\mid u \mid^{2}]
\end{equation}

\begin{equation}\label{lbm3.5}
g_{\alpha}^{eq}(r,t)= T \omega_{\alpha} [1+3(e_{\alpha}\cdot u)+\frac{9}{2}(e_{\alpha}\cdot u)^{2}-\frac{3}{2}\mid u \mid^{2}]
\end{equation}

\noindent where the value of weights are: $\omega_{0}=2/9$, $\omega_{\alpha}=1/9$ for $\alpha=1-6$ and 
$\omega_{\alpha}=1/72$ for $\alpha=7-14$. $u$
denotes the macro velocity at each lattice node which can be calculated by 
$u=(\sum\limits_{\alpha=0}^{14}f_{\alpha}e_{\alpha}+\dfrac{1}{2}(\Delta T(\dfrac{u_{c}}{c})^{2}+F_{B})\delta_{t})/\rho_{f}$, 
the macro fluid density is $\rho_{f}=\sum\limits_{\alpha=0}^{14}f_{\alpha}$ and the macro temperature can be 
calculated by $T=\sum\limits_{\alpha=0}^{14}g_{\alpha}e_{\alpha}+\dfrac{1}{2}Q_{B}\delta_{t}$.  
The discrepancy of the formulations of $u$ and $T$ here with the conventional ones is due to the fact that they are 
modified by the momentum and heat flux~\cite{Niu2006173}. For example, $F_{B}$ and $Q_{B}$ stand for the body force and heat source, 
respectively. $\Delta T(\dfrac{u_{c}}{c})^{2}$ represents the non-dimensional buoyancy caused by temperature 
gradients~\cite{hu2015study}. 

\subsection{PIBM}\label{GOPIBM}

 \begin{figure}
 \centering
 \includegraphics[width=0.54\textwidth]{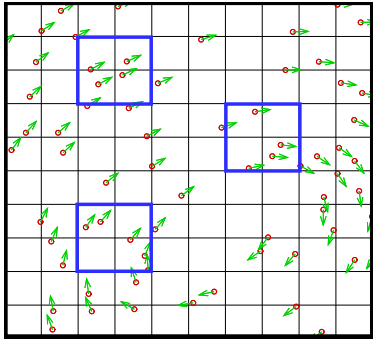}
 \vskip-0.2cm
 \caption{Schematic diagram of the PIBM.} \label{PIBM}
 \end{figure}

 It is worthwhile mentioning that the PIBMs for the no-slip velocity and temperature boundary 
conditions share a fairly similar logical manner. The common idea is to obtain an accurate expression of 
the velocity or temperature difference between the two phases and then use it to make further modification 
on the behavior of the fluid flow field and solid particles. In this study, the heat conduction between the 
solid particles or particle and wall is not considered.    
 For the sake of clarity, the two-dimensional schematic diagram of the PIBM is displayed with three-dimensional 
equation systems. As shown in Figure~\ref{PIBM}, the fluid is described using the Eulerian square lattices and the solid 
particles are denoted by the Lagrangian points moving in the flow field. The fluid density and temperature distribution 
functions on the solid particles are evaluated using the numerical extrapolation from the circumambient fluid points, 

\begin{equation}\label{lbm5}
f_{\alpha}(X_{l},t)=L(X_{l},r) \cdot f_{\alpha}(r,t)
\end{equation}

\begin{equation}\label{lbm5.5}
g_{\alpha}(X_{l},t)=L(X_{l},r) \cdot g_{\alpha}(r,t)
\end{equation}

\noindent where $X_{l}(X,Y,Z)$ is the coordinates of the solid particles, the subscript '$l$' denotes the variable at 
the location of the Lagrangian particles. $L(X_{l},r)$ is the three-dimensional polynomials, 

\begin{equation}\label{lbm6}
L(X_{l},r)=\sum\limits_{ijk}\left(\prod\limits_{l=1,l!=i}^{i_{max}}\frac{X-x_{ljk}}{x_{ijk}-x_{ljk}}\right)
\left(\prod\limits_{m=1,m!=j}^{j_{max}}\frac{Y-y_{imk}}{y_{ijk}-y_{imk}}\right)
\left(\prod\limits_{n=1,n!=k}^{k_{max}}\frac{Z-z_{ijn}}{z_{ijk}-z_{ijn}}\right)
\end{equation}

\noindent where $i_{max}$, $j_{max}$  and $k_{max}$ are the maximum numbers of the Eulerian points used 
in the extrapolation as shown by blocks in Figure~\ref{PIBM}. With the movement of the solid particle, 
$f_{\alpha}(X_{l},t)$ 
will be further affected by the particle velocity, $U_{p}$,

\begin{equation}\label{lbm7}
f_{\beta}(X_{l},t+\delta_{t})=f_{\alpha}(X_{l},t)-2\omega_{\alpha}\rho_{f}\frac{e_{\alpha}U_{p}}{c_{s}^{2}}
\end{equation}

\noindent and $g_{\alpha}(X_{l},t)$ 
will be further affected by the temperature difference between the solid particle and fluid, $\Delta T$~\cite{hu2015study},

\begin{equation}\label{lbm7.5}
g_{\beta}(X_{l},t+\delta_{t})=g_{\alpha}(X_{l},t)-2\omega_{\alpha}\Delta T\dfrac{h}{\delta_{t}}
\end{equation}

\noindent in Equations~\ref{lbm7} and ~\ref{lbm7.5}, the subscript $\beta$ represents the opposite direction of 
$\alpha$, and $h$ is the mesh spacing. Based on the momentum exchange 
between fluid and particles, the force density, $F_{f}(X_{l},t)$, at each solid particle can be calculated 
using $f_{\alpha}$ and $f_{\beta}$,

\begin{equation}\label{lbm8}
F_{f}(X_{l},t)=\sum\limits_{\beta}e_{\beta}[f_{\beta}(X_{l},t)-f_{\alpha}(X_{l},t)]
\end{equation}

\noindent The effect on the flow fields from the solid boundary is the body force term $F_{\alpha}\delta_{t}$ in 
Equation~\ref{lbm2}, $F_{\alpha}$ can be expressed by

\begin{equation}\label{lbm9}
F_{\alpha}=\left(1-\frac{1}{2\tau_{f}}\right)\omega_{\alpha}\left(3\frac{e_{\alpha}-u}{c^{2}}+9\frac{e_{\alpha}\cdot u}{c^{4}}e_{\alpha}\right)F_{B}(r,t)
\end{equation}

\noindent where

\begin{equation}\label{lbm10}
F_{B}(r,t)=\sum\limits_{l}F_{f}(X_{l},t)D_{ijk}(r_{ijk}-X_{l})A_{p}
\end{equation}

\noindent The heat source $G_{\alpha}\delta_{t}$ in Equation~\ref{lbm2.5} is one dimensional and can thus be directly given 
as  

\begin{equation}\label{lbm12.5}
G_{\alpha}=\left(1-\frac{1}{2\tau_{g}}\right)\omega_{\alpha}Q_{B}
\end{equation}

\noindent where

\begin{equation}\label{lbm12.6}
Q_{B}=\dfrac{u_{c}}{c\sqrt{PrRa}}(\sum\limits_{l}2\Delta T\dfrac{h}{\delta_{t}}D_{ijk}(r_{ijk}-X_{l})A_{p})
\end{equation}

\noindent in Equations~\ref{lbm10} and ~\ref{lbm12.6}, $A_{p}$ is the cross-sectional area of the particle which is given 
as $A_{p}=0.25\pi d_{p}^{2}$, $d_{p}$ is the diameter of the particle. $D_{ijk}$ is used to restrict the feedback force to 
only take effect on the lattice nodes close to the solid particle and is given by

\begin{equation}\label{lbm11}
D_{ijk}(r_{ijk}-X_{l})=\frac{1}{h^{3}}\delta_{h}\left(\frac{x_{ijk}-X_{l}}{h}\right)\delta_{h}\left(\frac{y_{ijk}-Y_{l}}{h}\right)\delta_{h}\left(\frac{z_{ijk}-Z_{l}}{h}\right)
\end{equation}

\noindent where

\begin{equation}\label{lbm12}
  \delta_{h}(a)= \left\{
   \begin{array}{cc}
   \frac{1}{4}(1+cos(\frac{\pi a}{2})), &  when \mid a \mid \leq 2  \\
   0, & otherwise \\
   \end{array}
   \right.
  \end{equation}
On the other hand, the fluid-solid interaction force exerted on the solid particle can
be obtained as the reaction force of $F_{f}(X_{l},t)$,

\begin{equation}\label{lbm13}
F_{fpi}=-F_{f}(X_{l},t) A_{p}
\end{equation}

\begin{figure}
\centering
\includegraphics[angle=  -0,width=0.48\textwidth]{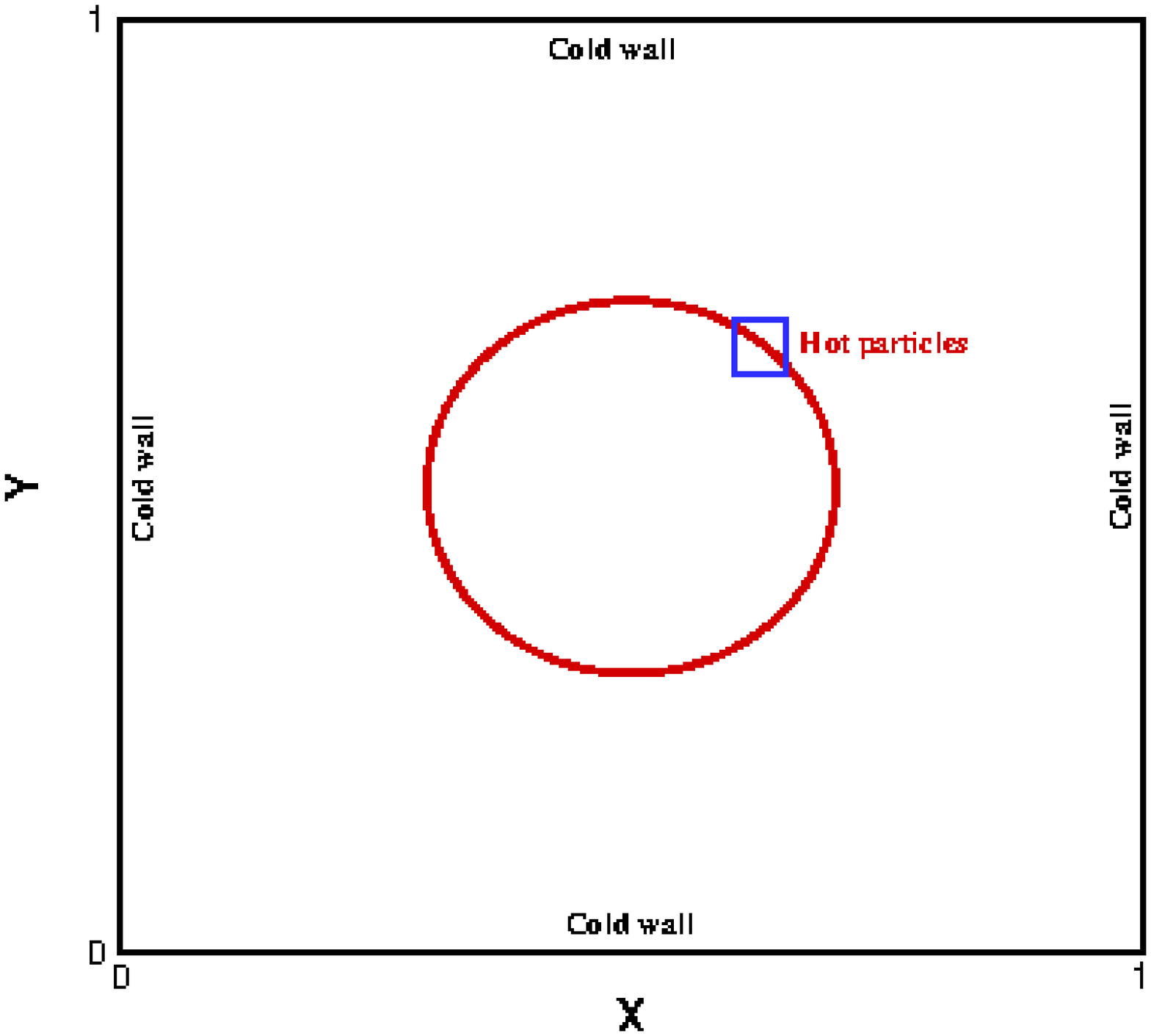}
\hspace{0mm}
\includegraphics[angle=  -0,width=0.48\textwidth]{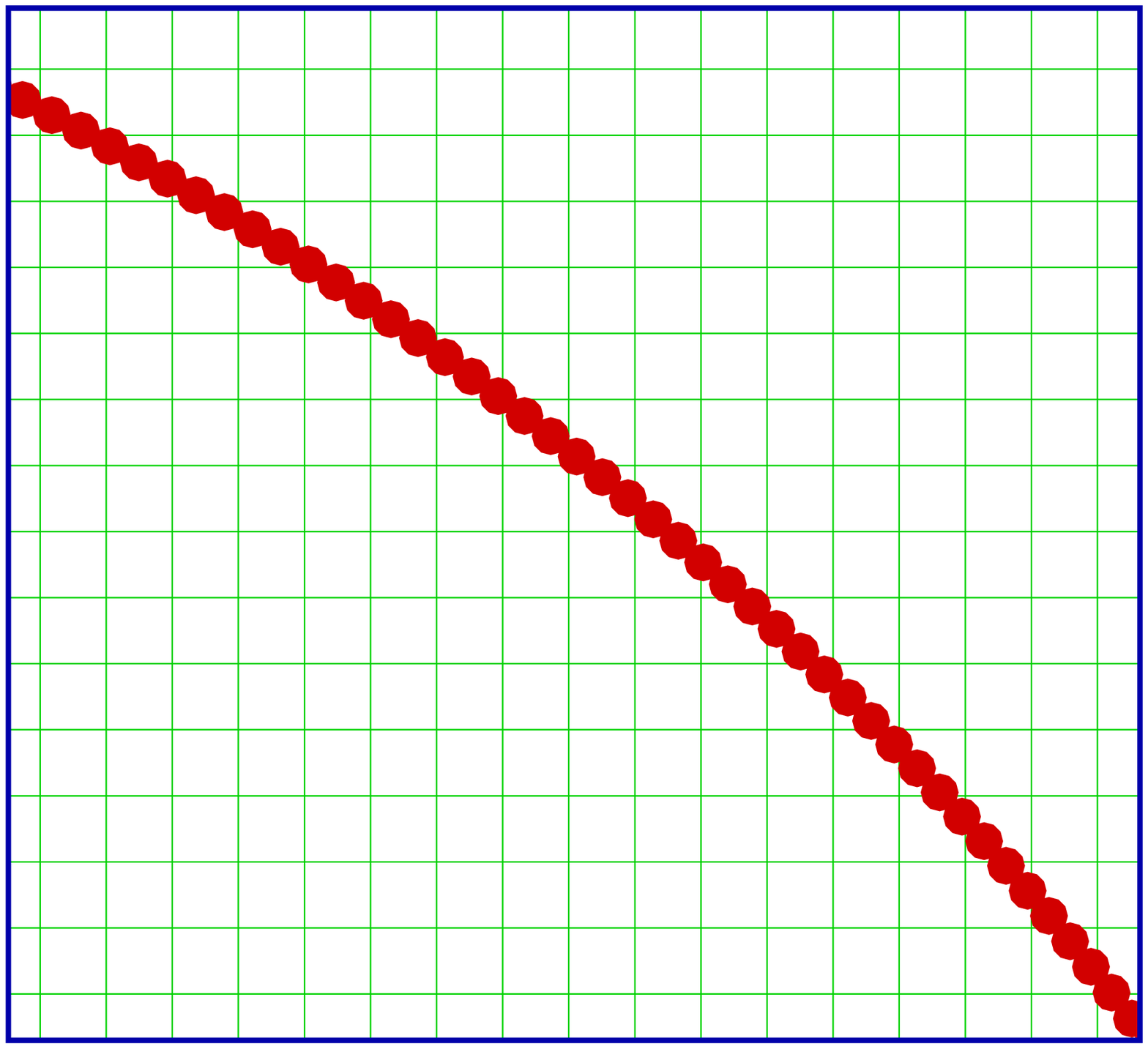}
\\
(a) \hspace{52mm} (b)
\caption{The schematic diagram of (a) square cavity with a concentric annulus and (b) local zoomed view.}
\label{schematicconcentricannulus}
\end{figure}

\begin{figure}[!ht]
\centering
\includegraphics[angle=  0,width=0.32\textwidth]{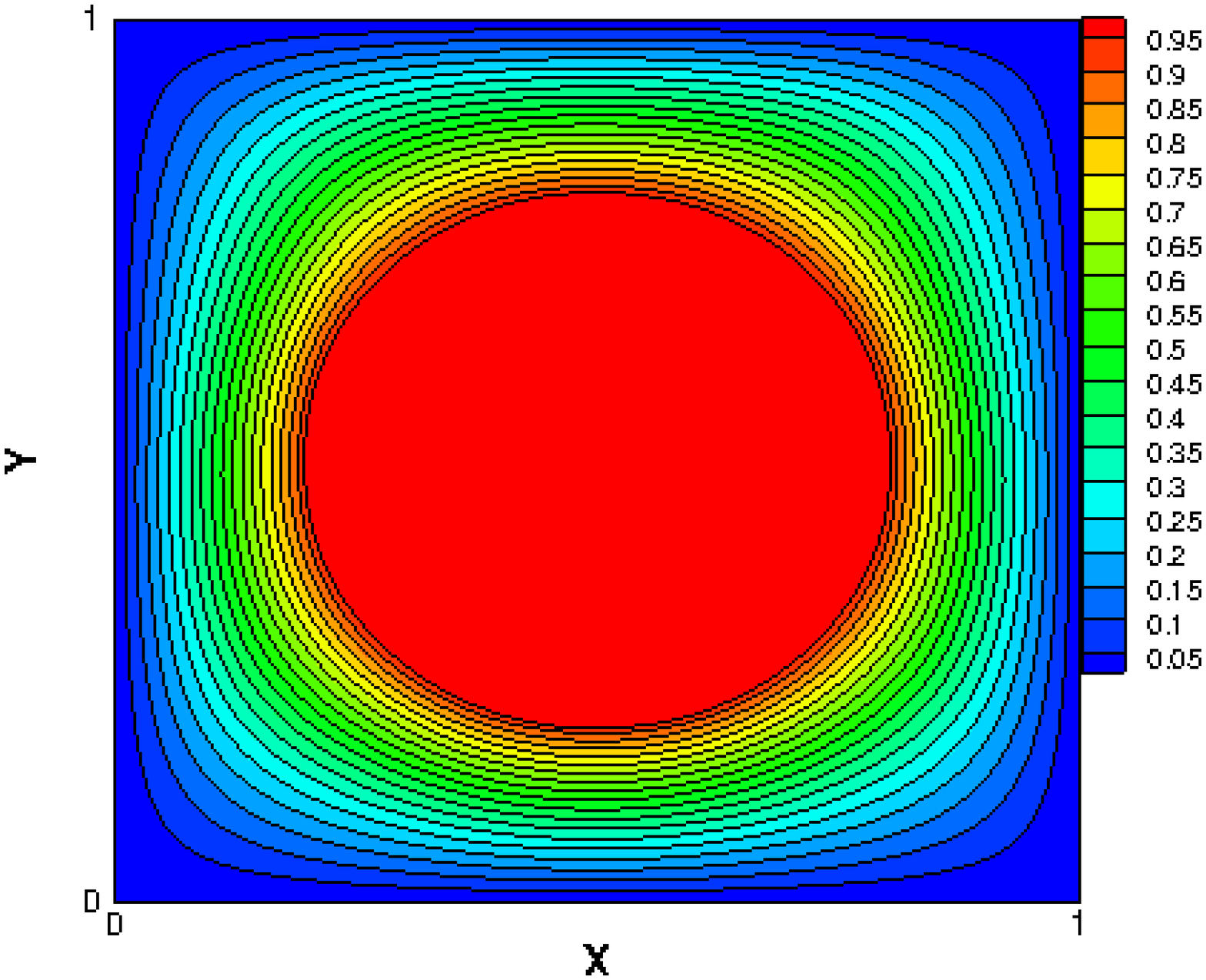}
\hspace{-2mm}
\includegraphics[angle=  0,width=0.32\textwidth]{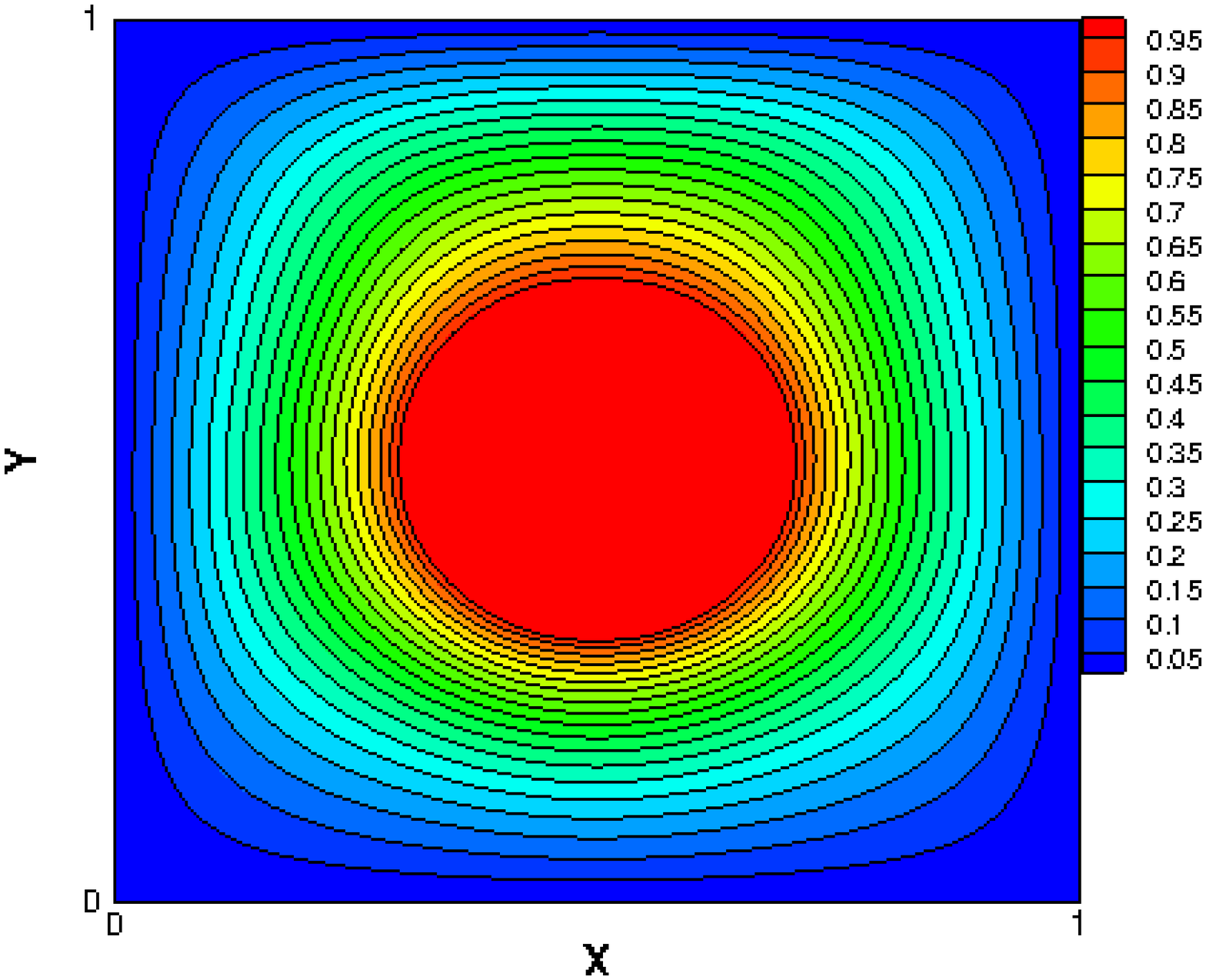}
\hspace{-2mm}
\includegraphics[angle=  0,width=0.32\textwidth]{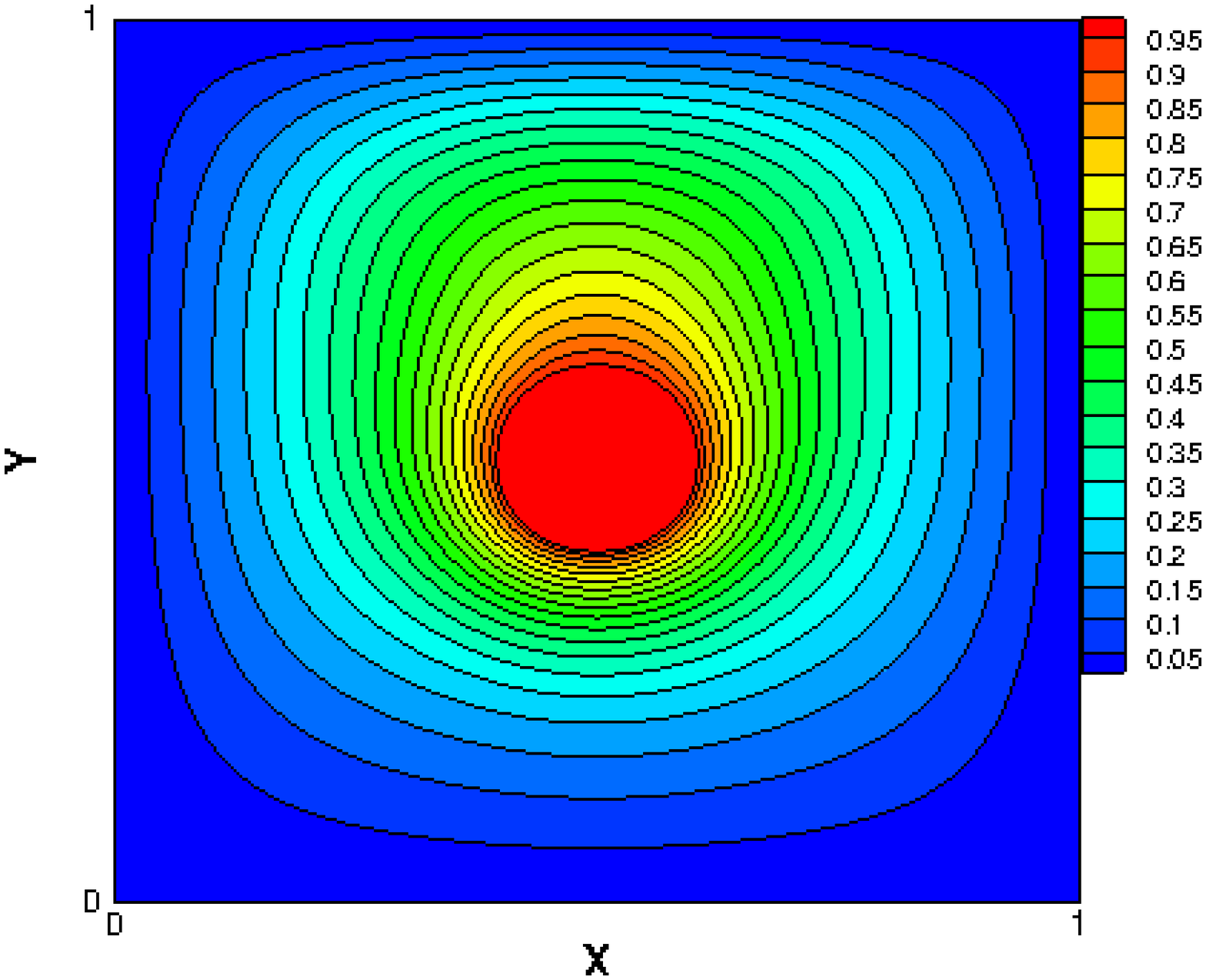}
\\
\includegraphics[angle=  0,width=0.32\textwidth]{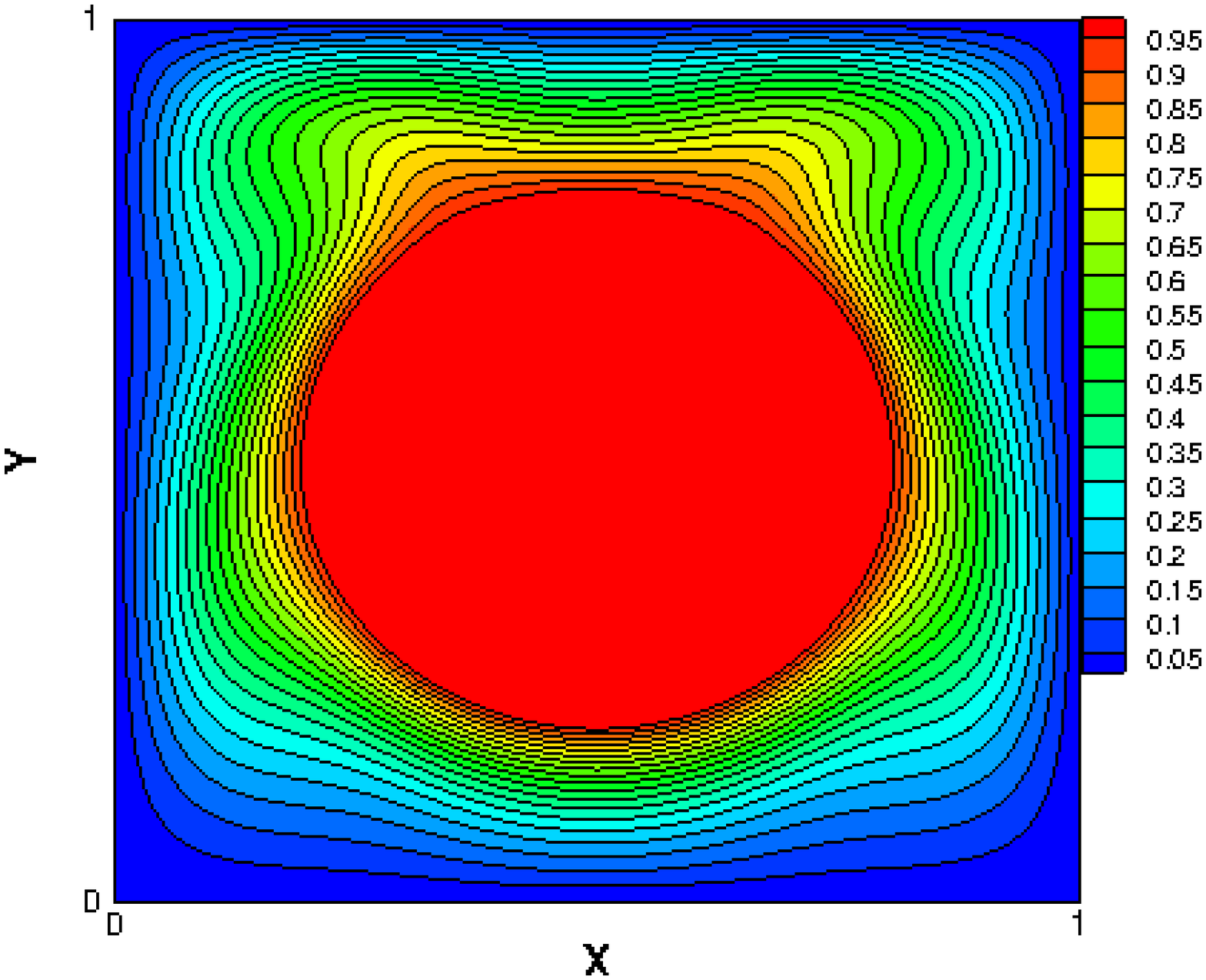}
\hspace{-2mm}
\includegraphics[angle=  0,width=0.32\textwidth]{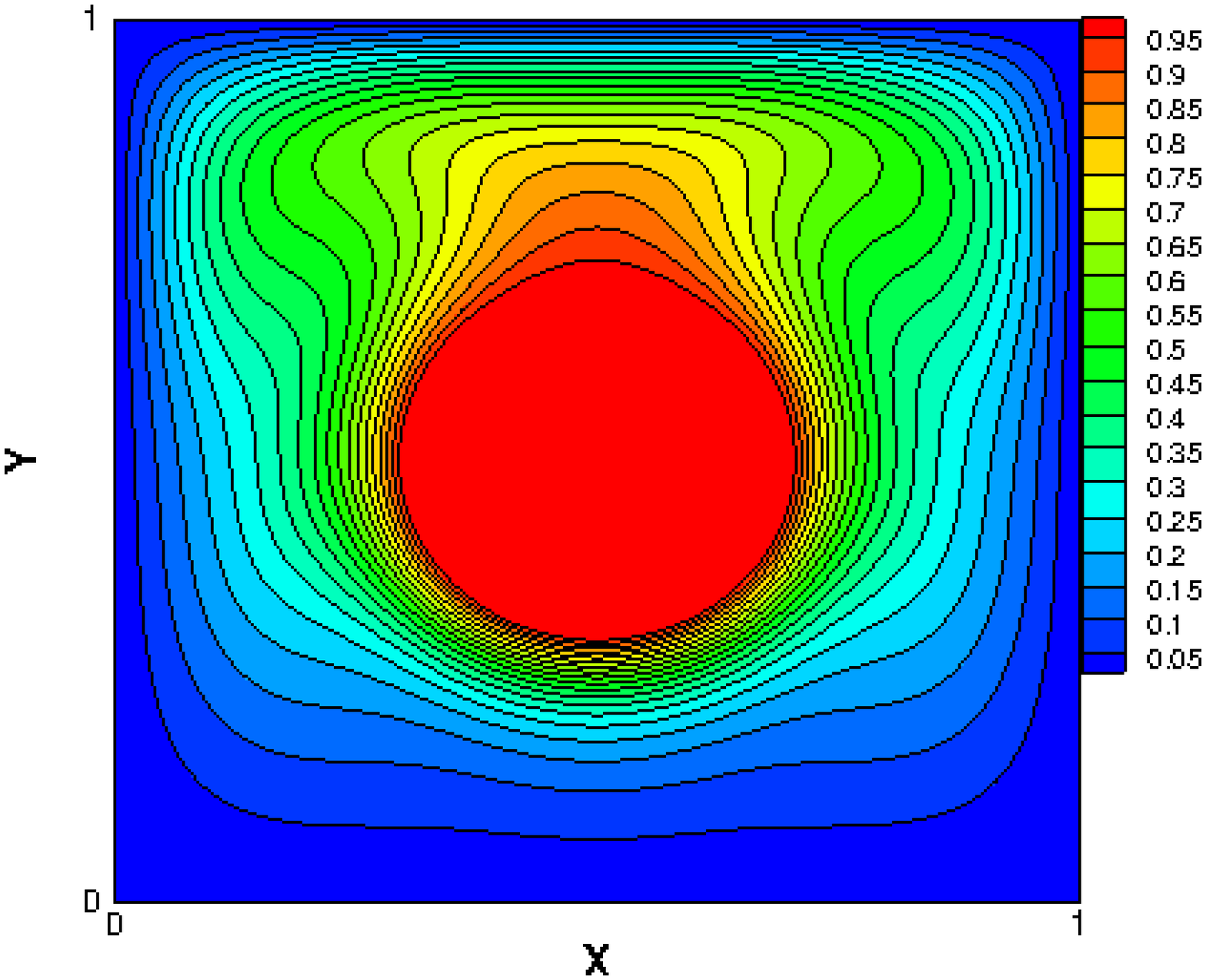}
\hspace{-2mm}
\includegraphics[angle=  0,width=0.32\textwidth]{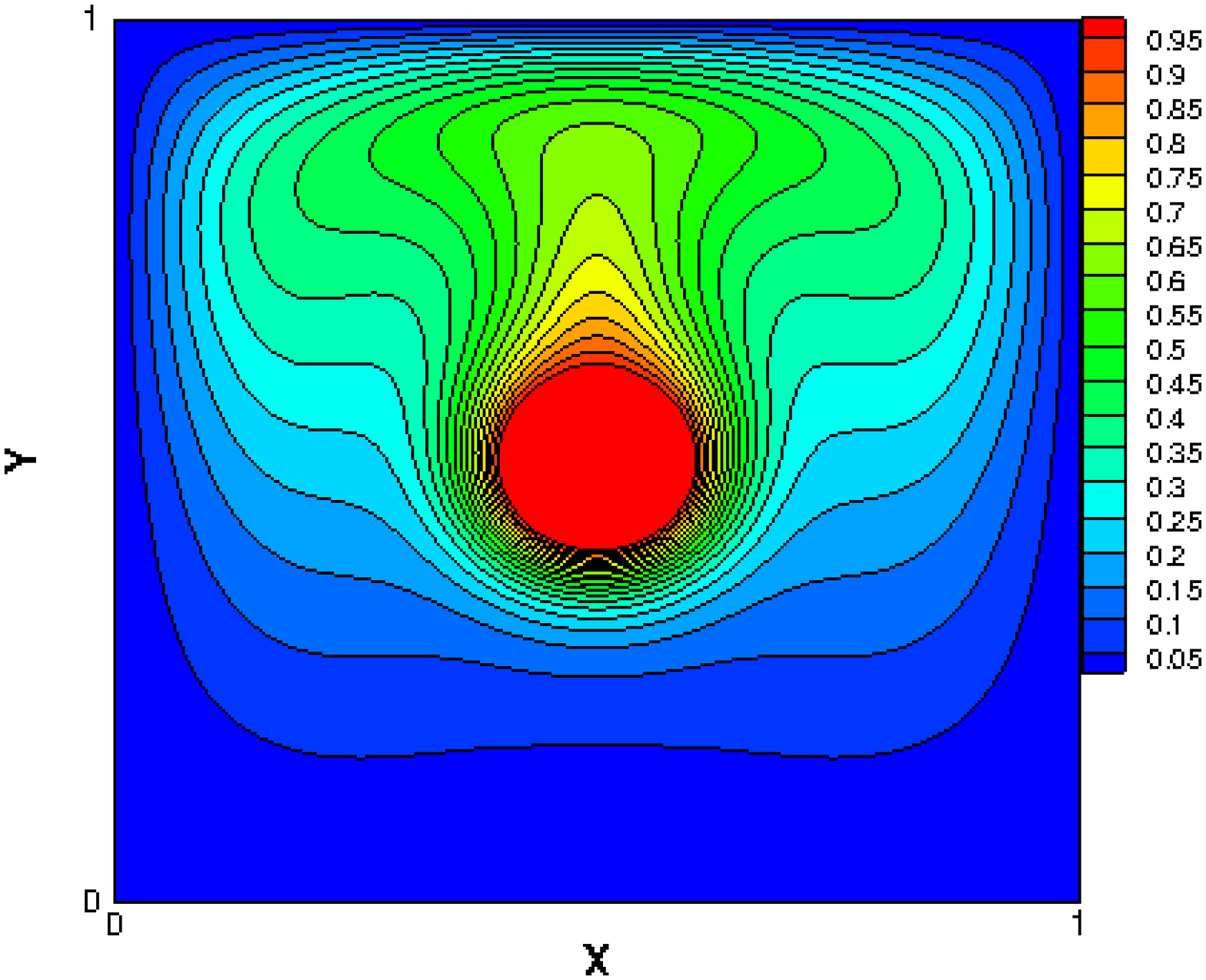}
\\
\includegraphics[angle=  0,width=0.32\textwidth]{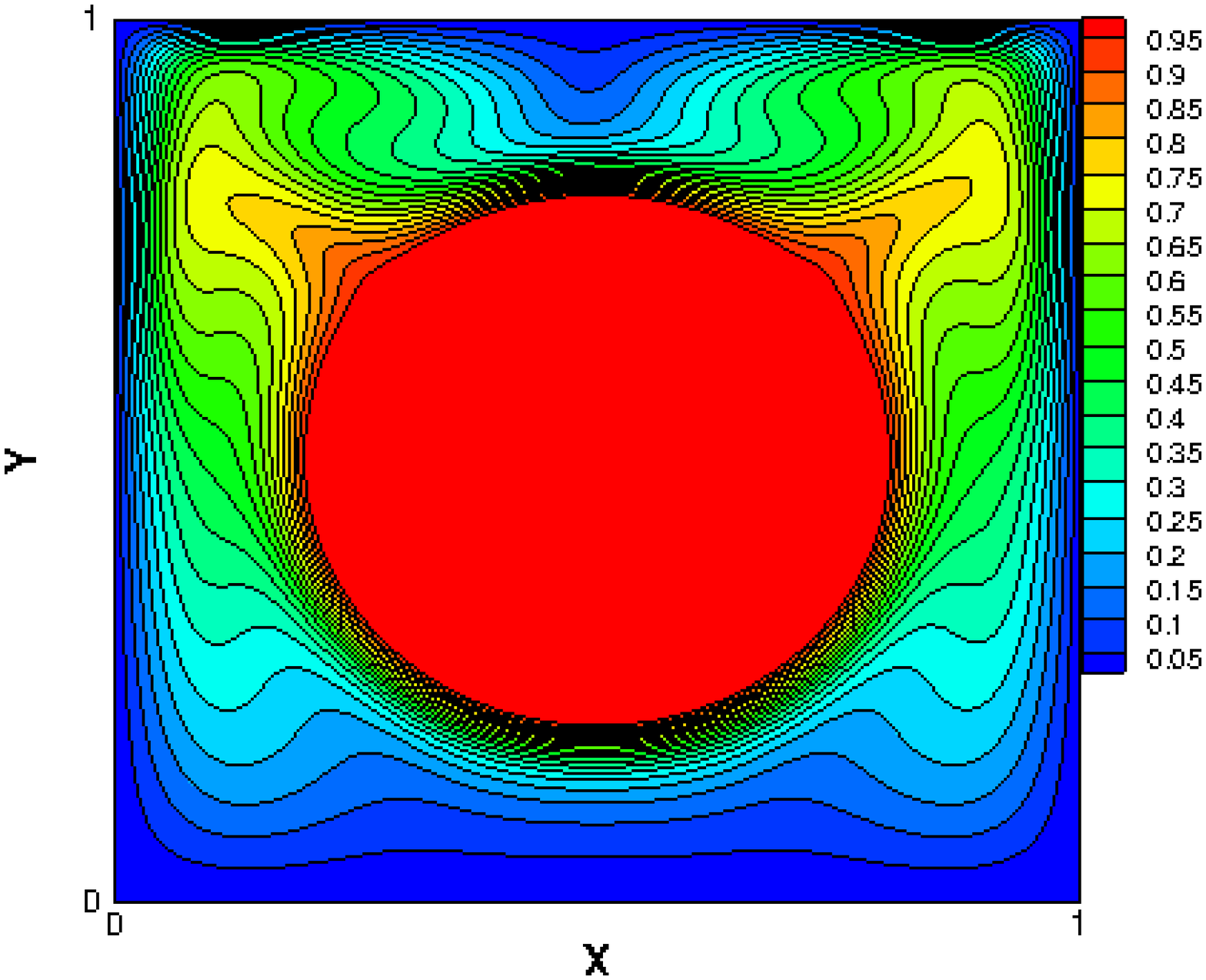}
\hspace{-2mm}
\includegraphics[angle=  0,width=0.32\textwidth]{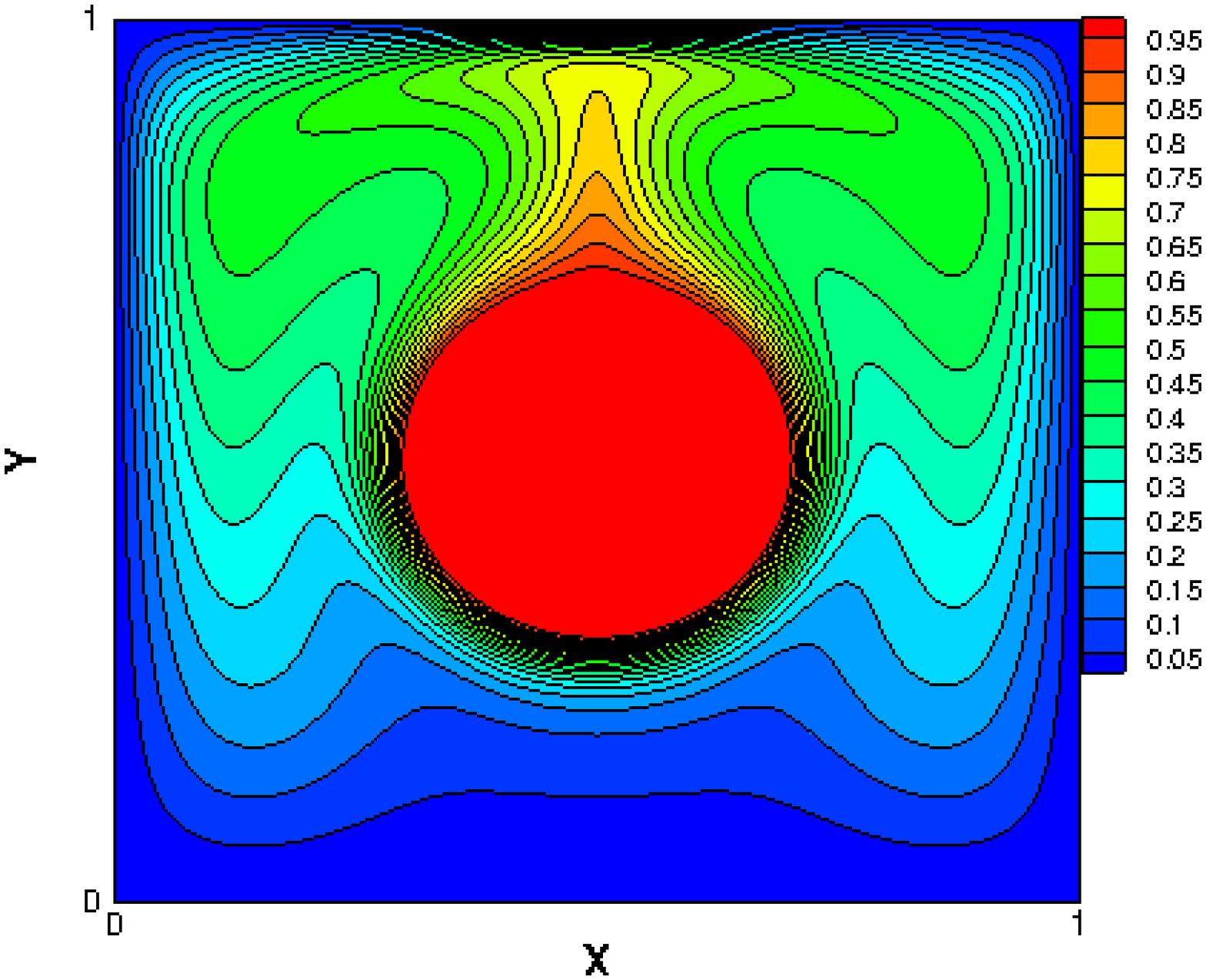}
\hspace{-2mm}
\includegraphics[angle=  0,width=0.32\textwidth]{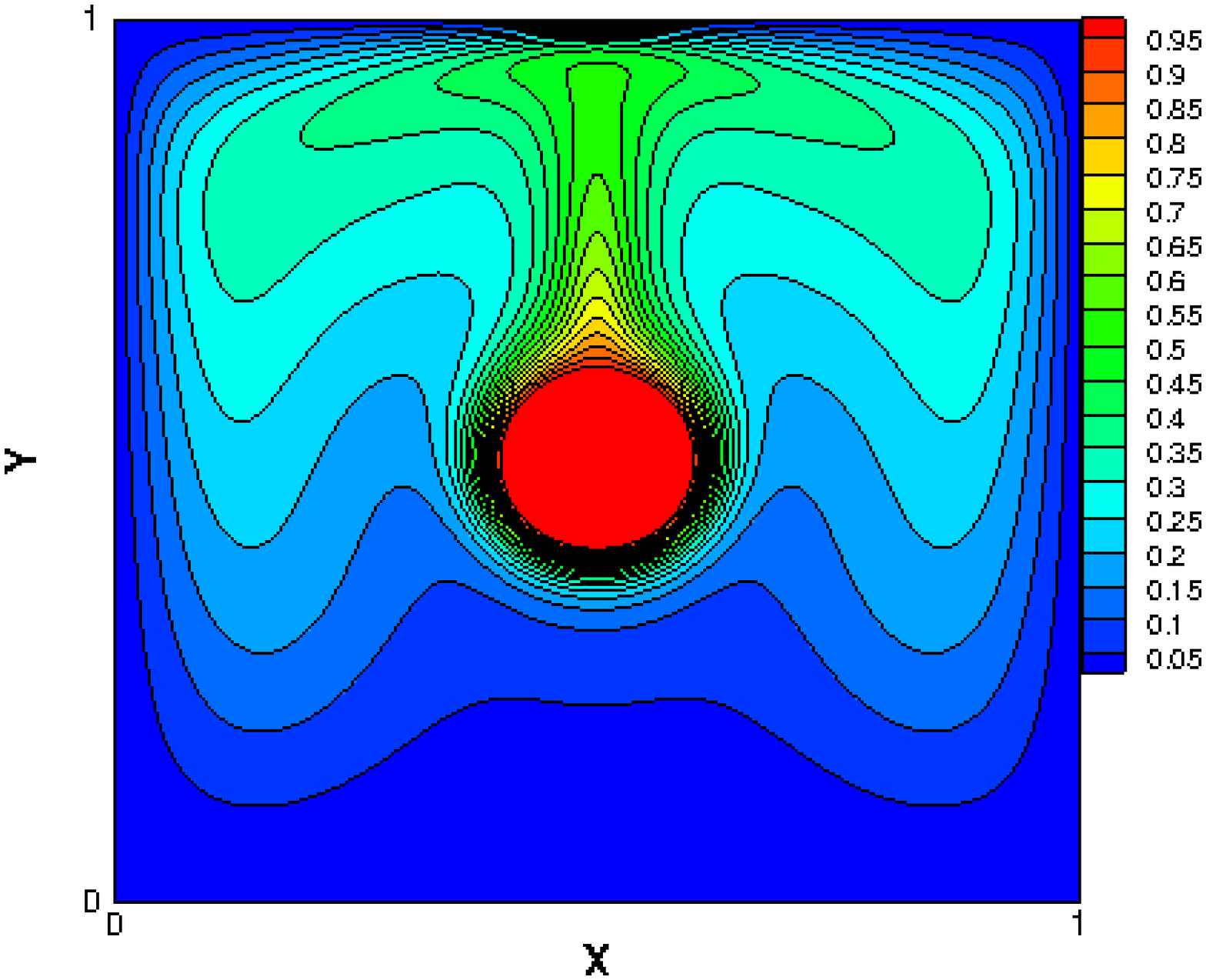}
\\
(a) \hspace{36mm} (b) \hspace{36mm} (c)
\caption{Isothermal contours of temperature distribution in the square cavity at different $Ra$ and $Ar$: 
(a)(top) $Ra=10^{4}$, $Ar=1.67$, (b)(top) $Ra=10^{4}$, $Ar=2.5$, (c)(top) $Ra=10^{4}$, $Ar=5$;
(b)(middle) $Ra=10^{5}$, $Ar=1.67$, (b)(middle) $Ra=10^{5}$, $Ar=2.5$, (b)(middle) $Ra=10^{5}$, $Ar=5$;
(c)(bottom) $Ra=10^{6}$, $Ar=1.67$, (c)(bottom) $Ra=10^{6}$, $Ar=2.5$, (c)(bottom) $Ra=10^{6}$, $Ar=5$.}
\label{annulustemperature}
\end{figure}

\begin{figure}[!ht]
\centering
\includegraphics[angle=  0,width=0.32\textwidth]{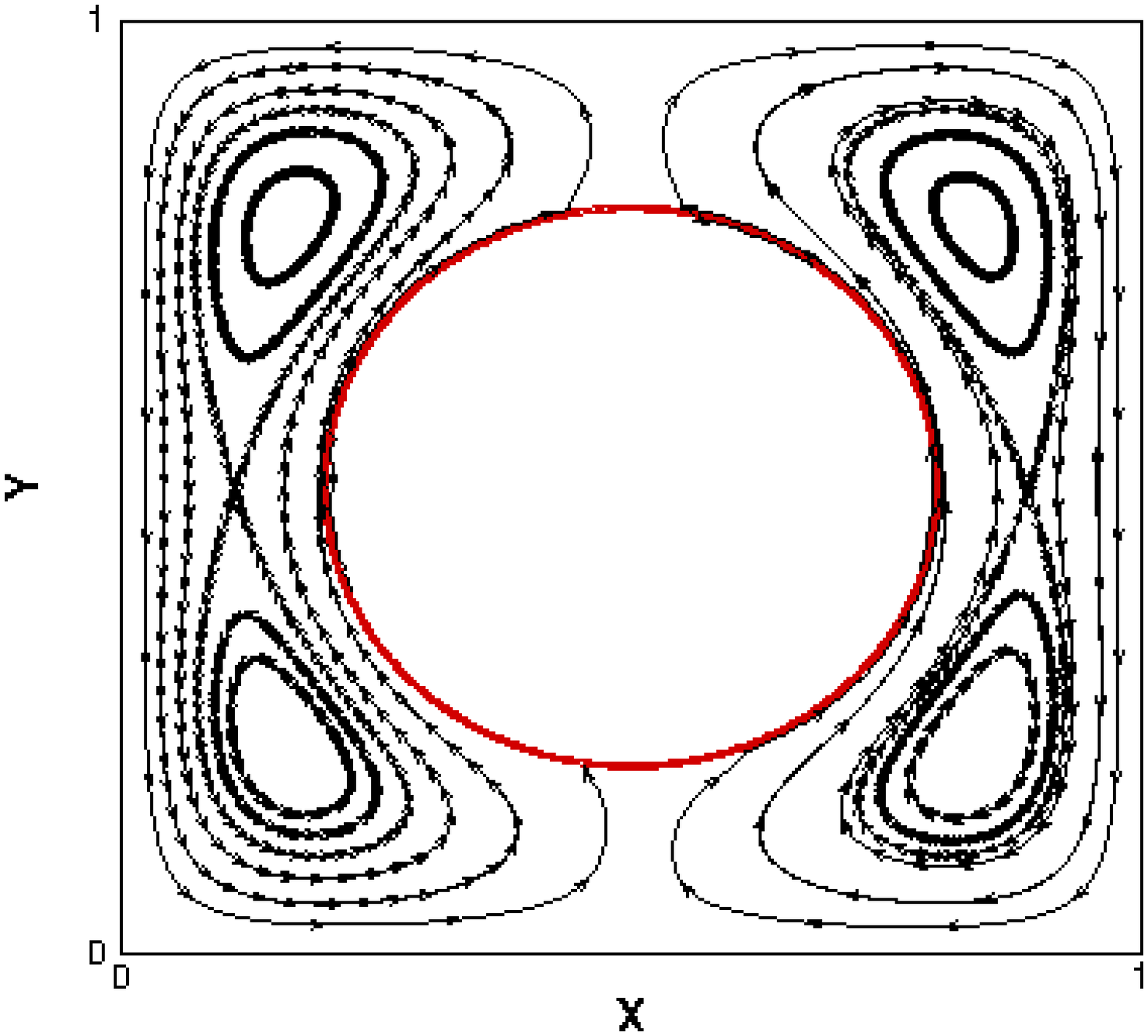}
\hspace{-2mm}
\includegraphics[angle=  0,width=0.32\textwidth]{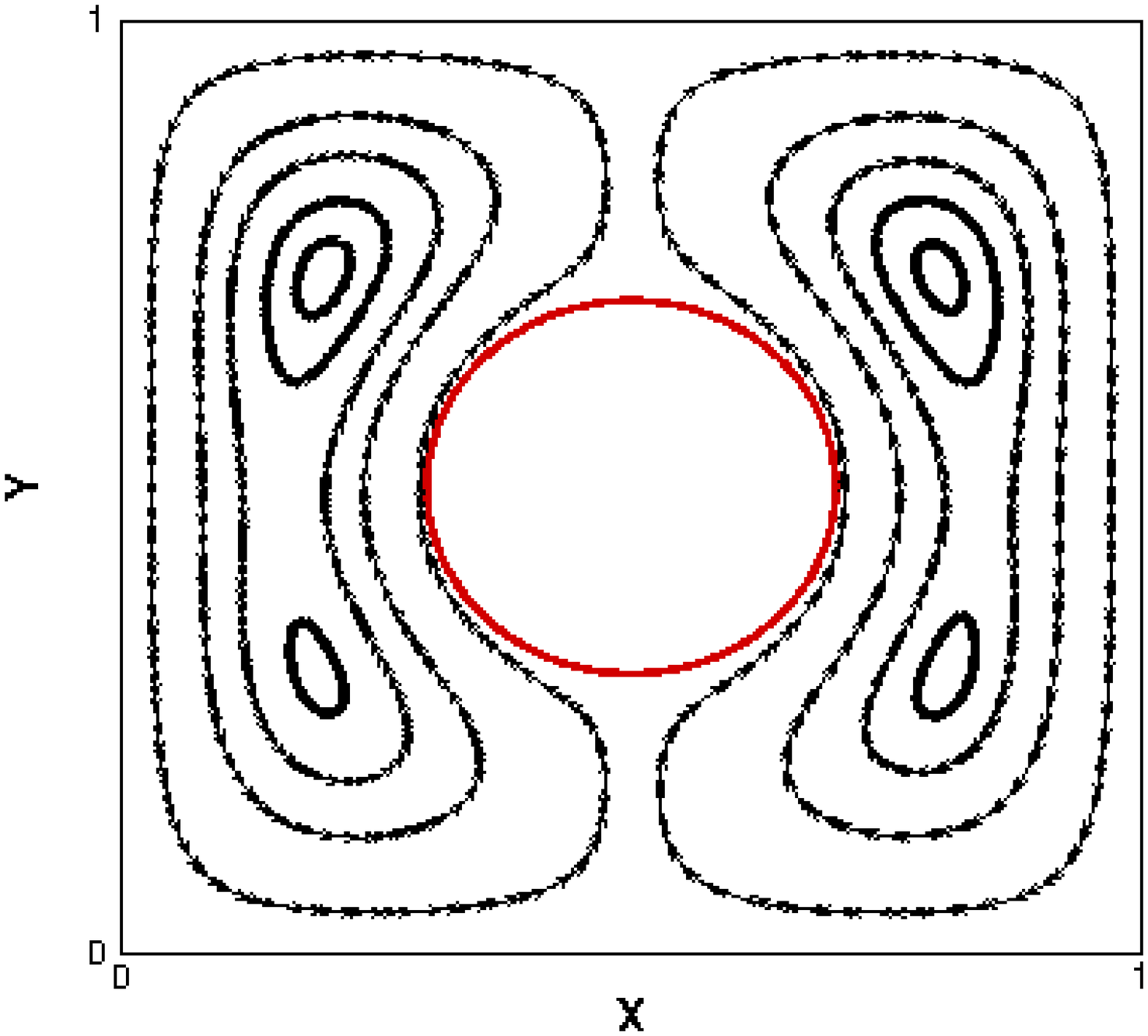}
\hspace{-2mm}
\includegraphics[angle=  0,width=0.32\textwidth]{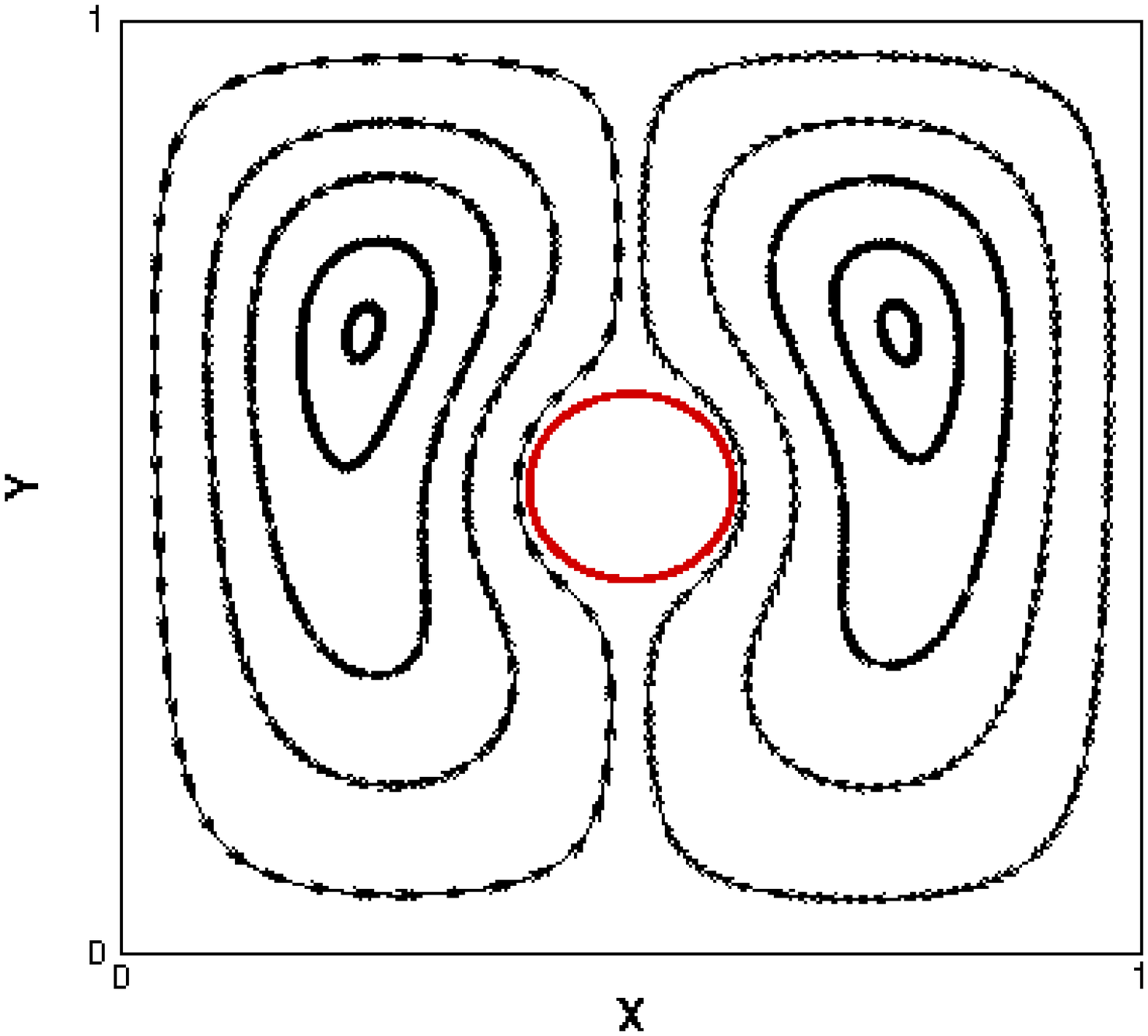}
\\
\includegraphics[angle=  0,width=0.32\textwidth]{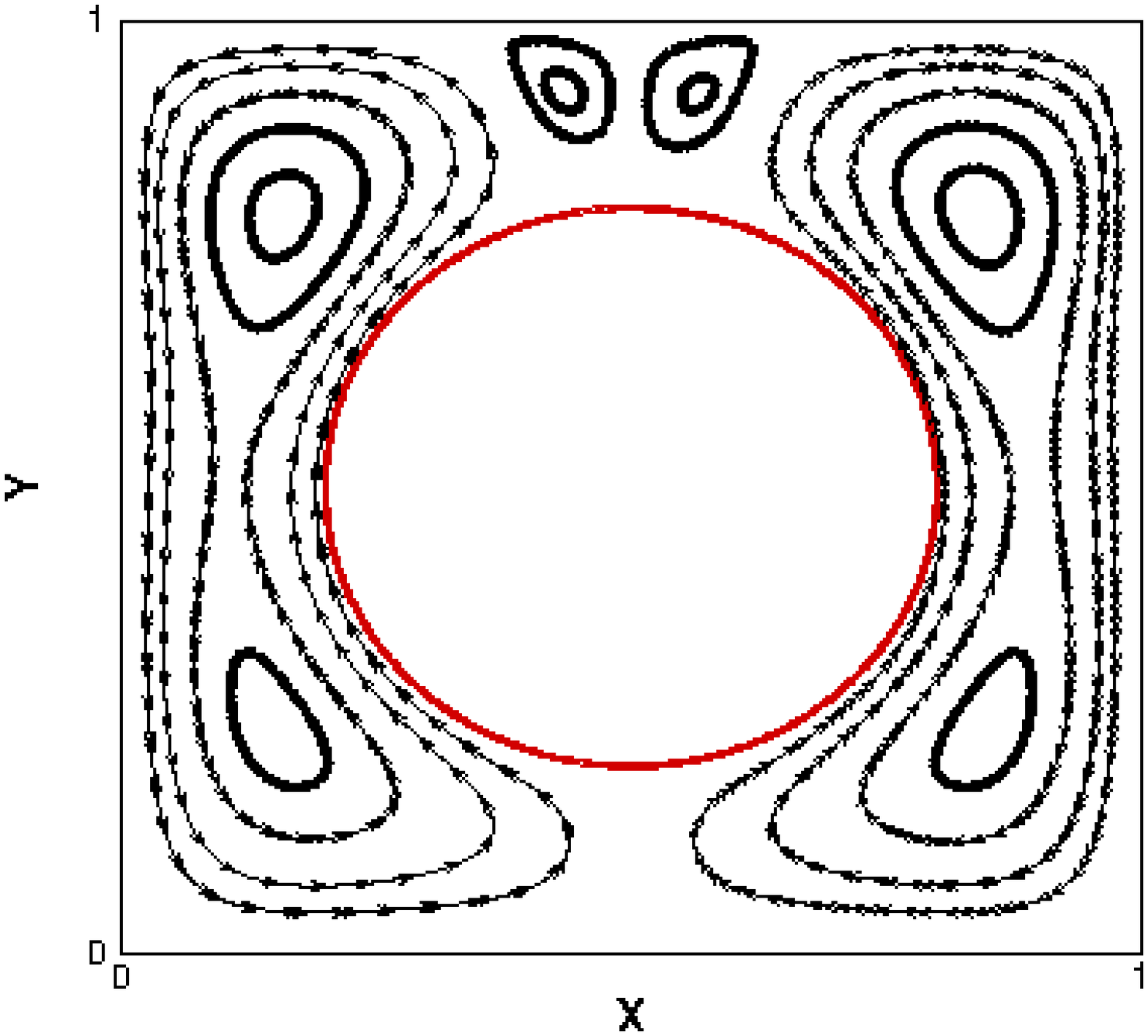}
\hspace{-2mm}
\includegraphics[angle=  0,width=0.32\textwidth]{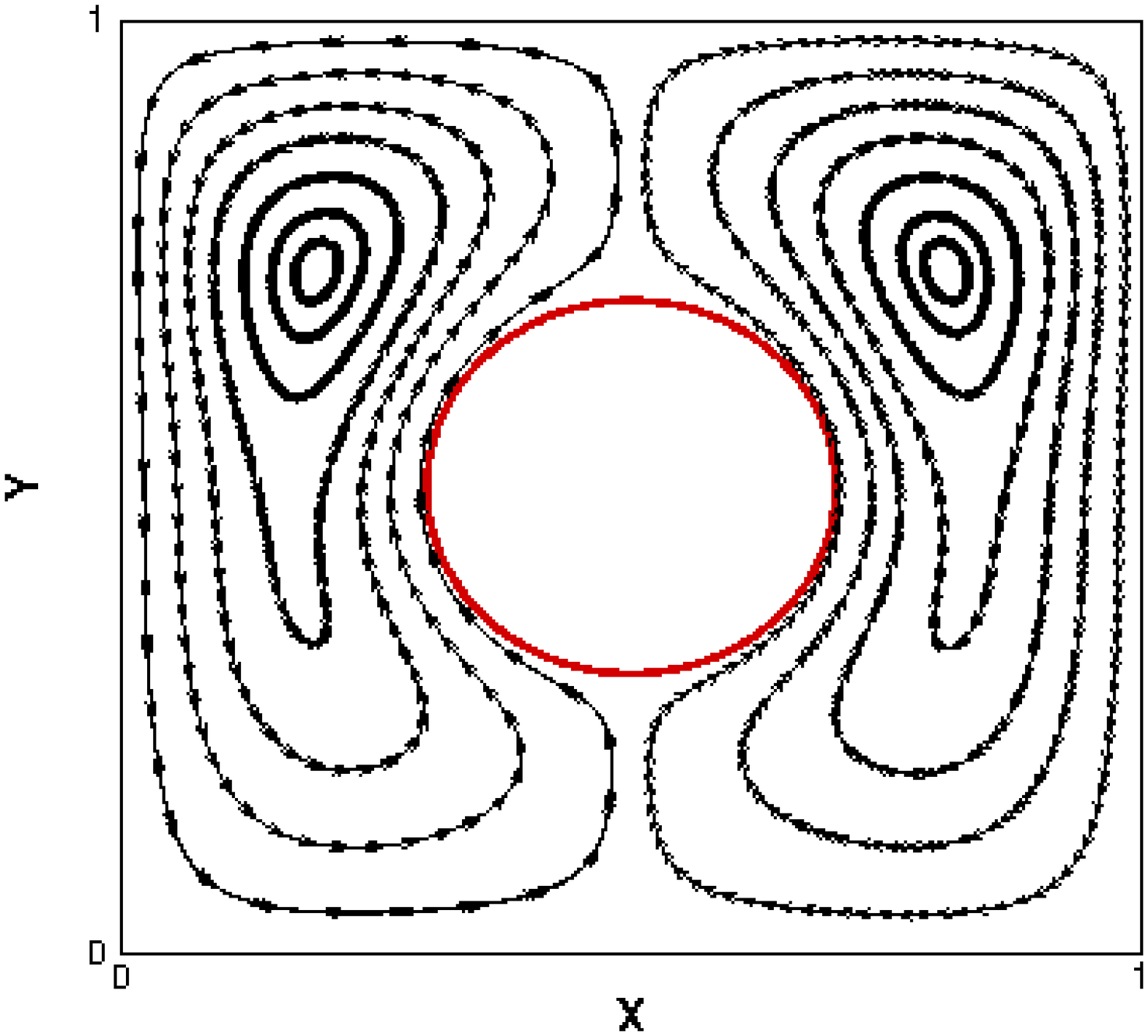}
\hspace{-2mm}
\includegraphics[angle=  0,width=0.32\textwidth]{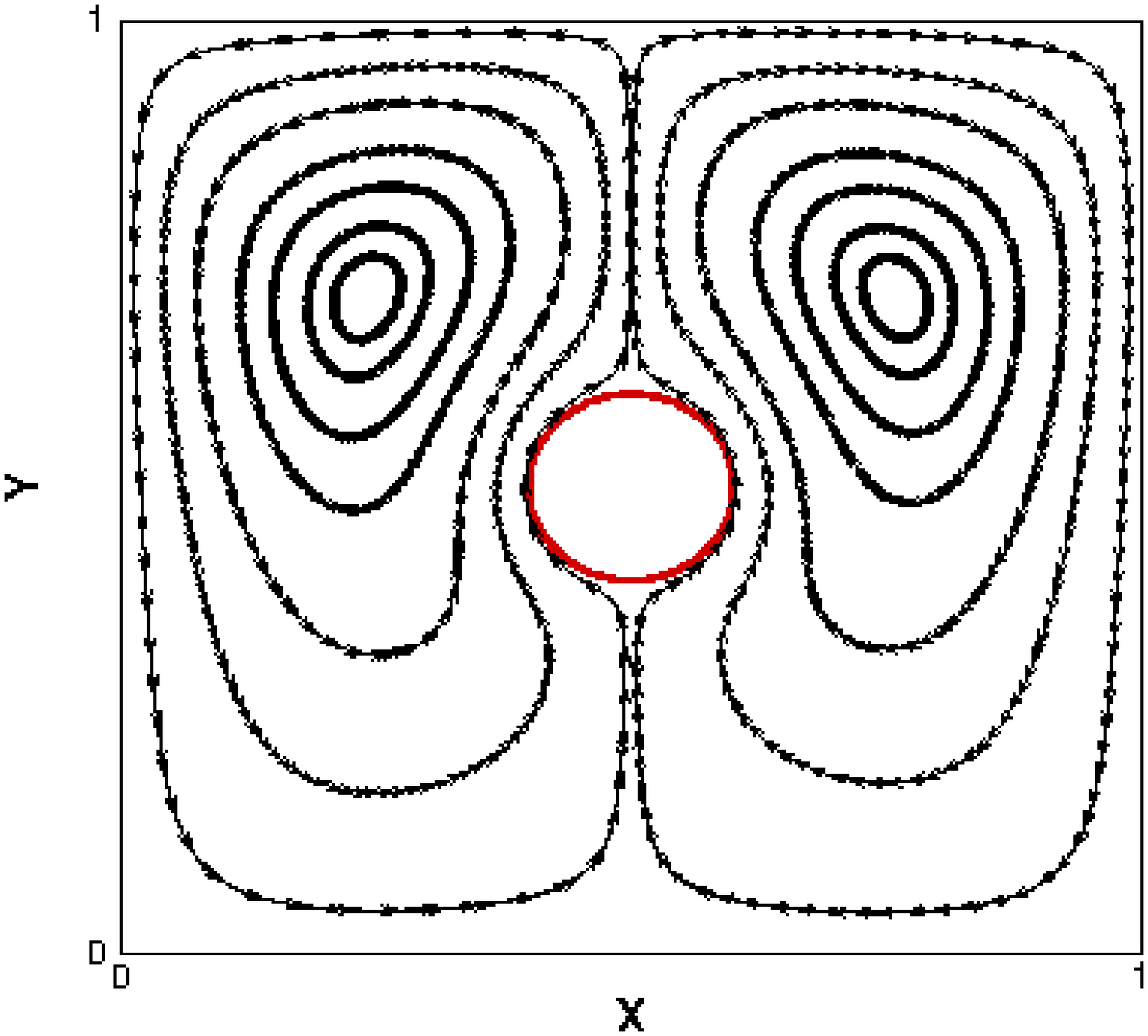}
\\
\includegraphics[angle=  0,width=0.32\textwidth]{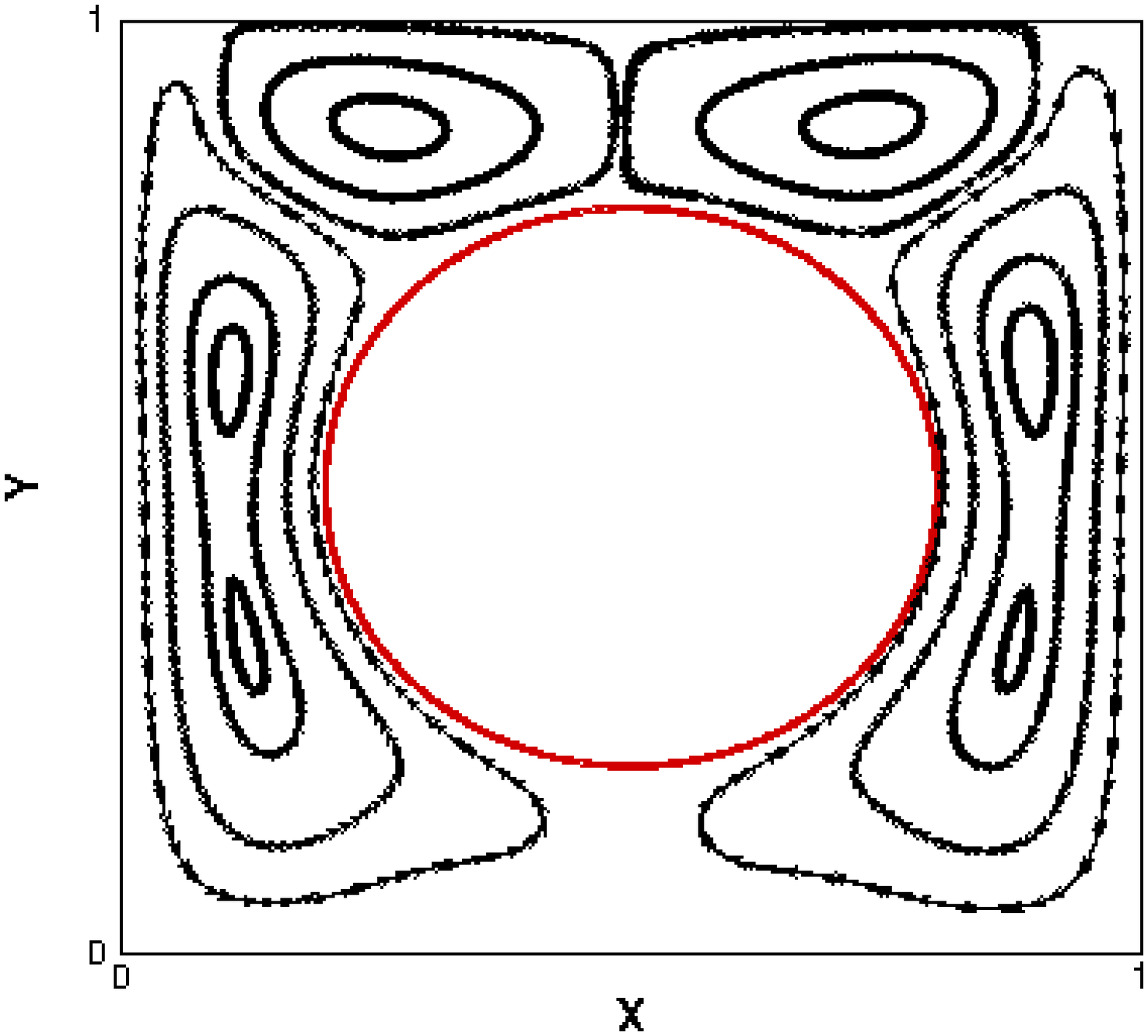}
\hspace{-2mm}
\includegraphics[angle=  0,width=0.32\textwidth]{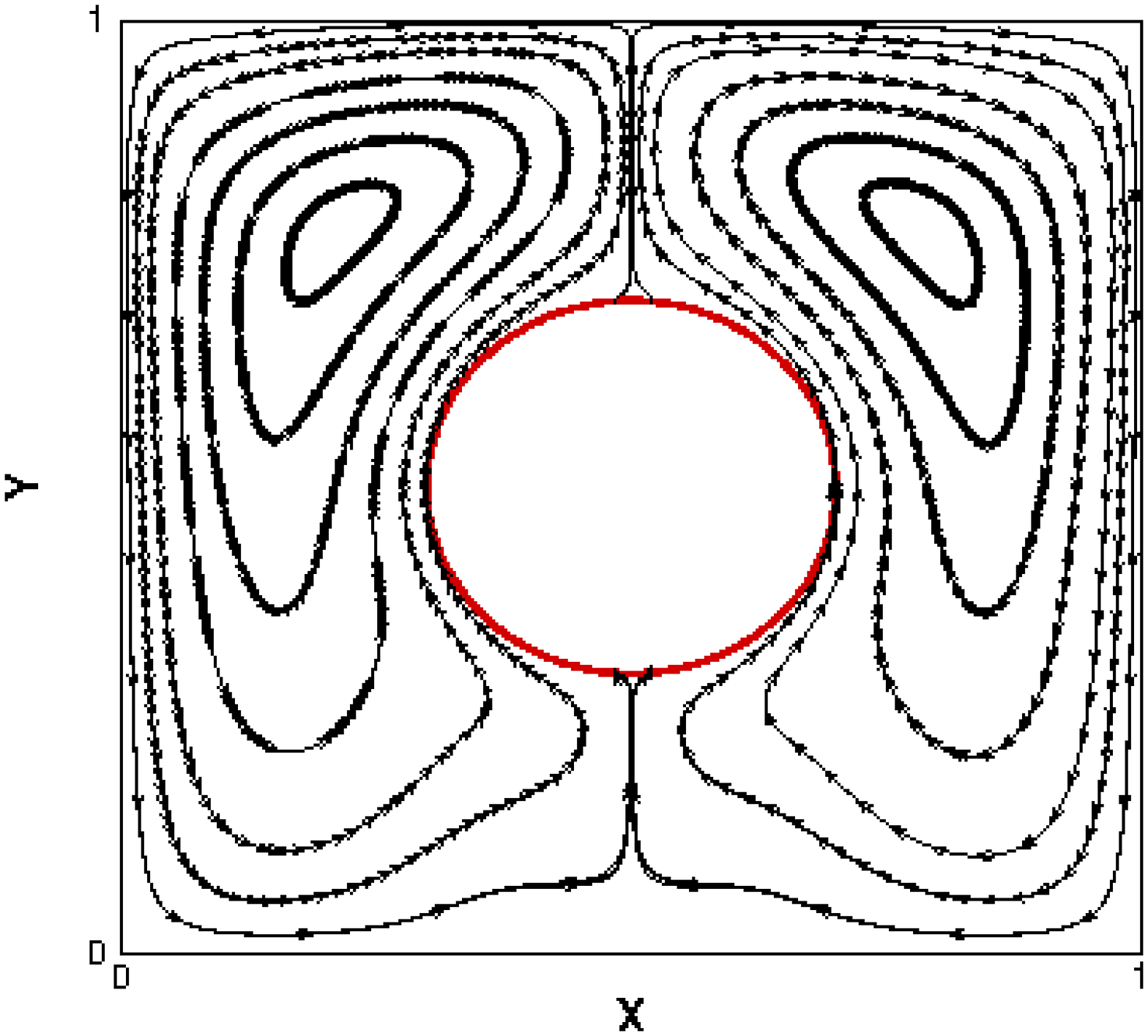}
\hspace{-2mm}
\includegraphics[angle=  0,width=0.32\textwidth]{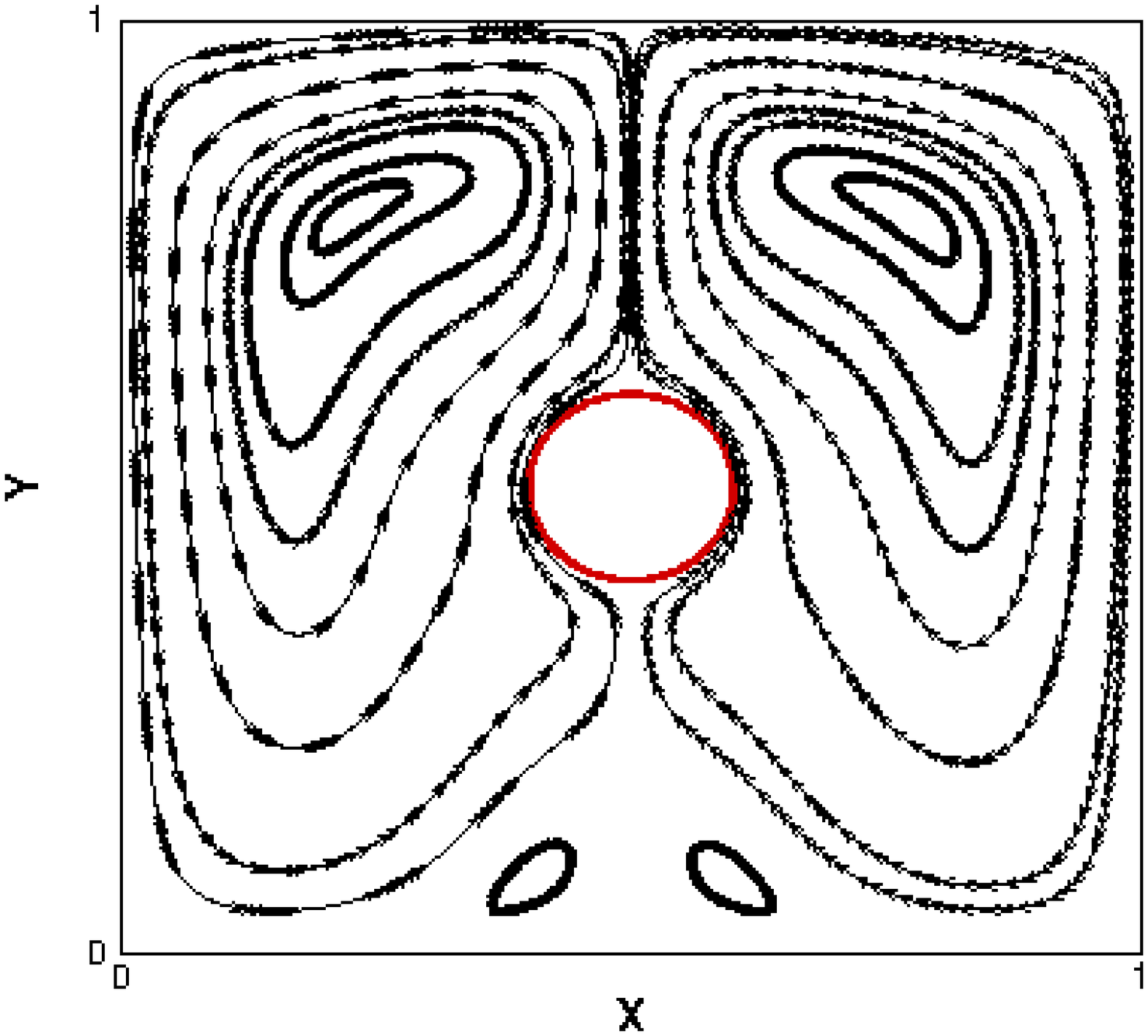}
\\
(a) \hspace{36mm} (b) \hspace{36mm} (c)
\caption{Isothermal contours of temperature distribution in the square cavity at different $Ra$ and $Ar$: 
(a)(top) $Ra=10^{4}$, $Ar=1.67$, (b)(top) $Ra=10^{4}$, $Ar=2.5$, (c)(top) $Ra=10^{4}$, $Ar=5$;
(b)(middle) $Ra=10^{5}$, $Ar=1.67$, (b)(middle) $Ra=10^{5}$, $Ar=2.5$, (b)(middle) $Ra=10^{5}$, $Ar=5$;
(c)(bottom) $Ra=10^{6}$, $Ar=1.67$, (c)(bottom) $Ra=10^{6}$, $Ar=2.5$, (c)(bottom) $Ra=10^{6}$, $Ar=5$.}
\label{annulusStreamlines}
\end{figure}

\subsection{Modeling of the particle-particle interactions}

Since heat transfer between particle-particle and particle-wall is not considered, the remaining job is to monitor 
the trajectories of the solid particles and treat the inter-particle collisions properly. Based on the Newton's second
law of motion, the dynamic equations of the solid particle can be expressed as
\begin{eqnarray}\label{demgov}
m\frac{d^{2}r}{d t^{2}} &=& (1-\frac{\rho_{f}}{\rho_{p}})g+F_{fpi} \\
I\frac{d^{2}\theta}{d t^{2}} &=& \tau_{p}
\end{eqnarray}
where $m$ and $I$ are the mass and the moment of inertia of the particle, respectively.
 $r$ is the particle position and $\theta$ is the angular position. $\rho_{f}$ and $\rho_{p}$
 are the densities of the fluid and particle, respectively. 
  $\tau_{p}$ is the torque. Another considered force on the right hand side of Equation~\ref{demgov} is 
the fluid-particle interaction force $F_{fpi}$. When the particles collide directly with other particles 
or the walls, DEM is employed to 
calculate the collision force. In this study, the particles and walls are directly specified by material 
properties in the simulation such as density, Young's modulus and friction coefficient.  
When the collisions take place, the theory of Hertz~\cite{Johnson} is used for modeling the force-displacement 
relationship while the theory of Mindlin and Deresiewicz~\cite{MD} is employed for the tangential force-displacement 
calculations. For particle of radius $R_{i}$ , Young's modulus $E_{i}$ and Poisson's ratios $\nu_{i}$, the normal 
force-displacement relationship between the colliding particles reads

\begin{eqnarray}
F_{n}=\frac{4}{3}E^{*}R^{*1/2}\delta_{n}^{3/2}
\end{eqnarray}

\noindent where the equivalent Young's modulus and radius can be calculated by 
$1/E^{*}=(1-\nu_{1}^{2})/E_{1}+(1-\nu_{2}^{2})/E_{2}$ and
$1/R^{*}=1/R_{1}+1/R_{2}$, respectively.

The incremental tangential force arising from an incremental tangential displacement depends on the
loading history as well as the normal force 

\begin{eqnarray}
\Delta T=8G^{*}r_{a}\theta_{k}\Delta \delta_{t}+(-1)^{k}\mu\Delta F_{n}(1-\theta_{k})
\end{eqnarray}

\noindent where $1/G^{*}=(1-\nu_{1}^{2})/G_{1}+(1-\nu_{2}^{2})/G_{2}$, $r_{a}=\sqrt{\delta_{n}R^{*}}$ is radius of the contact area. $\Delta \delta_{t}$ is the relative tangential 
incremental surface displacement, $\mu$ is the coefficient of friction, the value of $k$ and $\theta_{k}$ changes with the loading
history.

\section{Results and discussions}\label{Numericalresults}

\begin{table}
\centering
\begin{tabular}{cccccc}
\hline
 $Ra$   & $Ar$     & Present     & Hu et al.~\cite{hu2013natural}  & Ren et al.~\cite{ren2012boundary}    & Moukalled and Acharya~\cite{moukalled1996natural}    \\ 
\hline          
            & $5$    & $2.041$ & $2.038$  & $2.051$  & $2.071$   \\
$10^{4}$    & $2.5$  & $3.179$ & $3.184$  & $3.161$  & $3.331$   \\
            & $1.67$ & $5.213$ & $5.294$  & $5.303$  & $5.286$   \\
\hline
            & $5$    & $3.760$ & $3.778$  & $3.704$   & $3.825$        \\
$10^{5}$    & $2.5$  & $4.989$ & $4.917$  & $4.836$   & $5.080$      \\
            & $1.67$ & $6.193$ & $6.247$  & $6.171$   & $6.212$     \\
\hline      
            & $5$    & $6.025$ & $6.095$  & $5.944$  & $6.107$        \\
$10^{6}$    & $2.5$  & $8.831$ & $8.934$  & $8.546$  & $9.374$     \\
            & $1.67$ & $11.857$ & $11.995$  & $11.857$  & $11.620$      \\
\hline      
\end{tabular}
\caption{Comparison of computed average Nusselt numbers.}
\label{annulustable}
\end{table} 

\subsection{Natural convection in a two-dimensional square cavity with a concentric annulus}\label{Naturalconvectioncavityannulus}

Natural convection in a two-dimensional square cavity has been a popular benchmark case for the verification of 
one's numerical tools on heat transfer through fluid materials~\cite{moukalled1996natural,ren2012boundary,hu2013natural}. 
In this subsection, the case is employed to test the thermal LBM code coupling with the PIBM. An isothermal concentric 
annulus is planted in the center of the cavity as the heat source. The schematic diagram of the entire computational domain 
is shown in Figure~\ref{schematicconcentricannulus}(a). It should be stressed that the concentric annulus here is constructed 
by small isothermal particles with uniform size as shown in Figure~\ref{schematicconcentricannulus}(b). The positions of the 
solid particles are artificially set stagnant to prevent them following down under the action of gravity. The length and height 
of the cavity are $1$, respectively. We consider three sizes of concentric annulus. The ratios of the cavity length to the 
annulus diameter $Ar$ are $1.67$, $2.5$ and $5$. We use $250\times250$ meshes to cover the whole computational domain. The 
diameter of the solid particles is $h/2$. $Ra$ of interest is ranging from $10^{4}$ to $10^{6}$ and $Pr=0.71$.  The 
non-dimensional temperature is set $1$ and $0$ at the concentric annulus and the surrounding cold walls, respectively. 
The initial temperature of the stagnant fluid is $0.5$. The remainder parameters relevant to the simulations are $L_{c}=1$ 
and $u_{c}=0.25$.  

\begin{figure}
 \centering
 \includegraphics[width=0.54\textwidth]{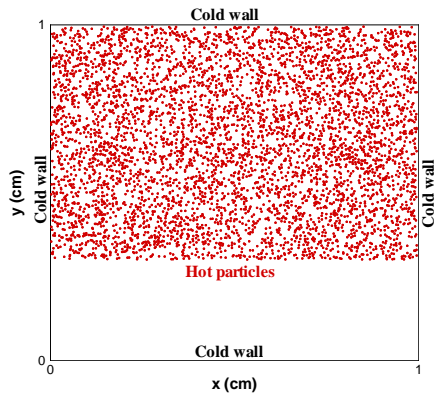}
 \vskip-0.2cm
 \caption{Schematic diagram of the PIBM.} \label{2Dinitial}
 \end{figure}
 
\begin{figure}[!ht]
\centering
\includegraphics[angle=  0,width=0.45\textwidth]{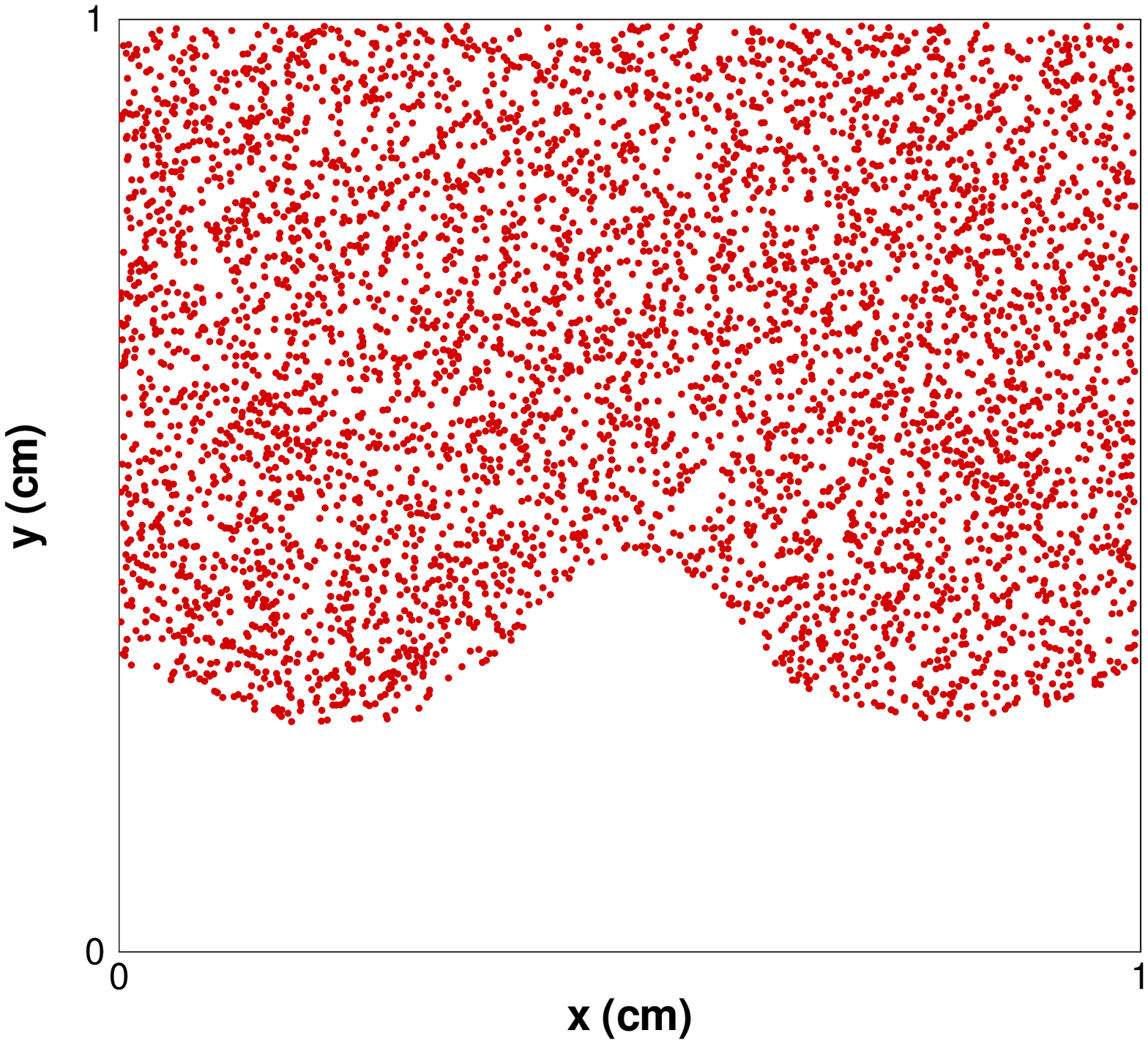}
\hspace{2mm}
\includegraphics[angle=  0,width=0.45\textwidth]{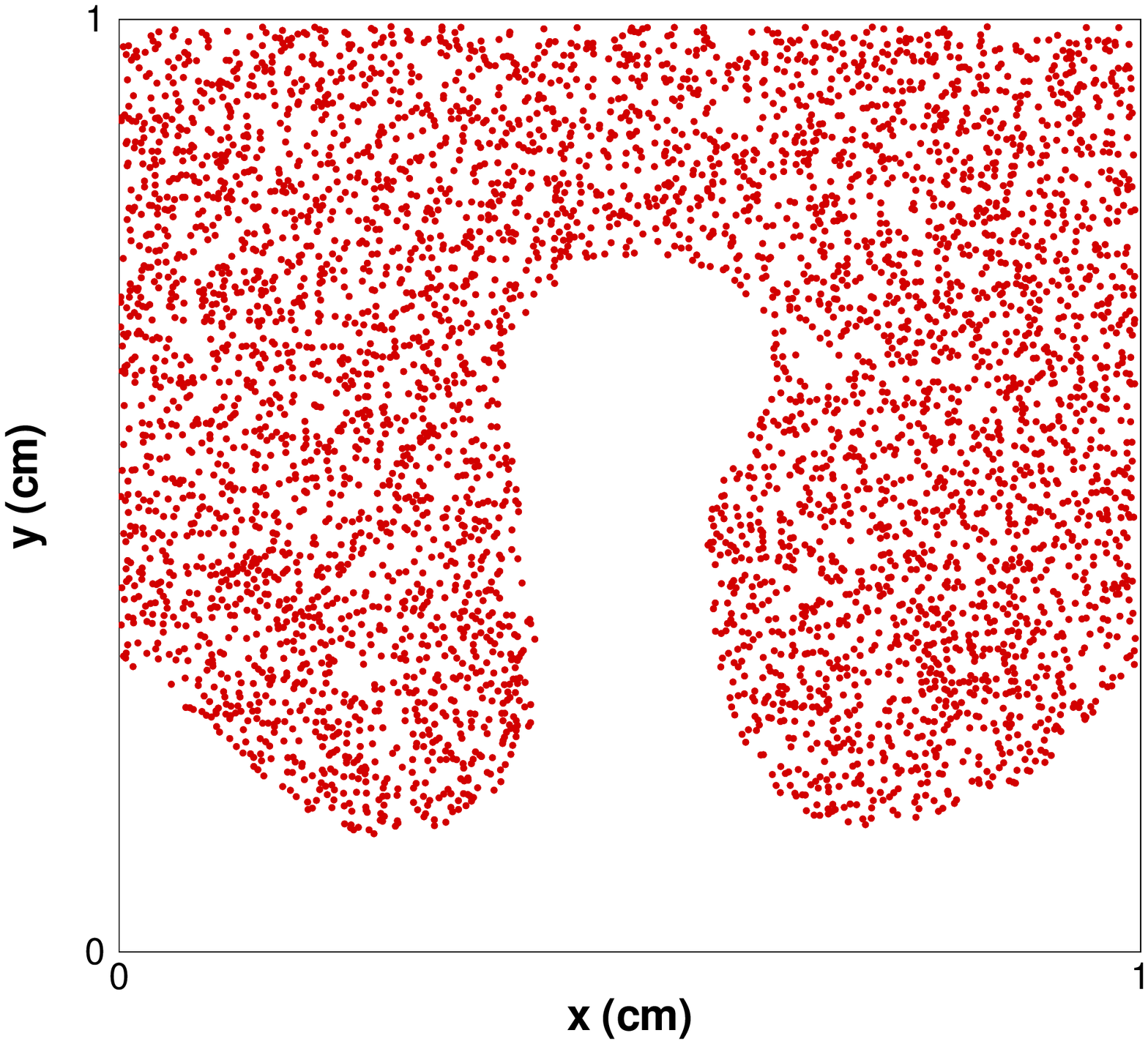}
\\
(a) \hspace{52mm} (b)
\\
\includegraphics[angle=  0,width=0.45\textwidth]{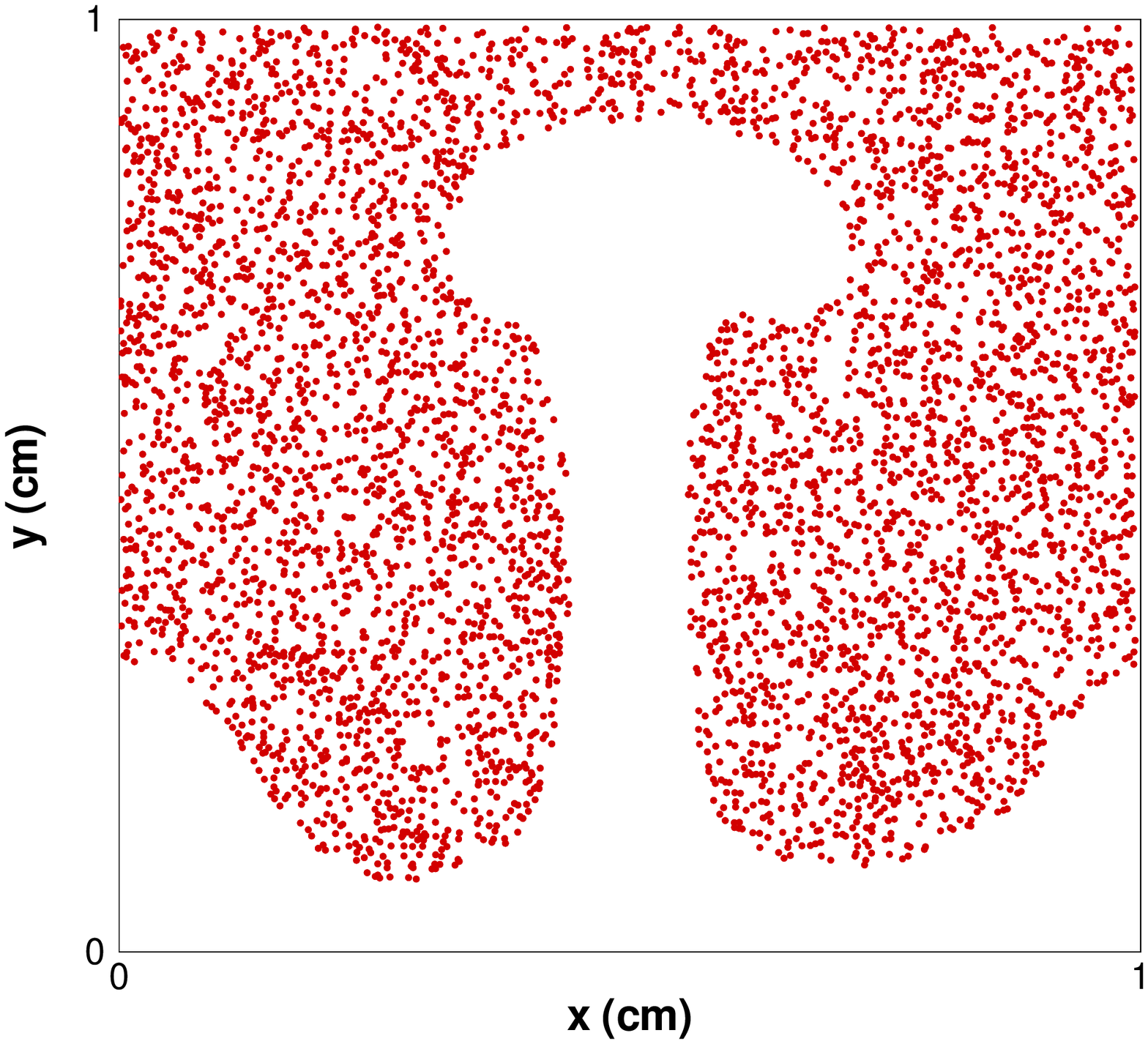}
\hspace{2mm}
\includegraphics[angle=  0,width=0.45\textwidth]{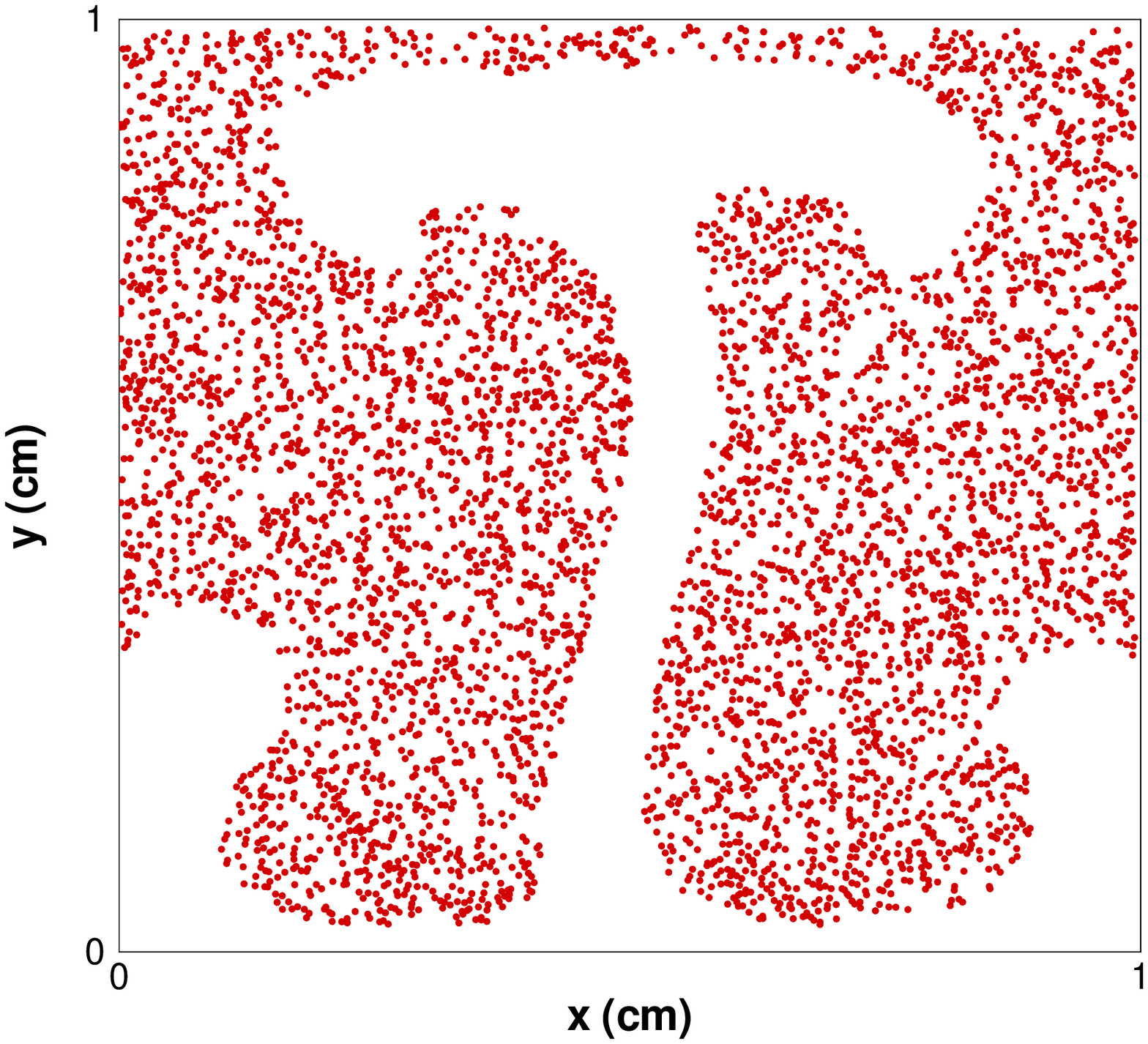}
\\
(c) \hspace{52mm} (d)
\caption{Positions of the $5000$ particles without heat transfer at time 
(a) $t=2.5 s$, (b) $t=5.0 s$, (c) $t=7.5 s$, (d) $t=10.0 s$.}
\label{2dparticleposition1}
\end{figure}
 
\begin{figure}[!ht]
\centering
\includegraphics[angle=  0,width=0.45\textwidth]{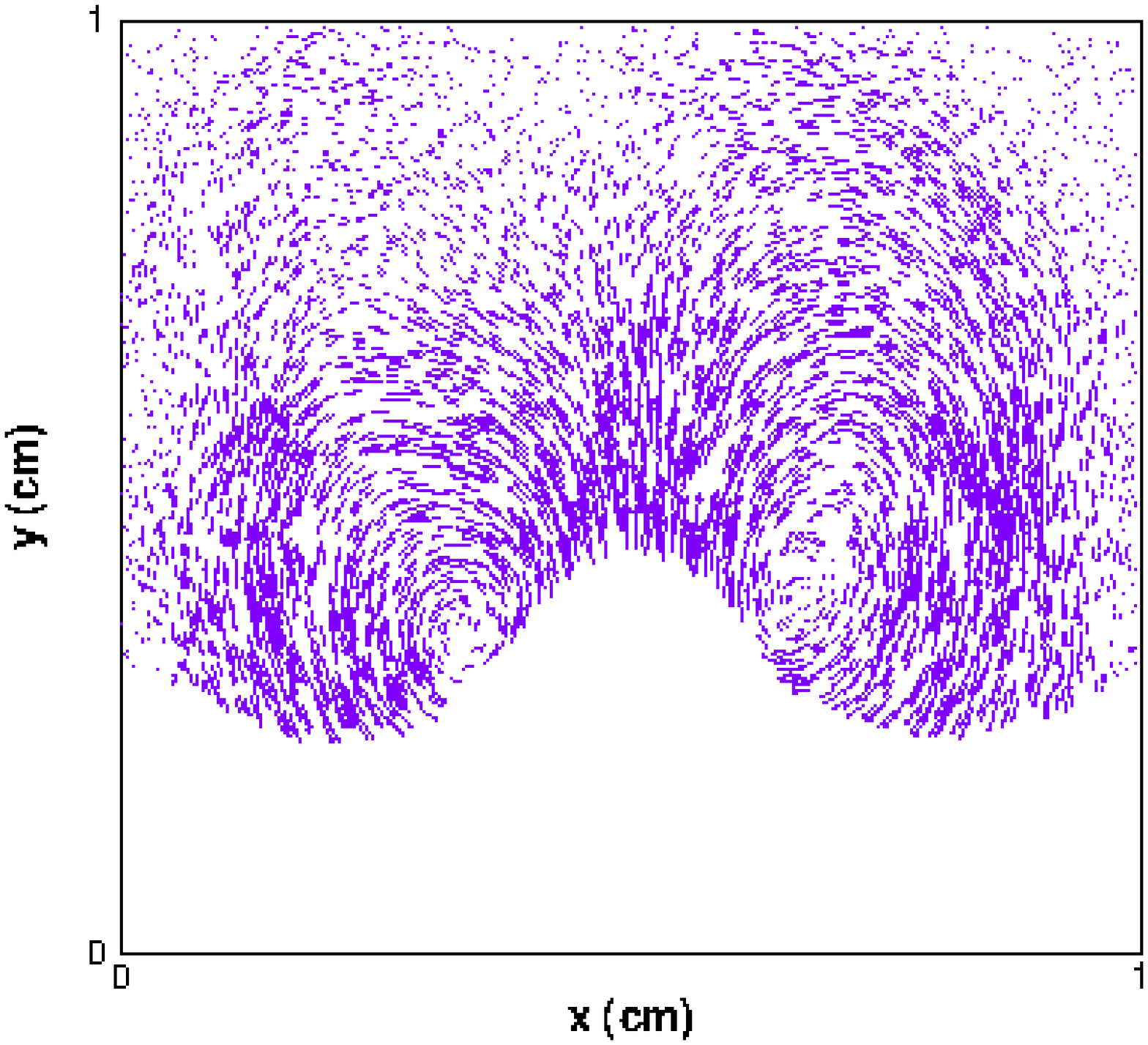}
\hspace{2mm}
\includegraphics[angle=  0,width=0.45\textwidth]{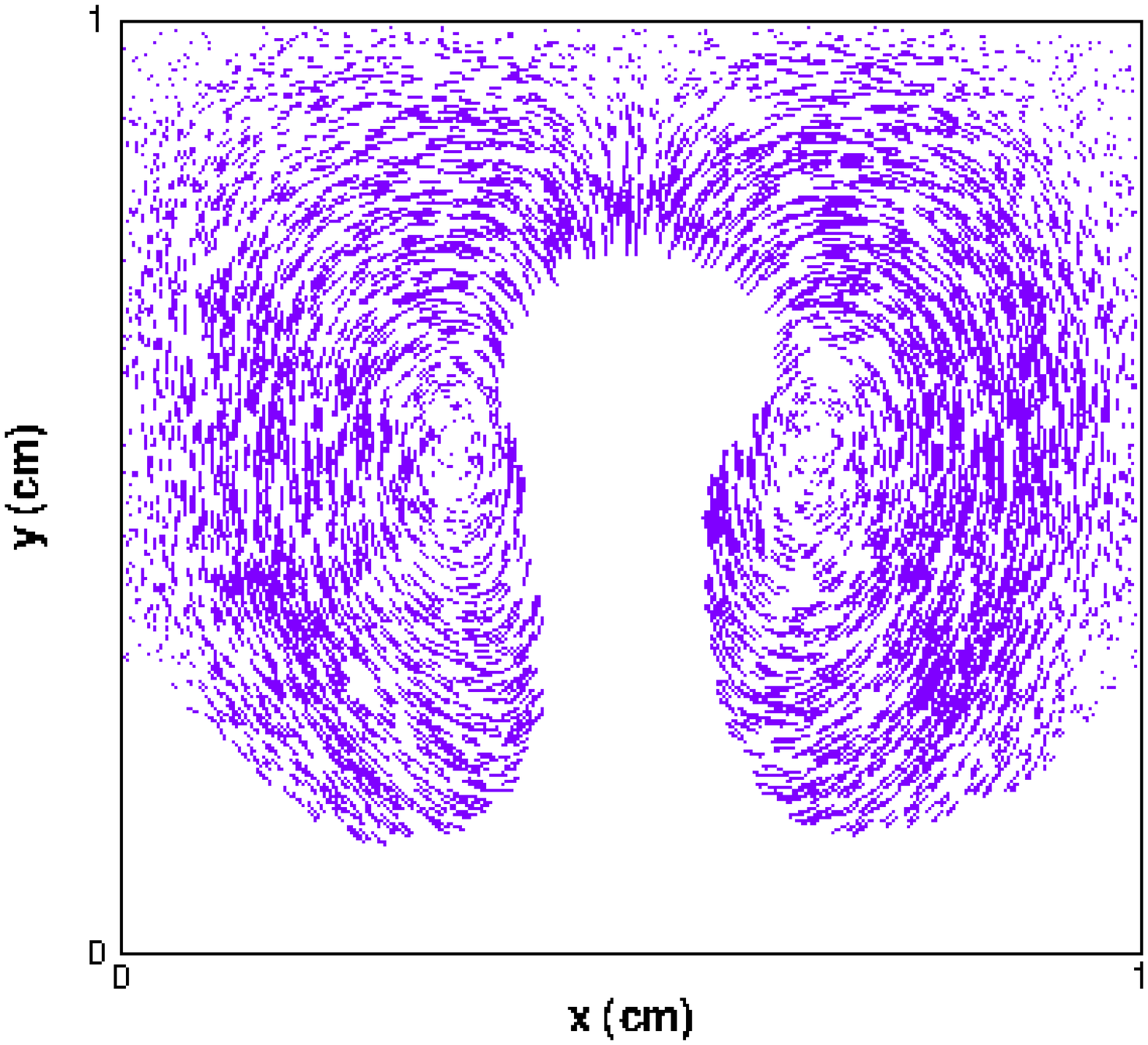}
\\
(a) \hspace{52mm} (b)
\\
\includegraphics[angle=  0,width=0.45\textwidth]{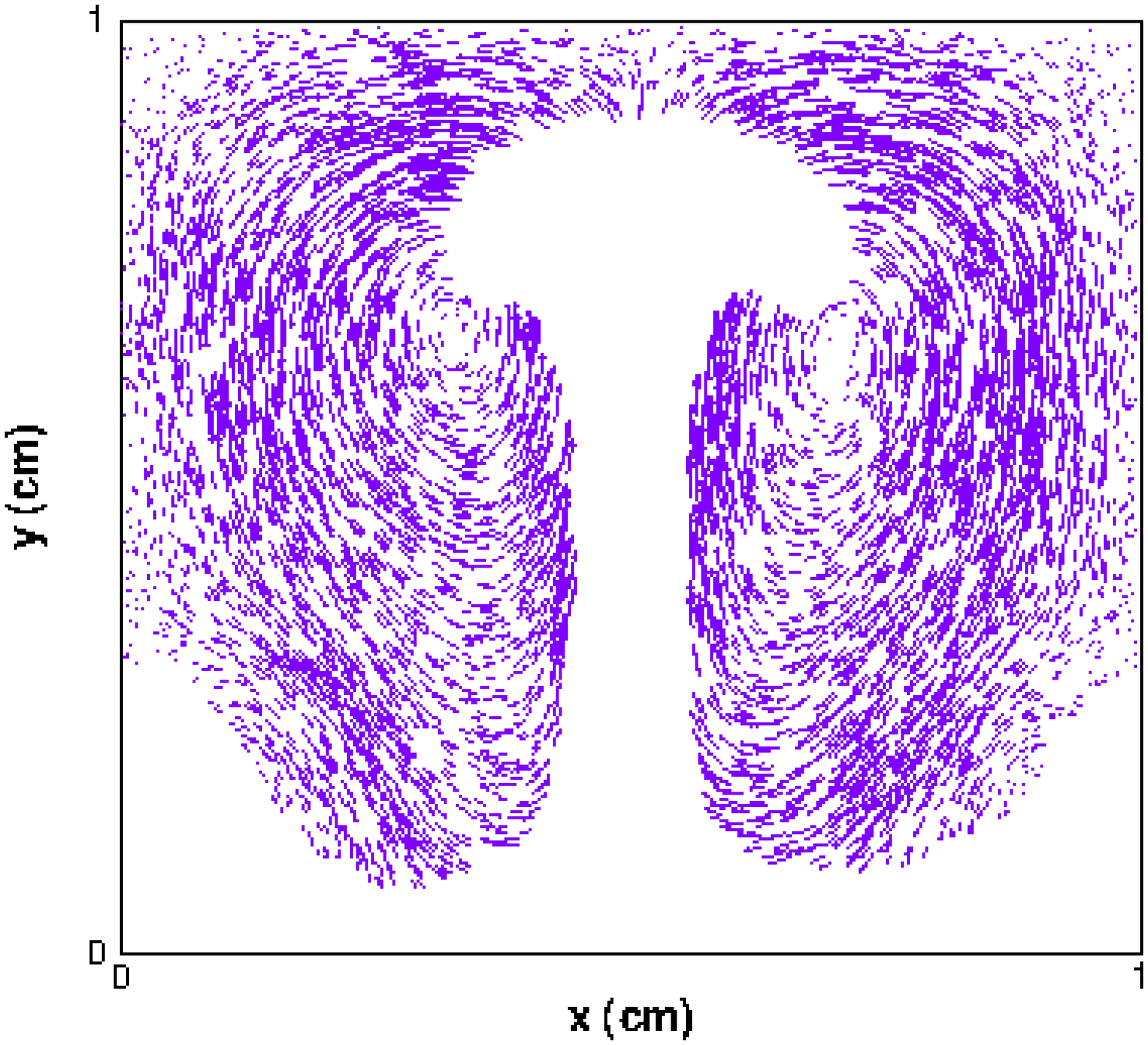}
\hspace{2mm}
\includegraphics[angle=  0,width=0.45\textwidth]{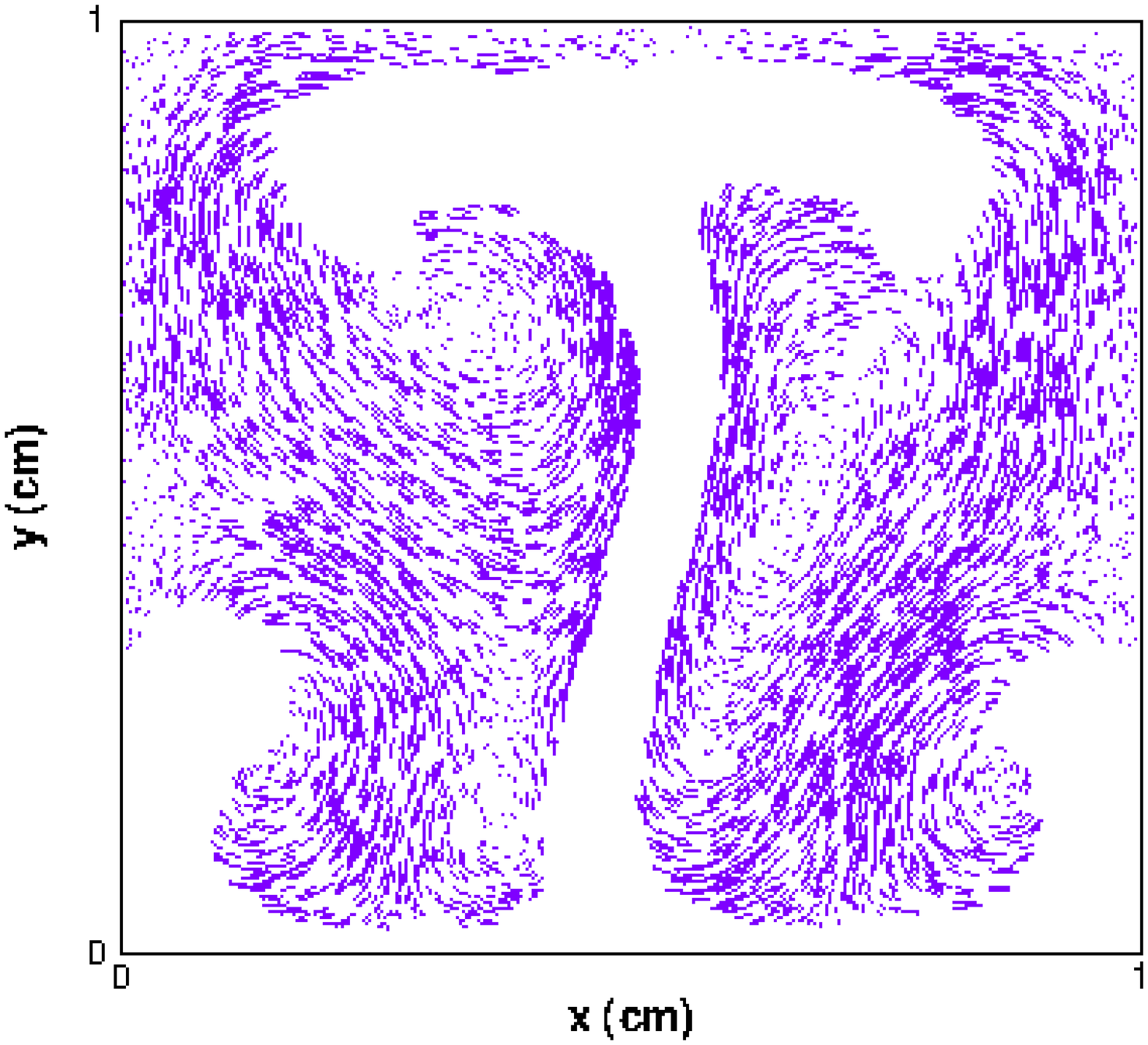}
\\
(c) \hspace{52mm} (d)
\caption{Velocity distribution of the $5000$ particles without heat transfer at time 
(a) $t=2.5 s$, (b) $t=5.0 s$, (c) $t=7.5 s$, (d) $t=10.0 s$.}
\label{2dparticlevelocity1}
\end{figure}

\begin{figure}[!ht]
\centering
\includegraphics[angle=  0,width=0.45\textwidth]{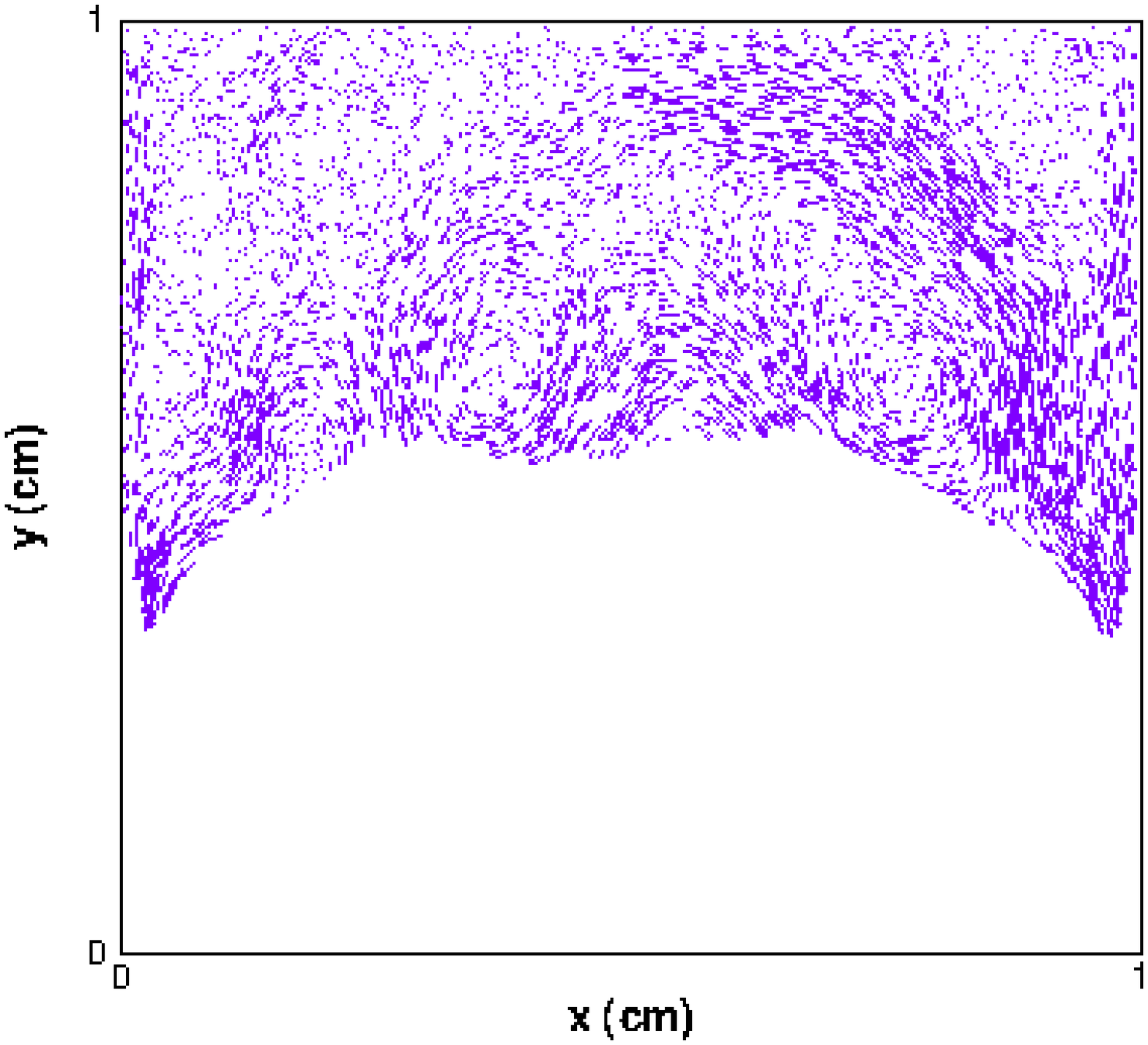}
\hspace{2mm}
\includegraphics[angle=  0,width=0.45\textwidth]{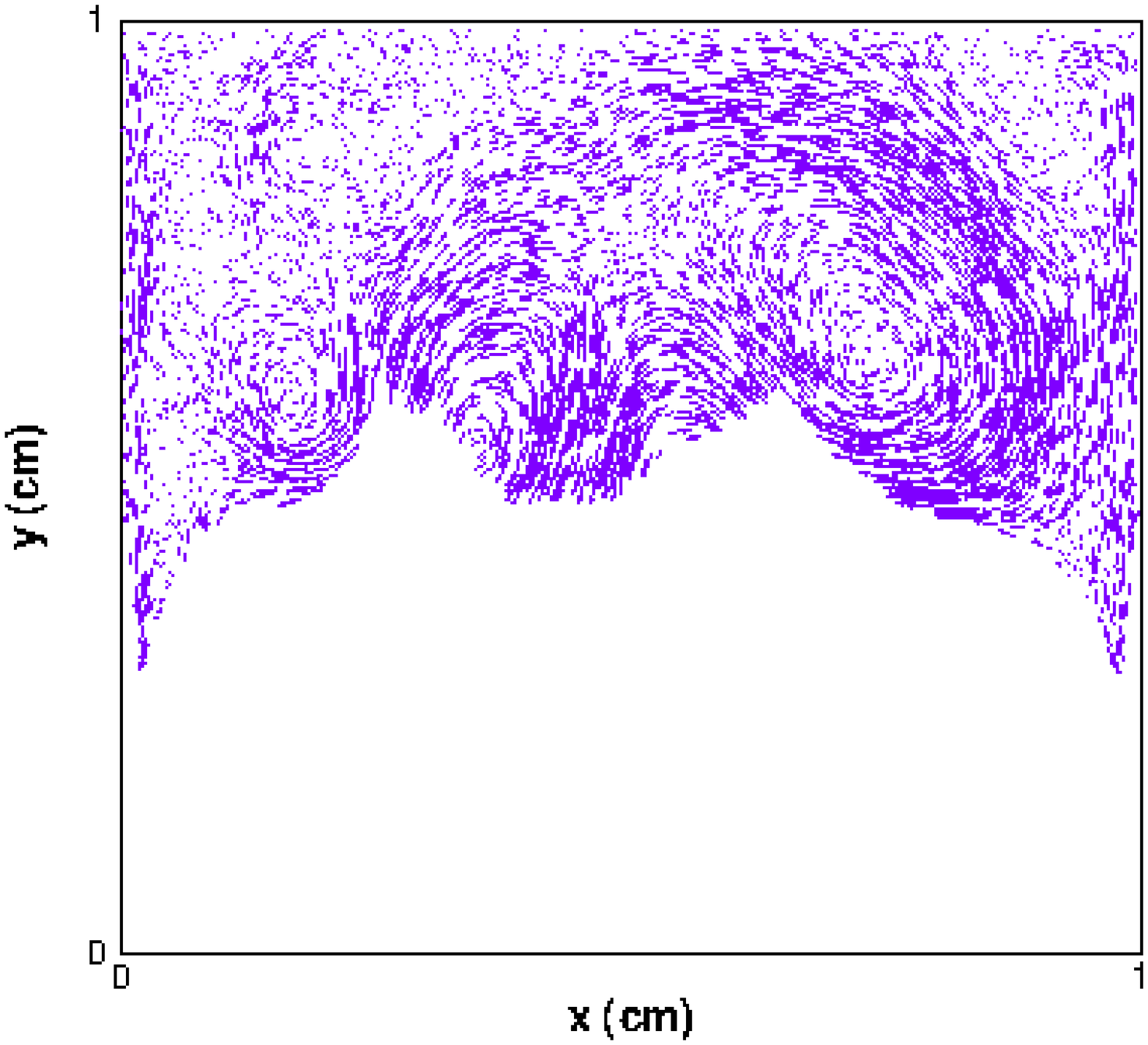}
\\
(a) \hspace{52mm} (b)
\\
\includegraphics[angle=  0,width=0.45\textwidth]{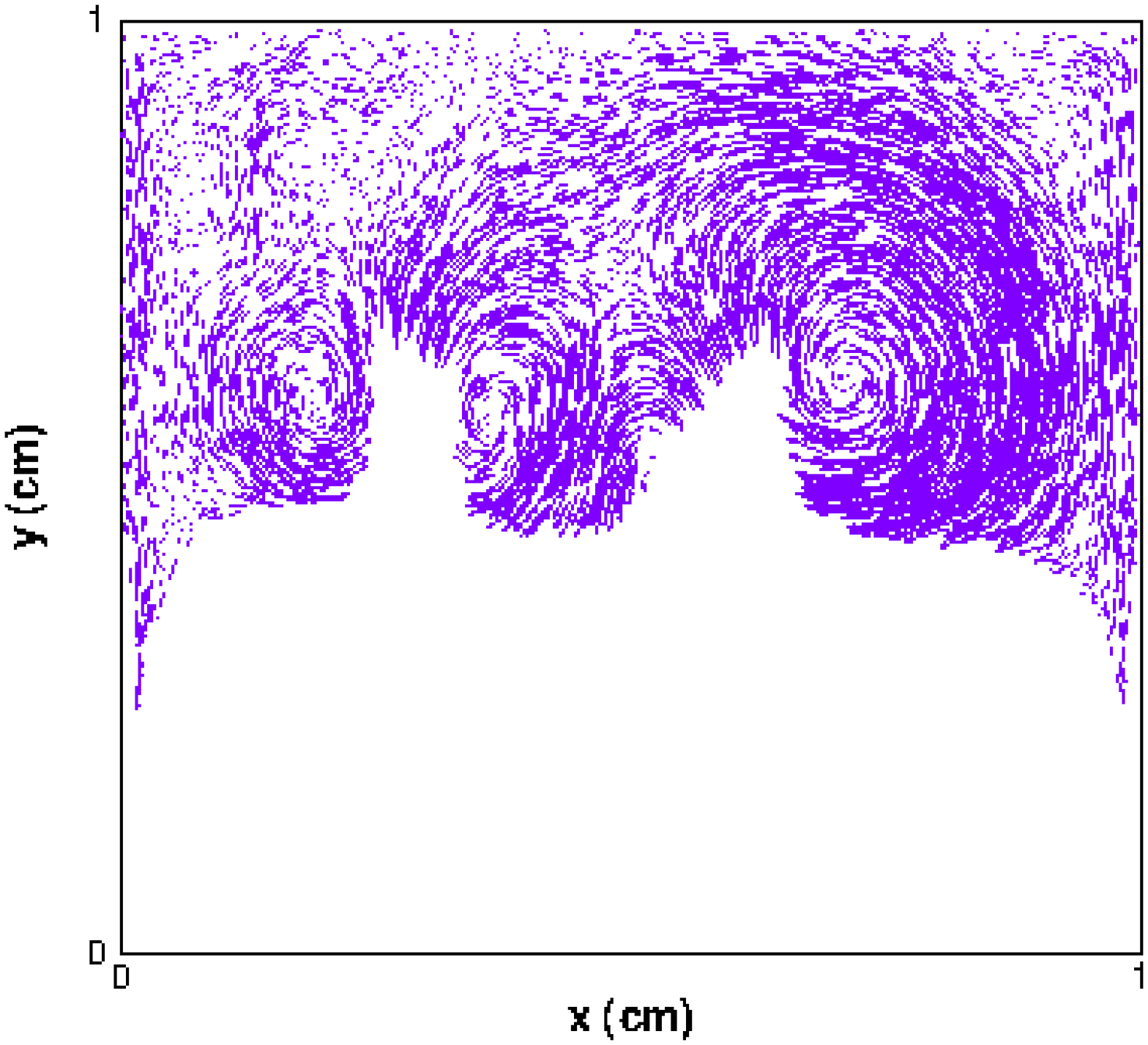}
\hspace{2mm}
\includegraphics[angle=  0,width=0.45\textwidth]{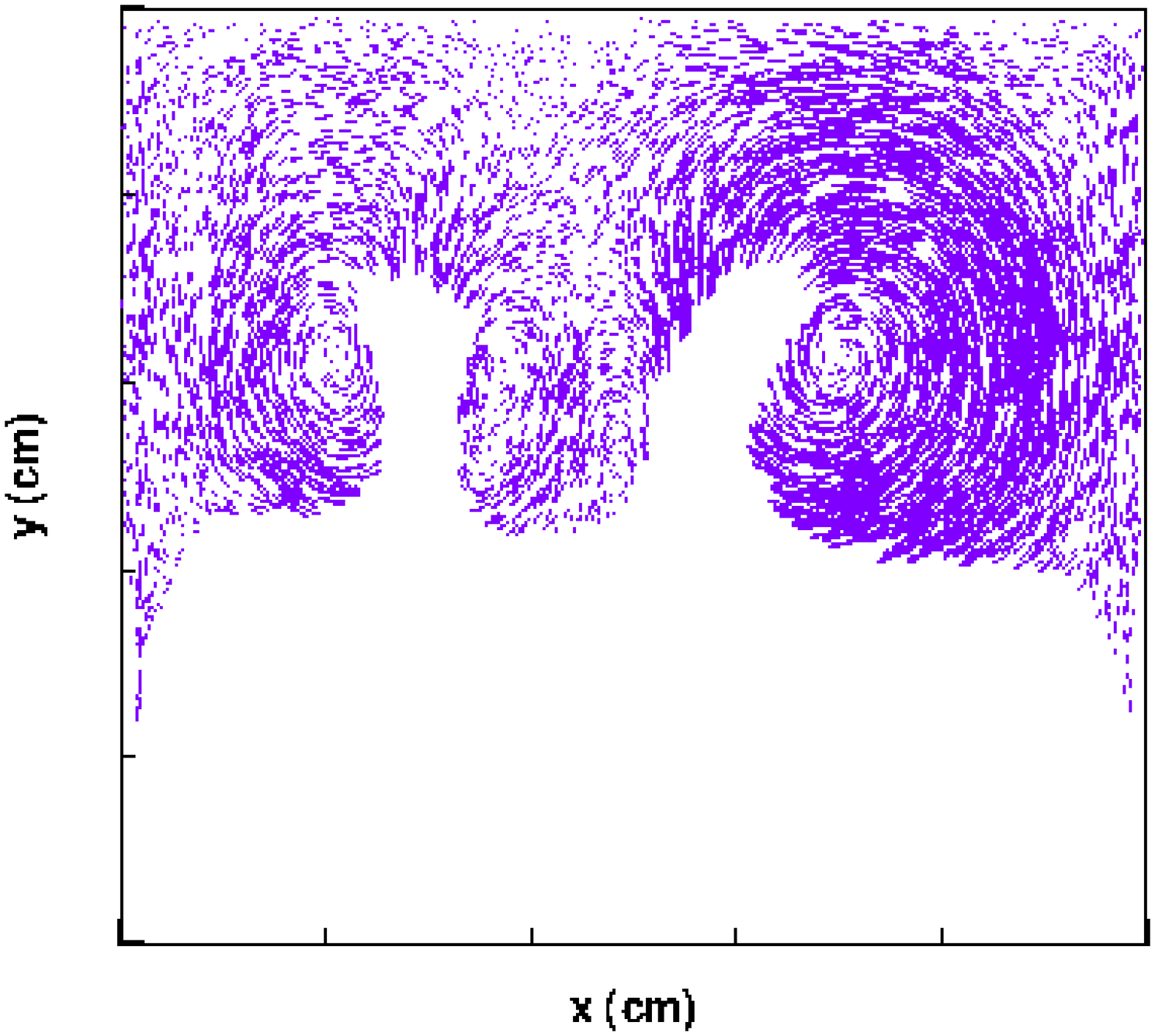}
\\
(c) \hspace{52mm} (d)
\caption{Velocity distribution of the $5000$ particles with heat transfer at time 
(a) $t=2.5 s$, (b) $t=5.0 s$, (c) $t=7.5 s$, (d) $t=10.0 s$.}
\label{2dparticlevelocity2}
\end{figure}

\begin{figure}[!ht]
\centering
\includegraphics[angle=  0,width=0.45\textwidth]{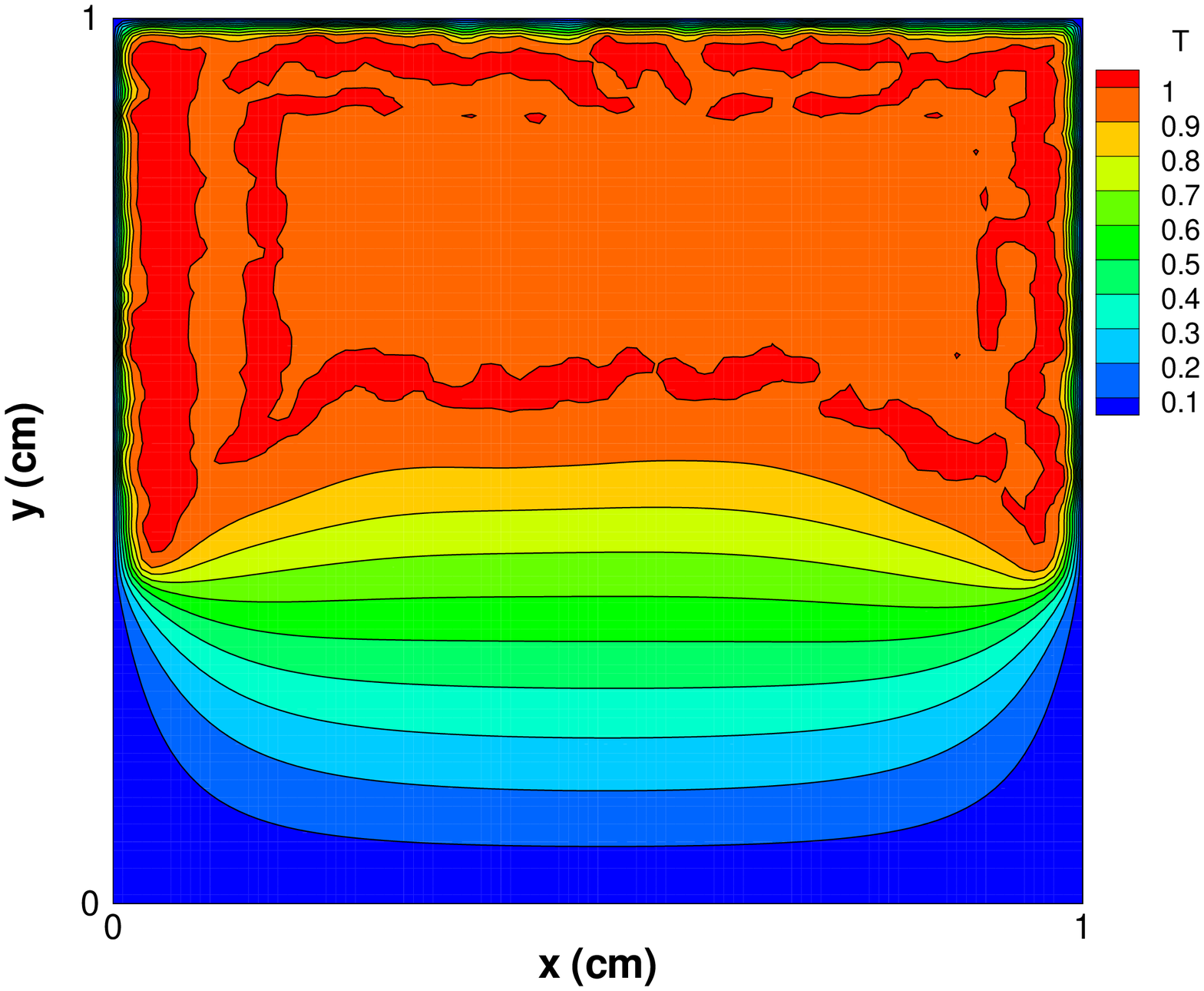}
\hspace{2mm}
\includegraphics[angle=  0,width=0.45\textwidth]{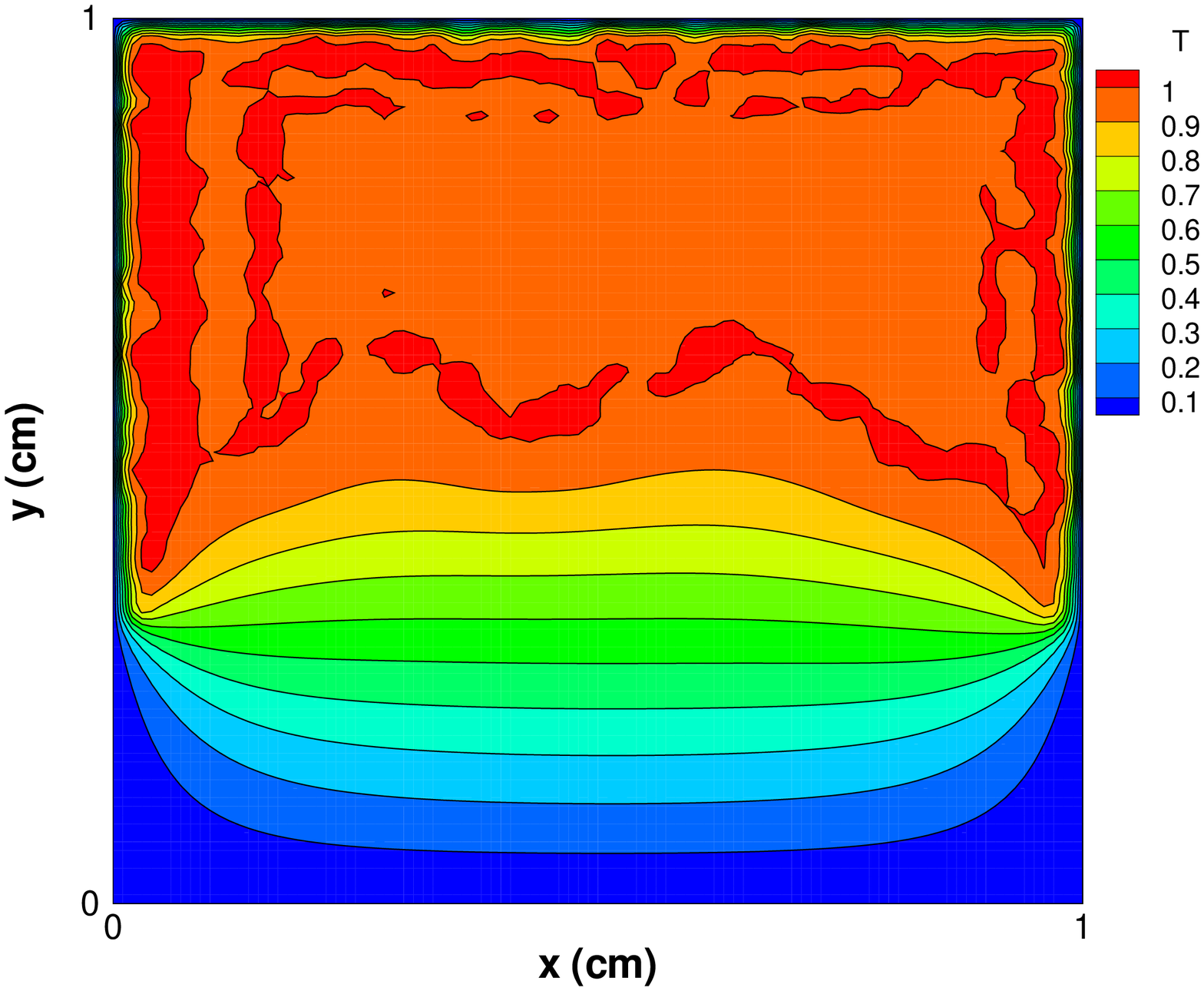}
\\
(a) \hspace{52mm} (b)
\\
\includegraphics[angle=  0,width=0.45\textwidth]{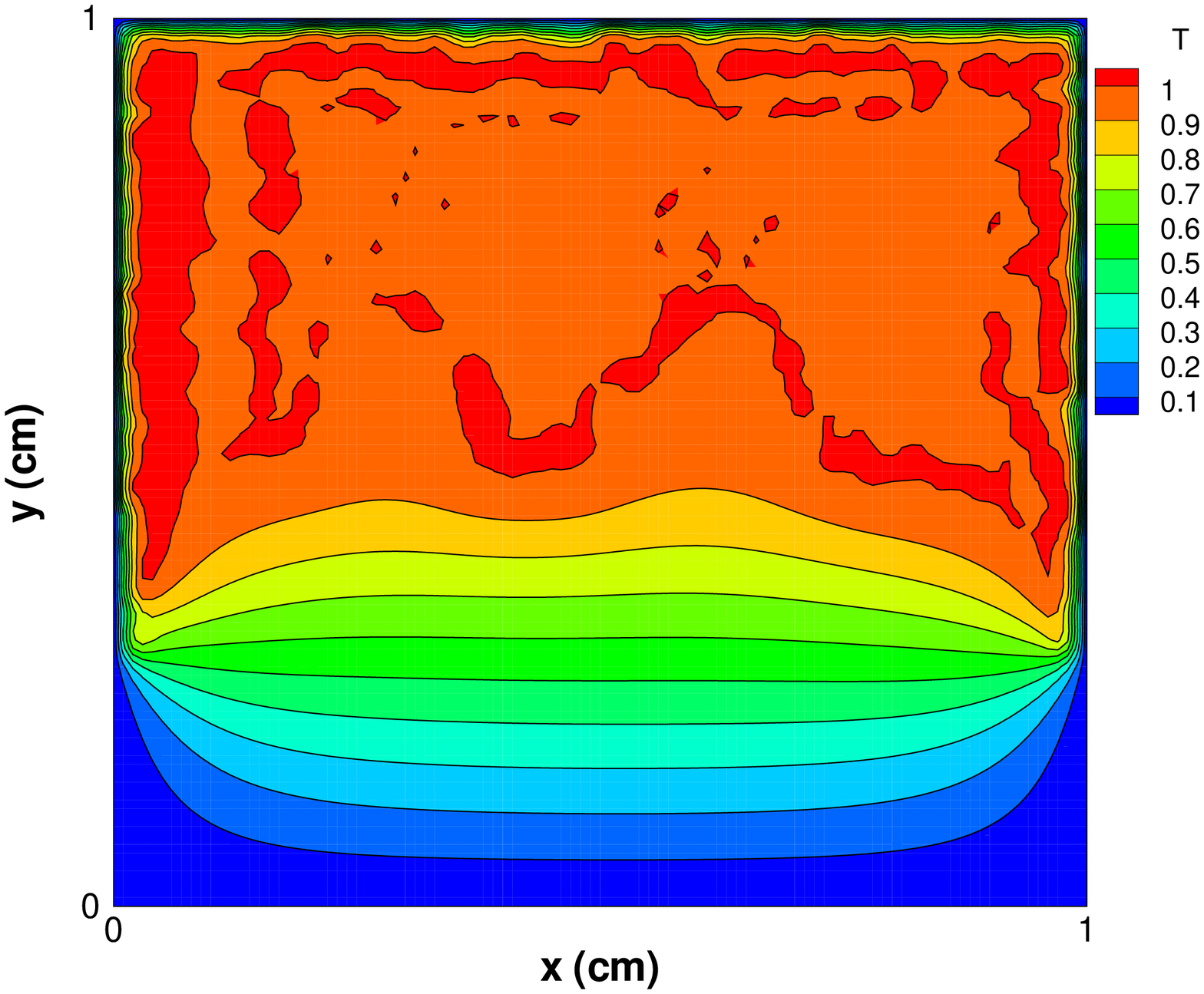}
\hspace{2mm}
\includegraphics[angle=  0,width=0.45\textwidth]{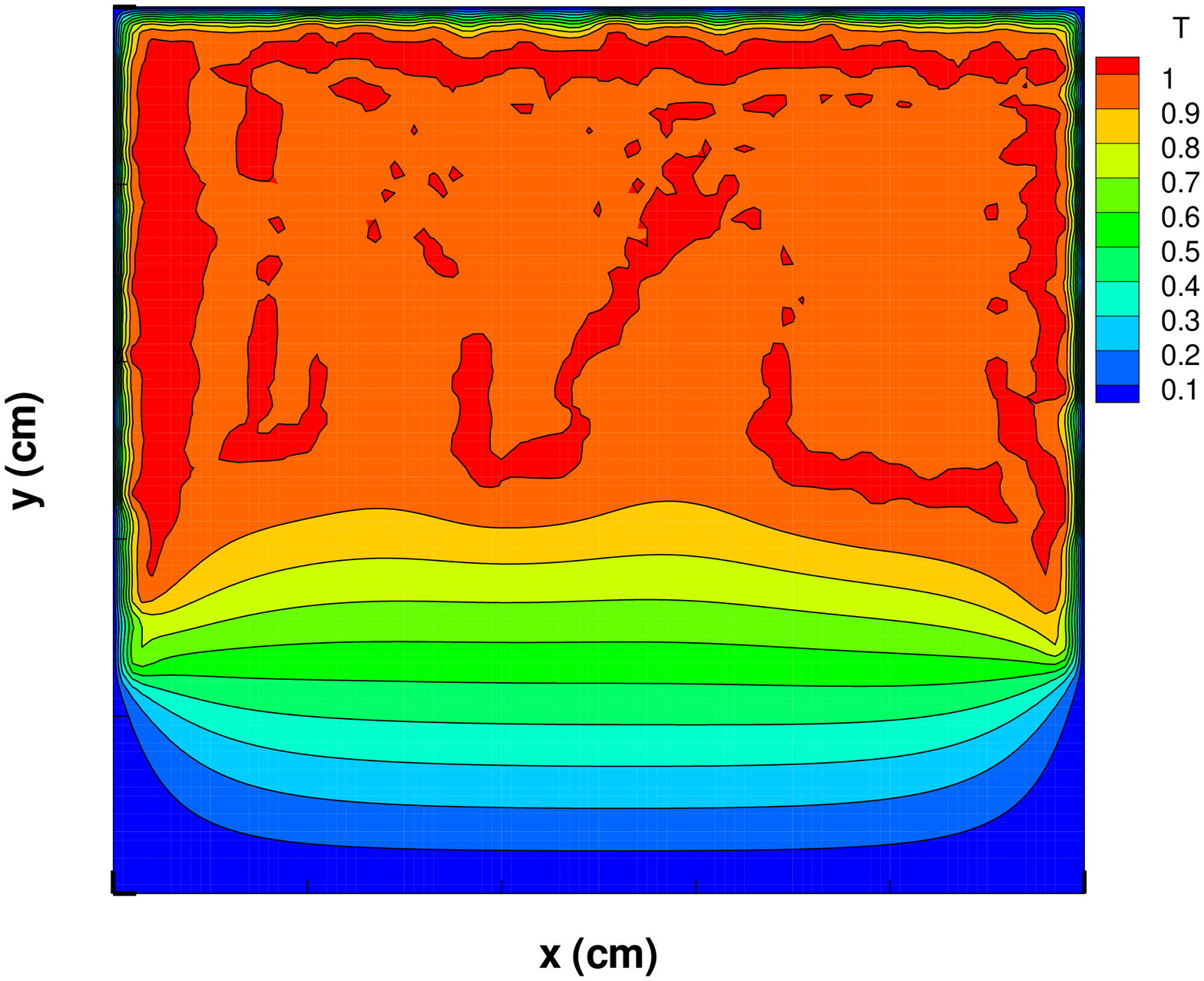}
\\
(c) \hspace{52mm} (d)
\caption{Temperature distribution in the cavity at time 
(a) $t=2.5 s$, (b) $t=5.0 s$, (c) $t=7.5 s$, (d) $t=10.0 s$.}
\label{Temperaturedistribution1}
\end{figure}

The isothermal contours of temperature distribution in the cavity at different $Ra$ and $Ar$ are displayed in 
Figure~\ref{annulustemperature}. As shown, the temperature distribution is affected by both $Ra$ and $Ar$. The strength 
of convection is stronger as $Ra$ increases which gives rise to much more complex isothermal patterns. The thermal 
boundary at the lower surface of concentric annulus is thinner at the higher $Ra$ for each $Ar$. As for the region 
just above the concentric annulus, a gradual changing of temperature can be observed at $Ra=10^{4}$ which means 
that heat conduction mainly dominates at this region at low $Ra$. However, when $Ra$ increases, a crown-like region with 
high temperature is formed due to the high effect of convection. Namely, the temperature distribution is more affected 
by the fluid velocity field at higher $Ra$. The streamlines in the cavity at different $Ra$ and $Ar$ are displayed in 
Figure~\ref{annulusStreamlines}. Similar to the temperature fields, the streamlines are symmetrically distributed about
the perpendicular bisector of the concentric annulus. The influence of $Ra$ on the flow field is obvious. As $Ra$ increases, 
two more counter-rotating vortexes are shown between the upper surface of the concentric annulus and the top wall of the 
cavity when $Ar=1.67$, the two vortexes besides the concentric annulus merge into a big one, respectively, when $Ar=2.5$ and 
two more counter-rotating vortexes can be observed below the the concentric annulus when $Ar=5$. All these observations 
have very good agreements with the previous numerical results under the same conditions~\cite{ren2012boundary,hu2013natural}.
Quantitative comparison of computed average Nusselt numbers is shown in Table~\ref{annulustable}. In this study, we calculate 
the average Nusselt number following Hu et al.~\cite{hu2013natural}: 

\begin{eqnarray}
\overline{Nu}=\dfrac{\sum\limits_{l}\Delta ThA_{p}}{\delta_{t}k(T_{hot}-T_{\infty})}
\end{eqnarray}

\noindent  where $k$ is the thermal conductivity. As can be seen from Table~\ref{annulustable}, the LBM-PIBM scheme also does a 
good job of predicting reasonable 
$\overline{Nu}$ and thus is a promising method for simulating the free convection problems. 

\begin{figure}[!ht]
\centering
\includegraphics[angle=  0,width=0.45\textwidth]{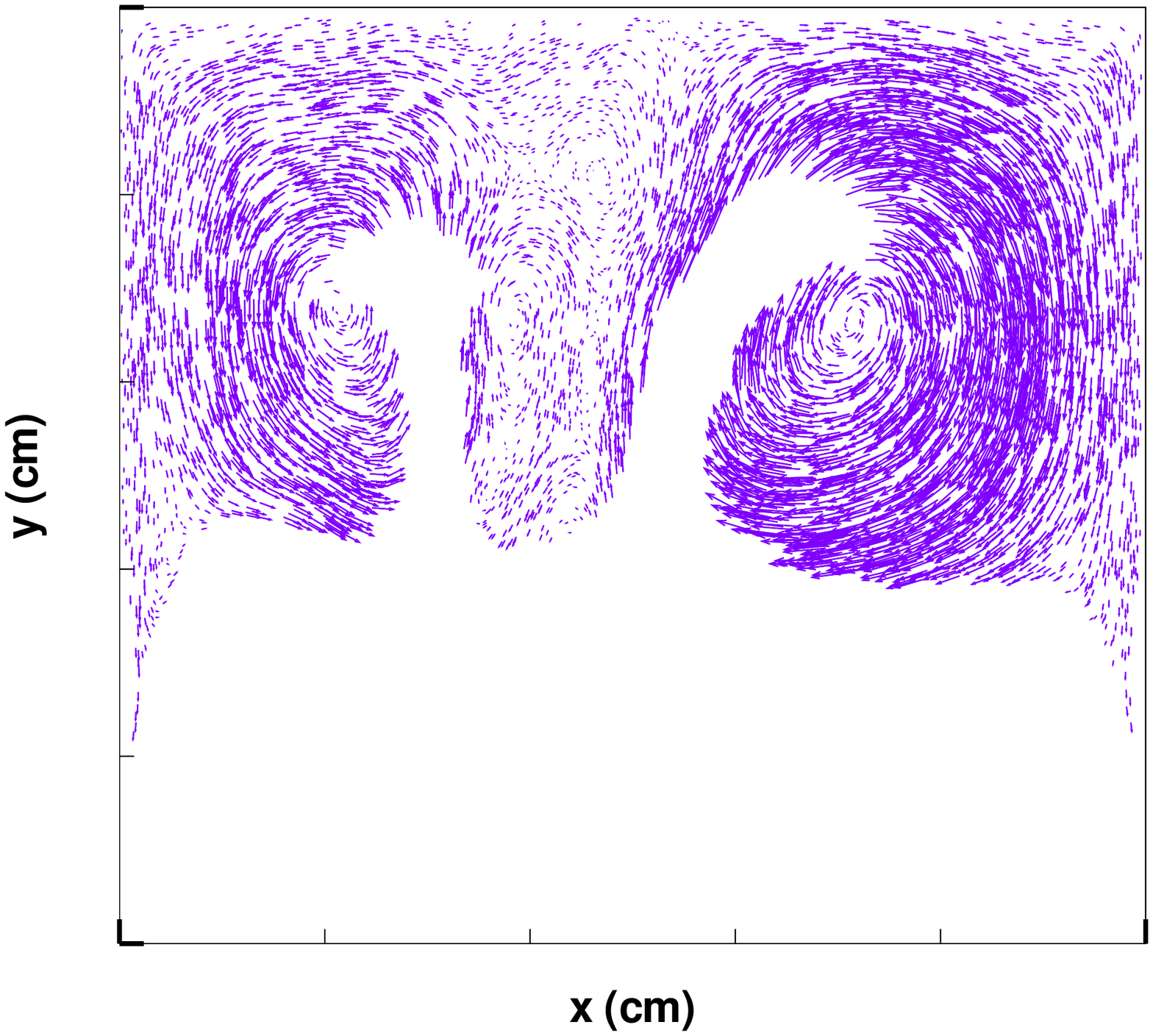}
\hspace{2mm}
\includegraphics[angle=  0,width=0.45\textwidth]{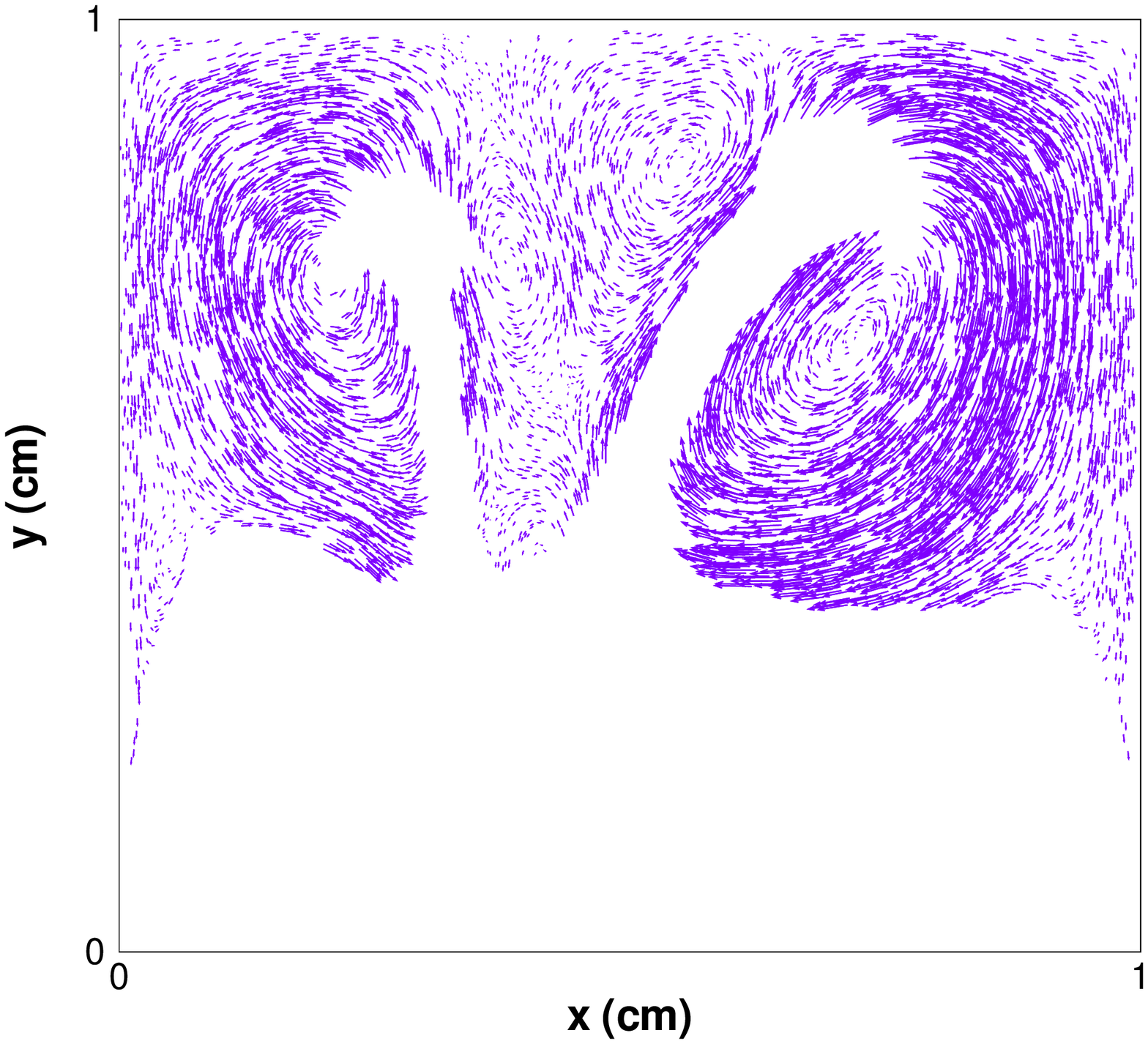}
\\
(a) \hspace{52mm} (b)
\\
\includegraphics[angle=  0,width=0.45\textwidth]{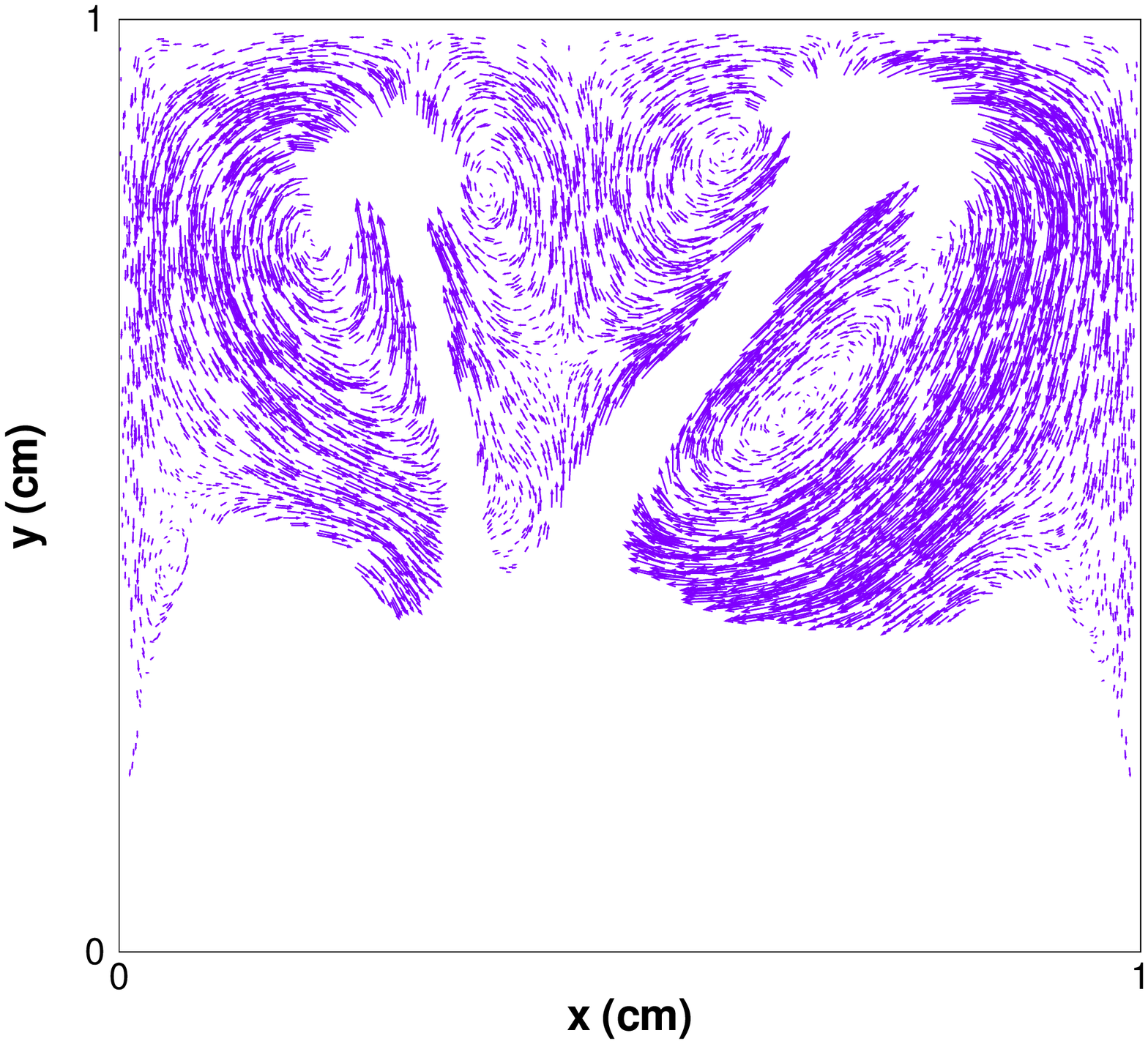}
\hspace{2mm}
\includegraphics[angle=  0,width=0.45\textwidth]{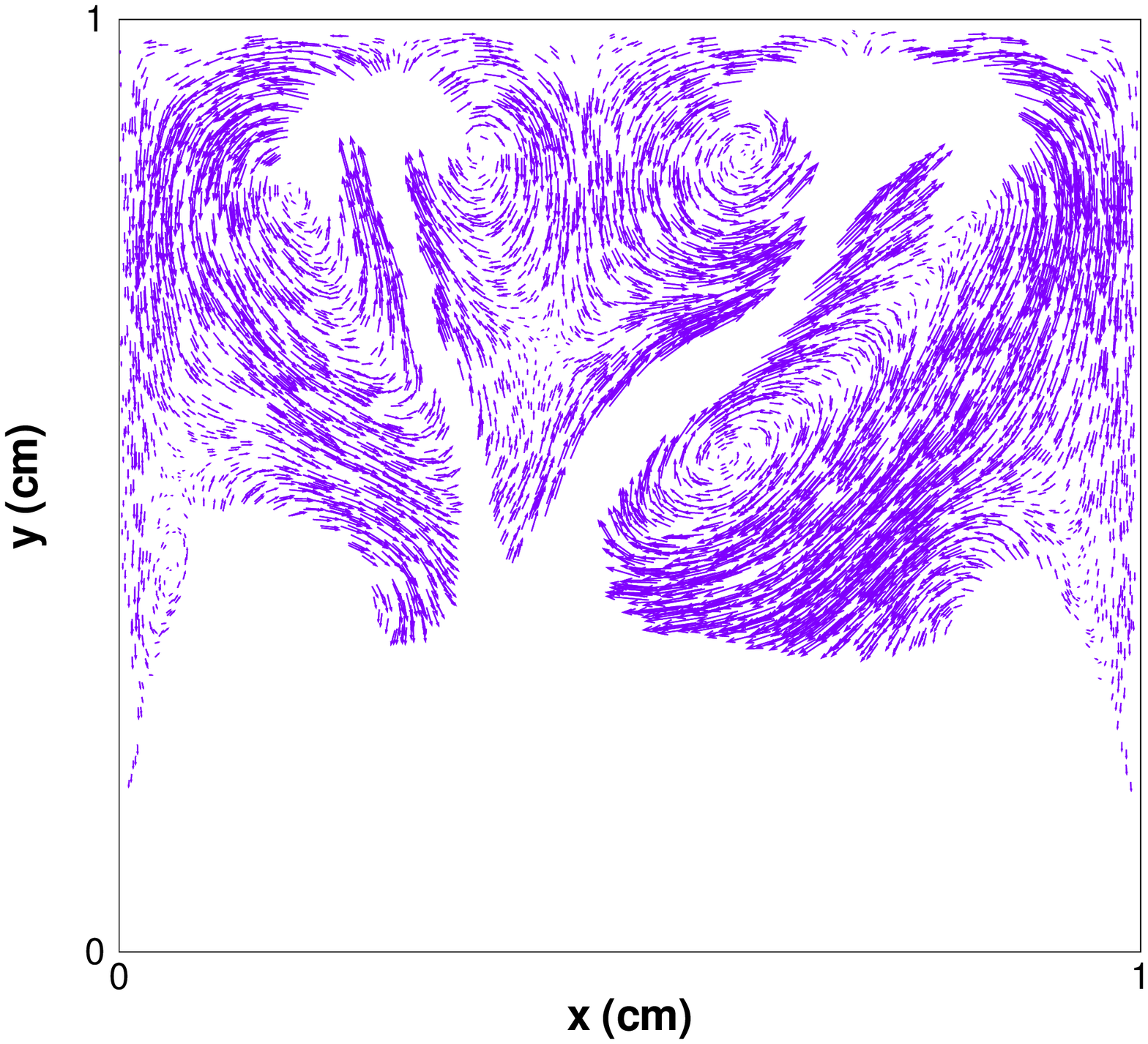}
\\
(c) \hspace{52mm} (d)
\caption{Velocity distribution of the $5000$ particles with heat transfer at time
(a) $t=12.5 s$, (b) $t=15.0 s$, (c) $t=17.5 s$, (d) $t=20.0 s$.}
\label{2dparticlevelocity3}
\end{figure}

\subsection{Sedimentation of two-dimensional isothermal particles in fluid}\label{Sedimentation2D}

From this subsection, fully coupled LBM-PIBM-DEM simulations are performed. The main target of this and next subsections
are to investigate the temperature distribution feature in the cavity during particle sedimentation and the effect of the 
thermal buoyancy on particle behaviors. The current numerical results are compared with the ready-made cases in our previous 
study without considering heat transfer~\cite{zhang2015powderEffect}.

In the two-dimensional simulation, we consider 
a $1$ $cm$ $\times$ $1$ $cm$ cavity with $5000$ two-dimensional particles. The calculating mesh for the LBM is $100\times100$. 
The diameter of the particles is $0.25\times10^{-2}$ $cm$ or $h/d_{p}=4$. The initially spacial condition is exactly 
the same as the two-dimensional case in~\cite{zhang2015powderEffect}. Firstly, the $5000$ particles are randomly generated 
in the upper three-fifths 
domain and then deposit under the effect of the gravitational force, $g=9.8$ $m\cdot s^{-2}$. The density ratio 
between solid and fluid is $1.01$, $Ra=10^{4}$, $Pr=0.71$, $L_{c}=1$ and $u_{c}=0.25$. The non-dimensional 
temperature is set $1$ and $0$ at the solid particles and the four surrounding cold walls, respectively. 
The initial temperature of the stagnant fluid is $0.5$. At last, the parameters 
responsible for the collision between the solid particles are $E=68.95$ $G Pa$ and $\nu=0.33$. 
For the sake of investigating the effect of the thermal buoyancy on particle behavior, two parallel simulations
are carried out with and without considering heat transfer. In other words, the heat-introduced buoyancy would be 
ignored in the case without considering heat transfer.

\begin{figure}
 \centering
 \includegraphics[width=0.54\textwidth]{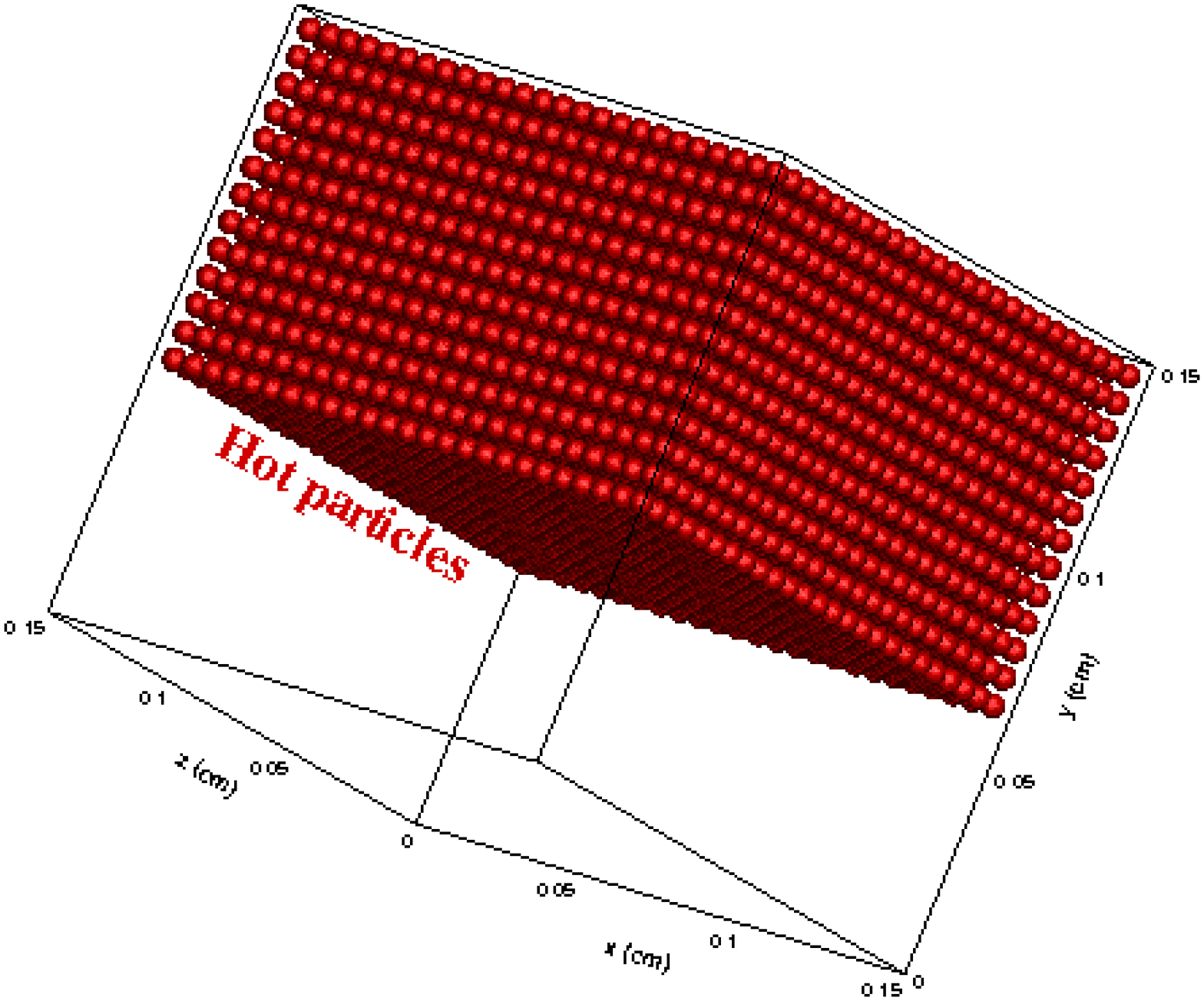}
 \vskip-0.2cm
 \caption{Positions of the 8125 isothermal particles at time $t=0.0 s$.} \label{3Dthermalinitial}
 \end{figure}
 
\begin{figure}[!ht]
\centering
\includegraphics[angle=  0,width=0.45\textwidth]{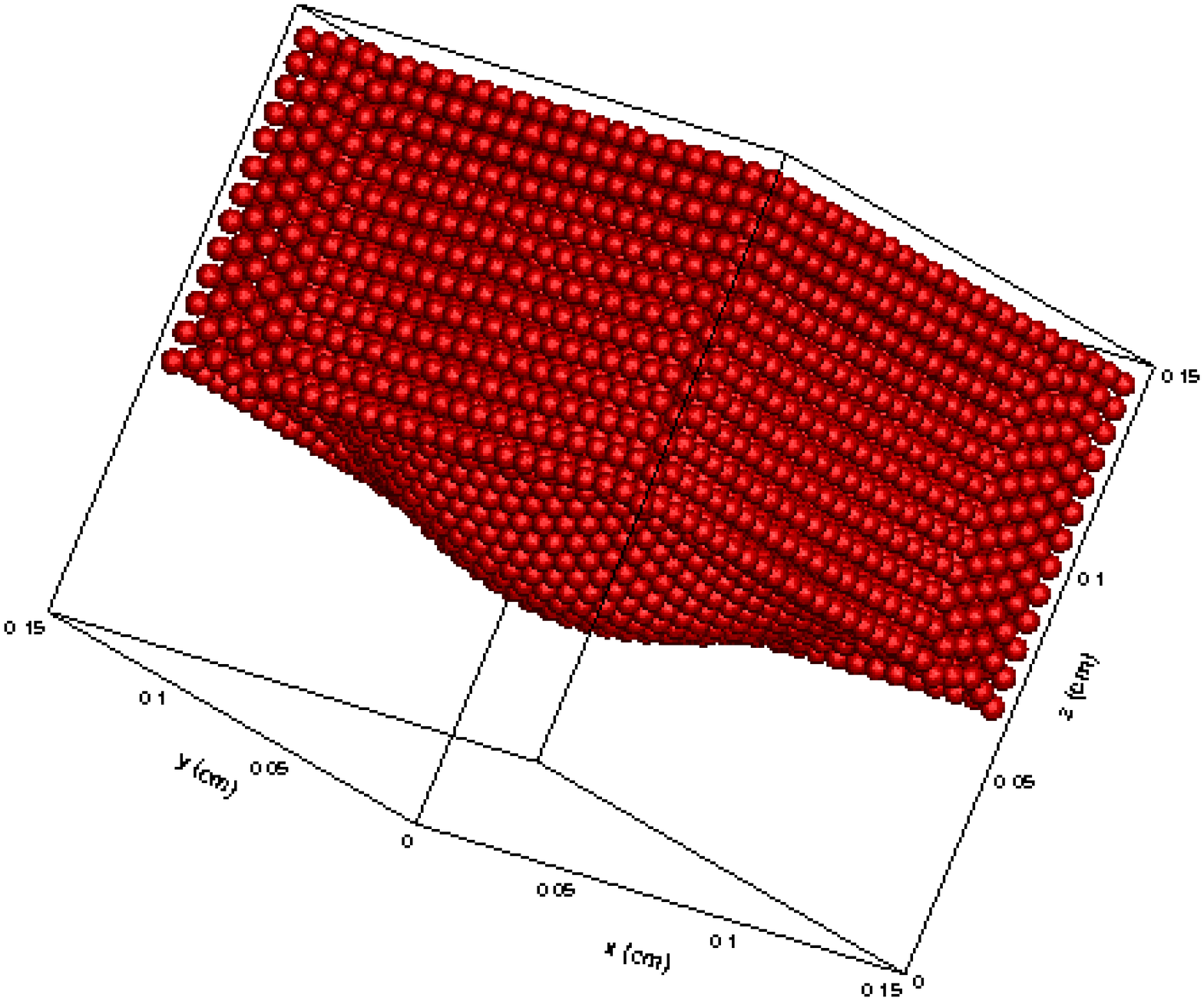}
\hspace{2mm}
\includegraphics[angle=  0,width=0.45\textwidth]{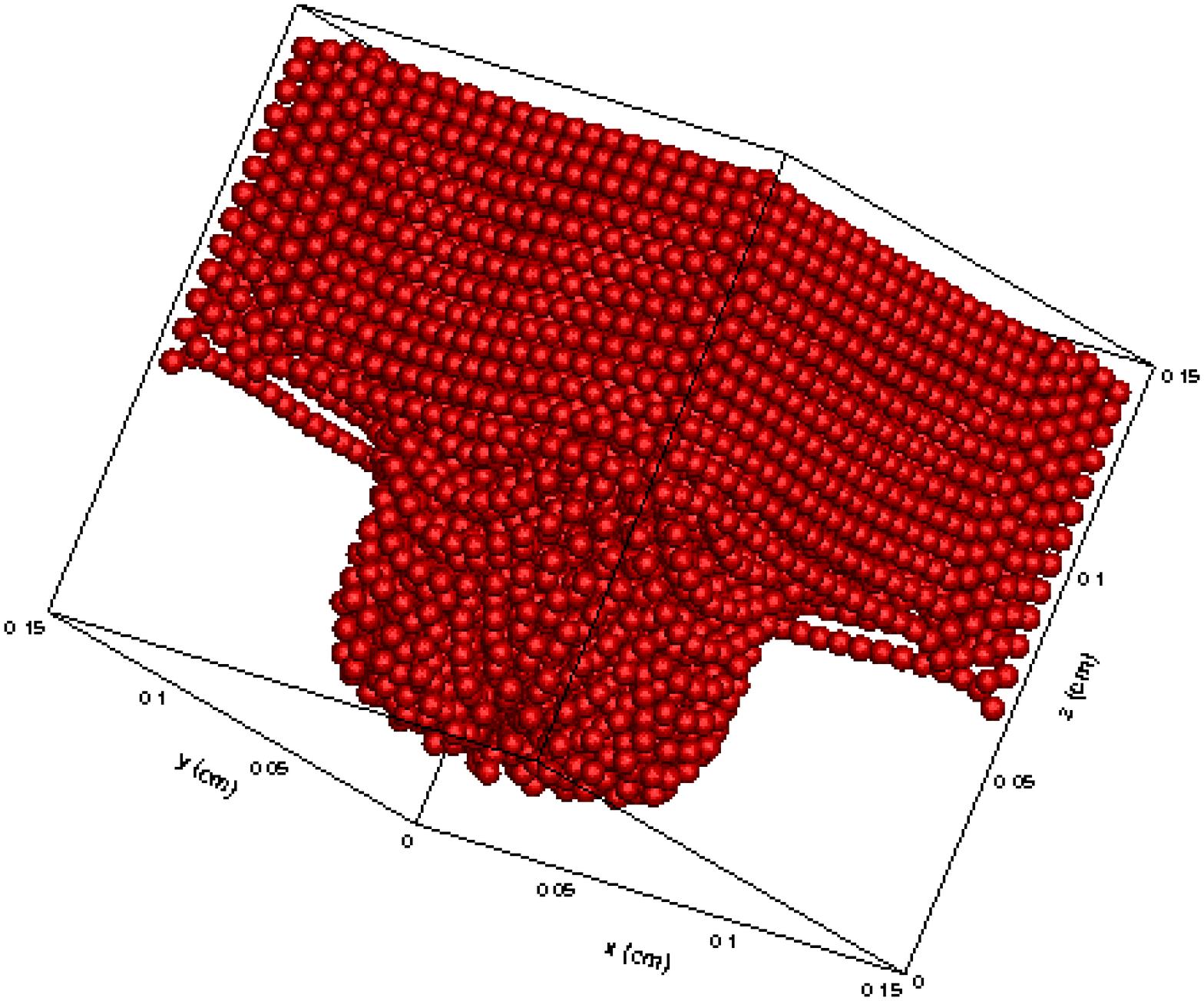}
\\
(a) \hspace{52mm} (b)
\\
\includegraphics[angle=  0,width=0.45\textwidth]{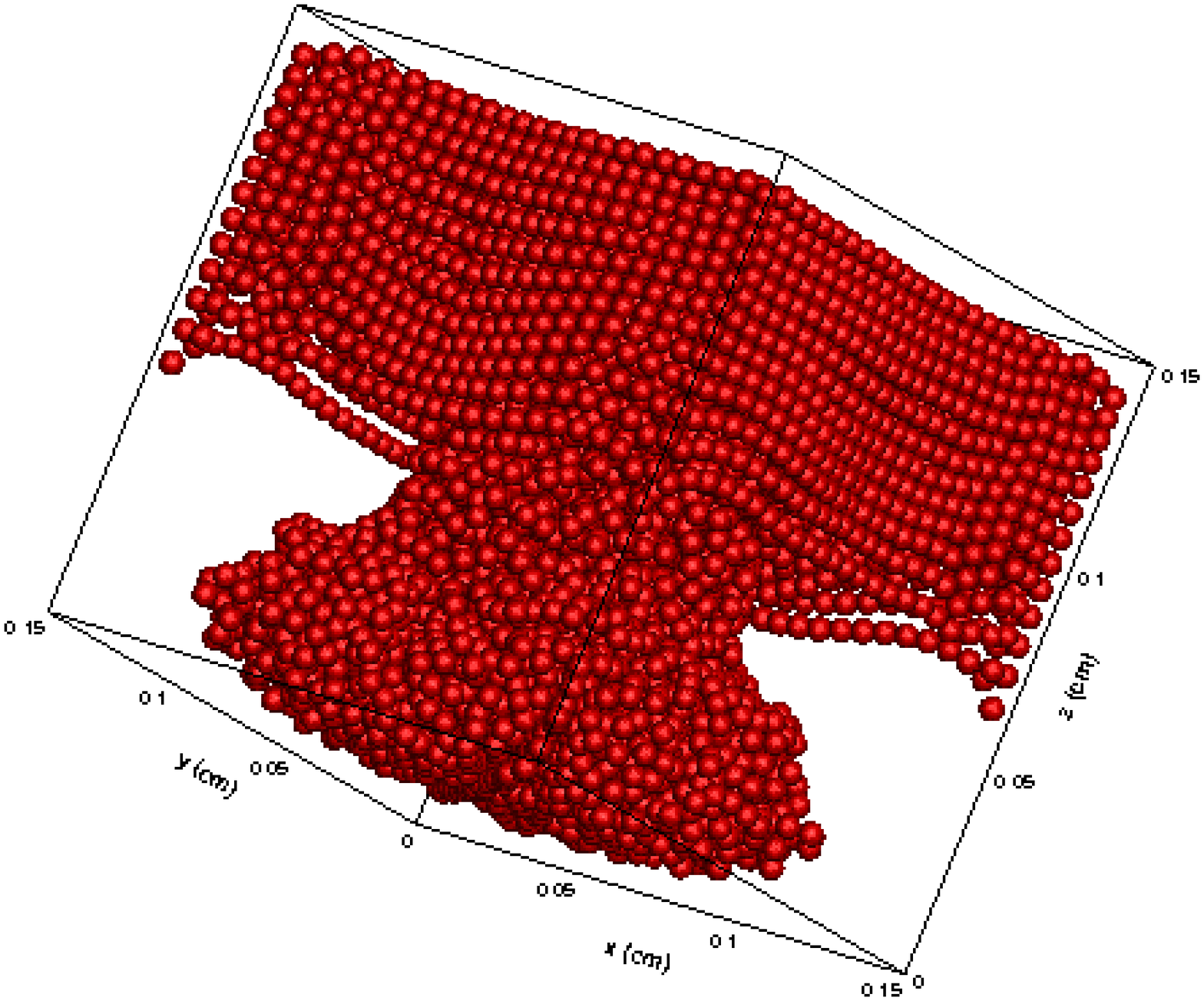}
\hspace{2mm}
\includegraphics[angle=  0,width=0.45\textwidth]{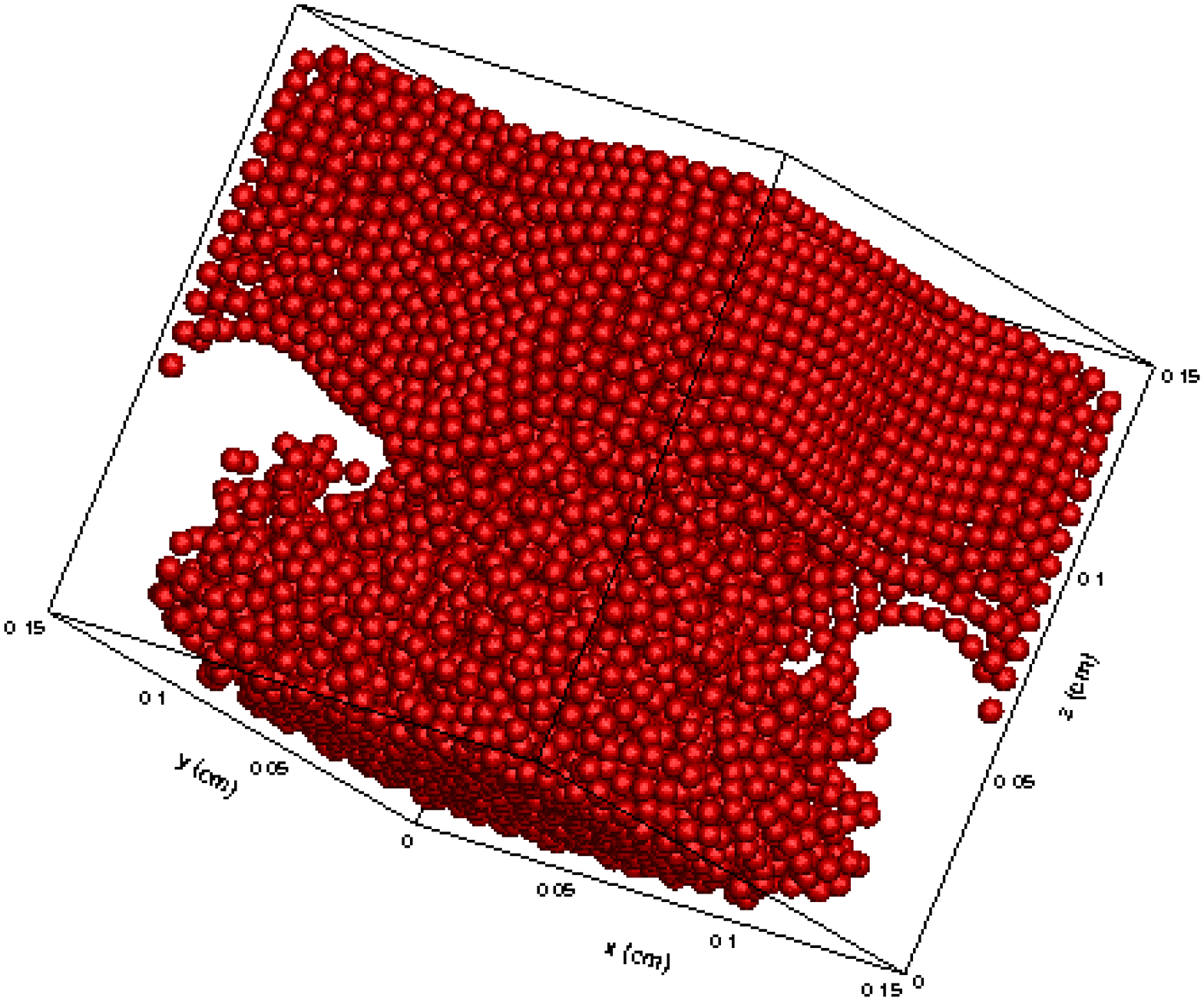}
\\
(c) \hspace{52mm} (d)
\caption{Positions of the $8125$ isothermal particles without heat transfer at time 
(a) $t=2.5 s$, (b) $t=5.0 s$, (c) $t=7.5 s$, (d) $t=10.0 s$.}
\label{3dparticleposition1}
\end{figure}

\begin{figure}[!ht]
\centering
\includegraphics[angle=  0,width=0.45\textwidth]{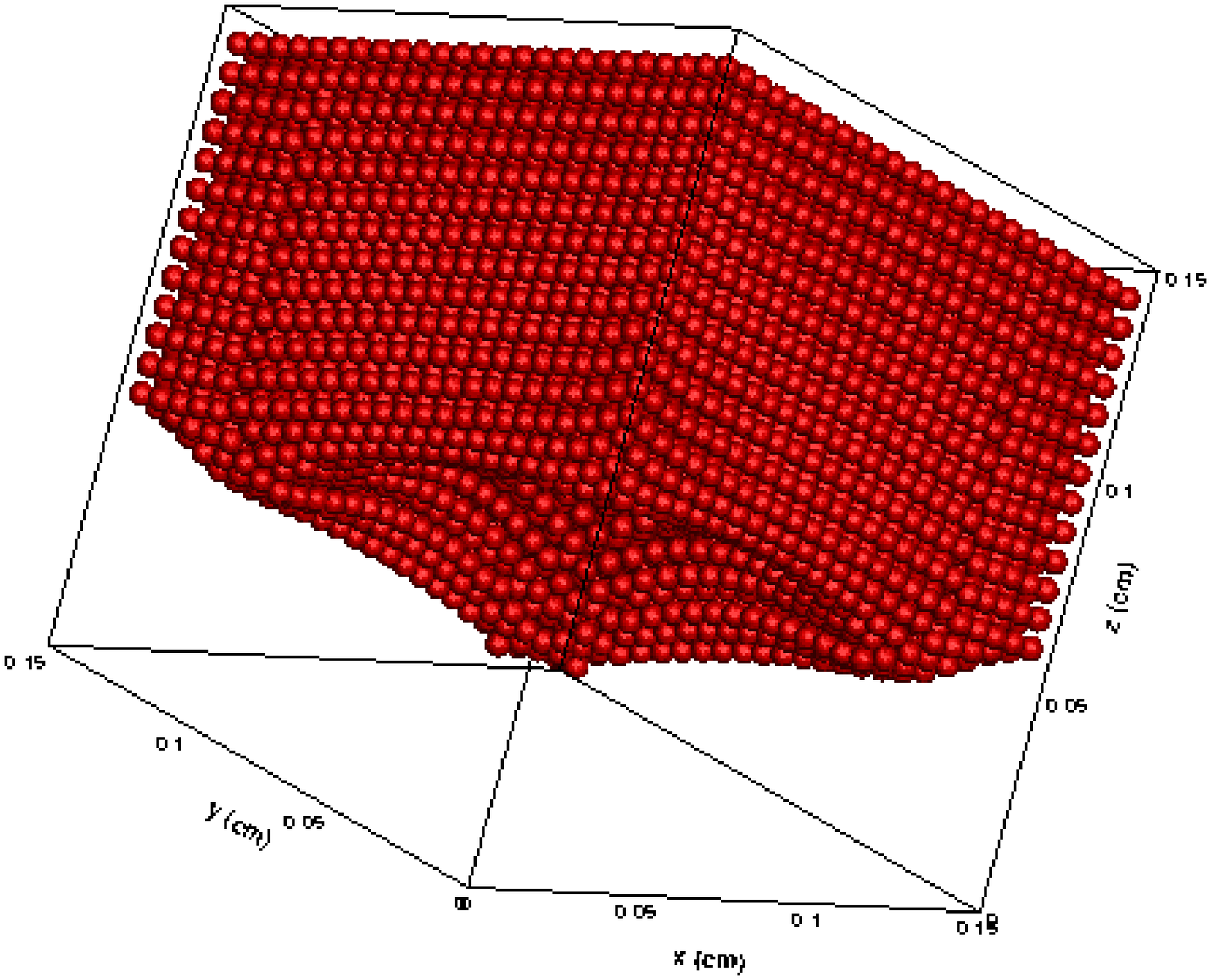}
\hspace{2mm}
\includegraphics[angle=  0,width=0.45\textwidth]{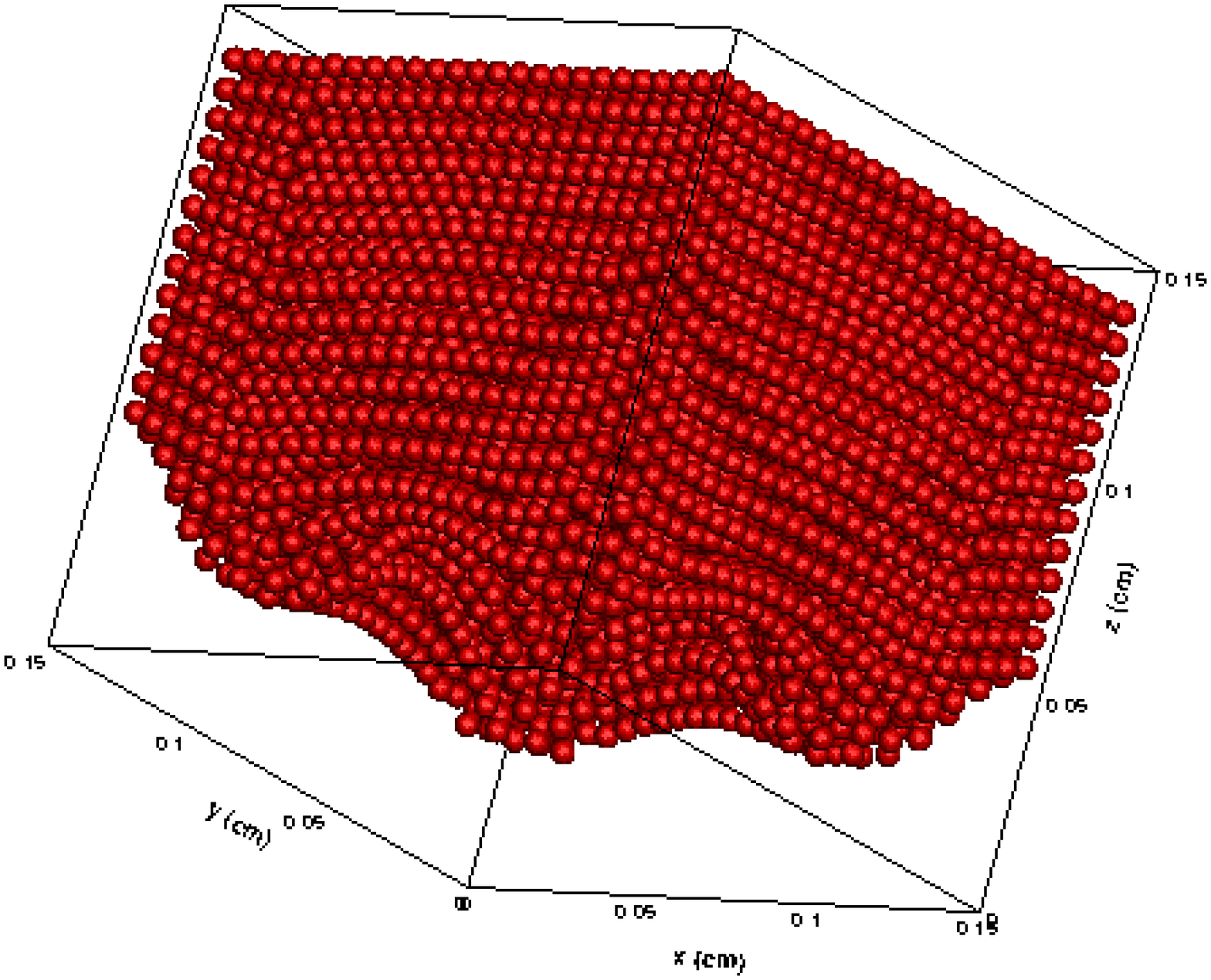}
\\
(a) \hspace{52mm} (b)
\\
\includegraphics[angle=  0,width=0.45\textwidth]{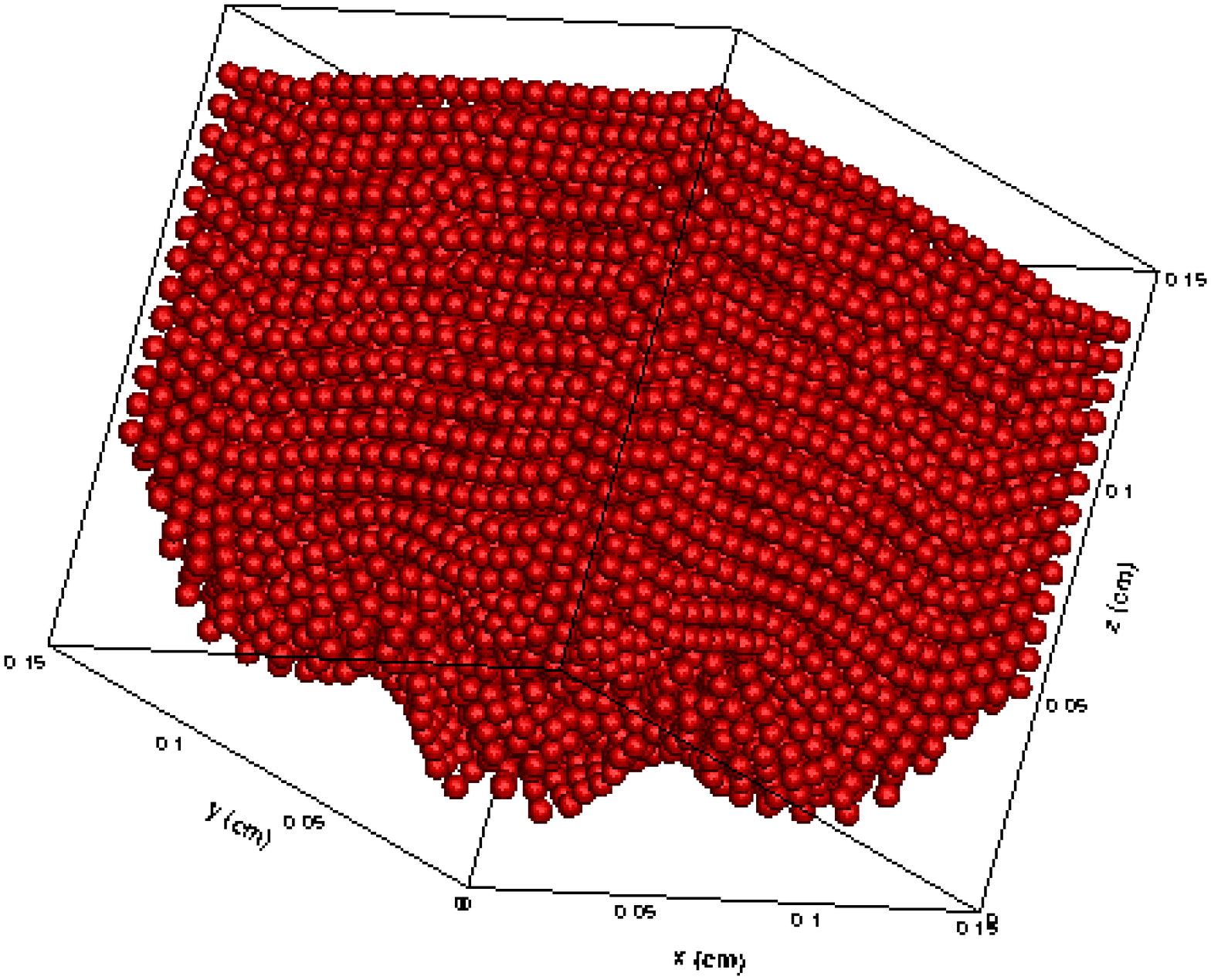}
\hspace{2mm}
\includegraphics[angle=  0,width=0.45\textwidth]{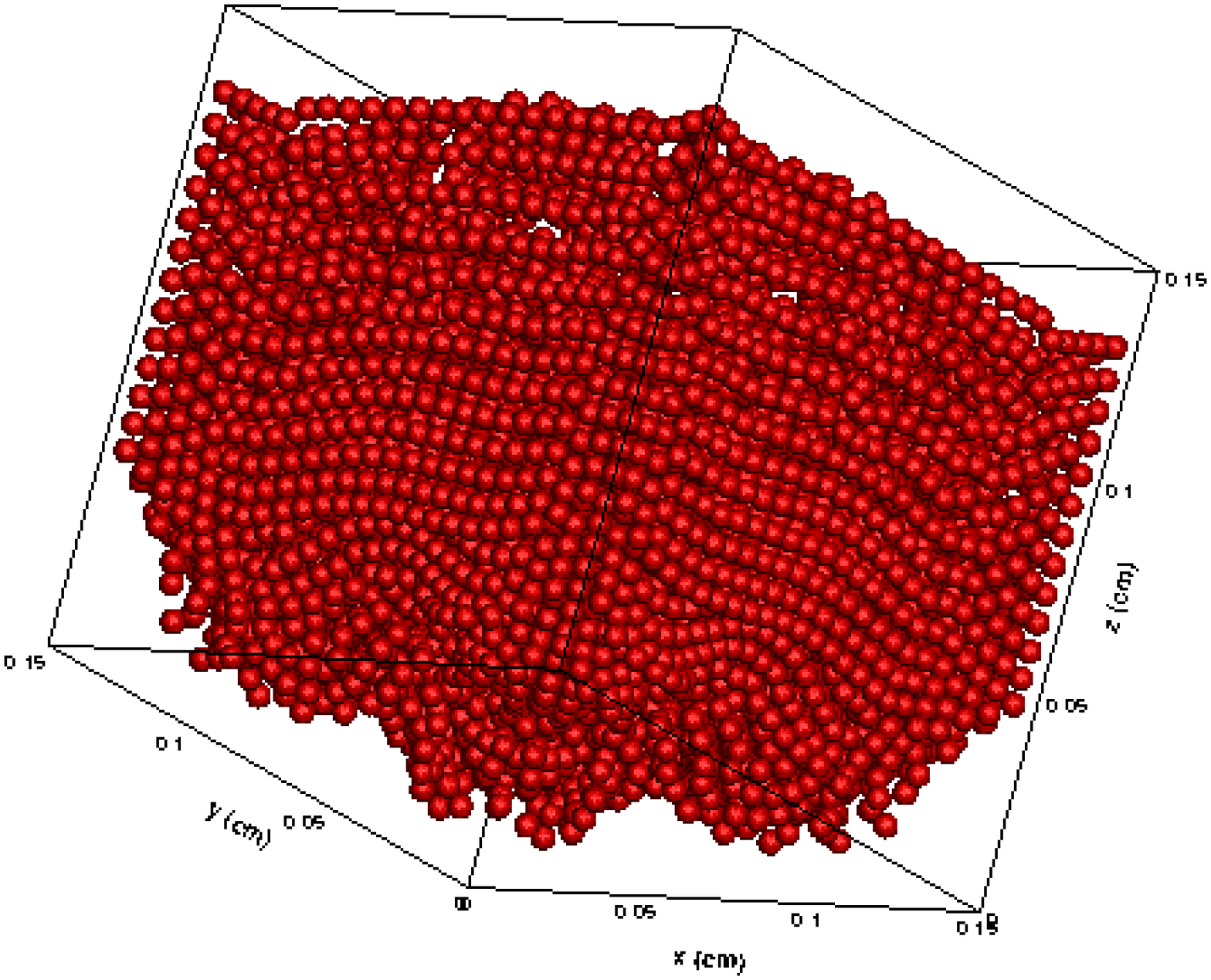}
\\
(c) \hspace{52mm} (d)
\caption{Positions of the $8125$ isothermal particles with heat transfer at time
(a) $t=2.5 s$, (b) $t=5.0 s$, (c) $t=7.5 s$, (d) $t=10.0 s$.}
\label{3dparticleposition2}
\end{figure}
 
 Figure~\ref{2dparticleposition1} displays the particle distribution during the beginning $10.0 s$ when heat transfer 
 is not considered. As shown, the flow fluid at lower position is swallowed into the particle aggregate forming a 
 fluid pocket with a mushroom shape. This typical two-dimensional phenomenon is regarded as the so-called Rayleigh-Taylor 
 instability which has been well investigated by several 
 studies~\cite{glowinski1999distributed,Feng2004602,ZHANGCAF2014,zhang2015powderEffect}. The solid particles prefer to 
 settle around the fluid column until being shot up by the upthrust flow field. This process can be clearly seen 
 from the distribution map of particle velocities as shown in Figure~\ref{2dparticlevelocity1}. 
 Firstly, two vortexes with solid particles are formed in the interior of the cavity. Then, 
 the height of these two vortexes rises meanwhile the downthrust of the particle streams introduces two more small 
 vortexes close to the lower corners of the cavity due to the intimate interaction between the particle and fluid 
 as shown in Figure~\ref{2dparticlevelocity1}(d).

However, the sedimentation is fairly delayed when heat transfer is considered. Figure~\ref{2dparticlevelocity2} shows
the velocity distribution of the $5000$ particles with heat transfer at the same moments as Figure~\ref{2dparticlevelocity1}.
Comparing with the case without heat transfer, the settling velocities of the thermal particles are significantly lower.
This is due to the fact that the particle temperature is higher than the surrounding fluid. The fluid receives heat 
from the solid particles meanwhile the wall temperature 
is the lowest. Therefore, the fluid temperature increases at the location close to the solid particles whereas decreases 
close to the cold walls as shown in Figure~\ref{Temperaturedistribution1}. It can be seen that the high temperature region is 
in conformity with the particle position. The temperature gradients make the density differences in the fluid occur. 
This gives rise to that the fluid in the cavity interior becomes less dense and rises. The solid particles at this region
also rise with the ascending fluid which prevents the evolution in Figure~\ref{2dparticleposition1}.  Besides the overall 
settling velocity, there is also large discrepancy on the particle distribution patterns when heat transfer is considered. 
Instead of one fluid pocket with a mushroom shape in Figure~\ref{2dparticleposition1}, two branches are generated at the 
interface of solid and fluid as shown in Figure~\ref{2dparticlevelocity2}. The whole system is more unstable when 
heat transfer is introduced.

Figure~\ref{2dparticlevelocity3} displays the further evolution of the velocity distribution of the $5000$ particles when heat 
transfer is considered. As shown, the above-mentioned two branches in Figure~\ref{2dparticlevelocity2} grow to two fluid 
pockets also with mushroom shapes. The two heads do not rise vertically any longer but towards the upper corners of the cavity.
This phenomenon is not observed in the simulation without heat transfer before
~\cite{glowinski1999distributed,Feng2004602,ZHANGCAF2014,zhang2015powderEffect}. It seems that the thermal buoyancy totally 
breaks the original rule and introduces more intensive fluid-particle interactions.

\begin{figure}
 \centering
 \includegraphics[width=0.54\textwidth]{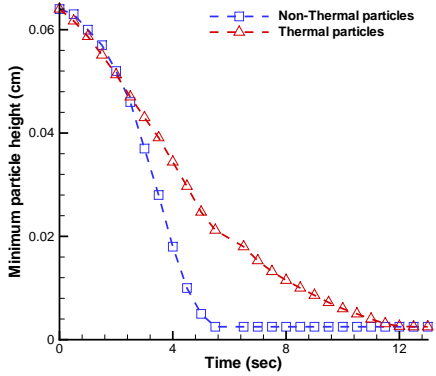}
 \vskip-0.2cm
 \caption{Minimum particle height versus time with and without heat transfer.} \label{droptime}
 \end{figure}
 
\subsection{Sedimentation of three-dimensional isothermal particles in fluid}\label{Sedimentation3D}

For the sake of conducting further investigation on the effect of thermal buoyancy on the particle behaviors, in this 
subsection, a three-dimensional $0.15$ $cm$ $\times$ $0.15$ $cm$ $\times$ $0.15$ $cm$ cubic cavity is considered with 
six cold walls. The calculating mesh for the LBM is $15\times15\times15$. The diameter of the solid particles is 
$0.5\times10^{-2}$ $cm$ or $h/d_{p}=2$. The initial spacial set of the three-dimensional case is the same as the 
three-dimensional case in ~\cite{zhang2015powderEffect} which also can be seen in Figure~\ref{3Dthermalinitial}.  
$8125$ particles are regularly planted in the upper three-fifths domain of the cubic cavity with an identical 
separation distance, $0.001$ $cm$. This leads to the local porosity, $0.719$. 
The non-dimensional 
temperature is set $1$ and $0$ at the solid particles and the six surrounding cold walls, respectively. 
The initial temperature of the stagnant fluid is $0.5$.
Other physical parameters of the fluid and 
particles are the same as the two-dimensional case in Section~\ref{Sedimentation2D}. Similarly, two parallel simulations 
are carried out with and without considering heat transfer.

\begin{figure}[!ht]
\centering
\includegraphics[angle=  0,width=0.45\textwidth]{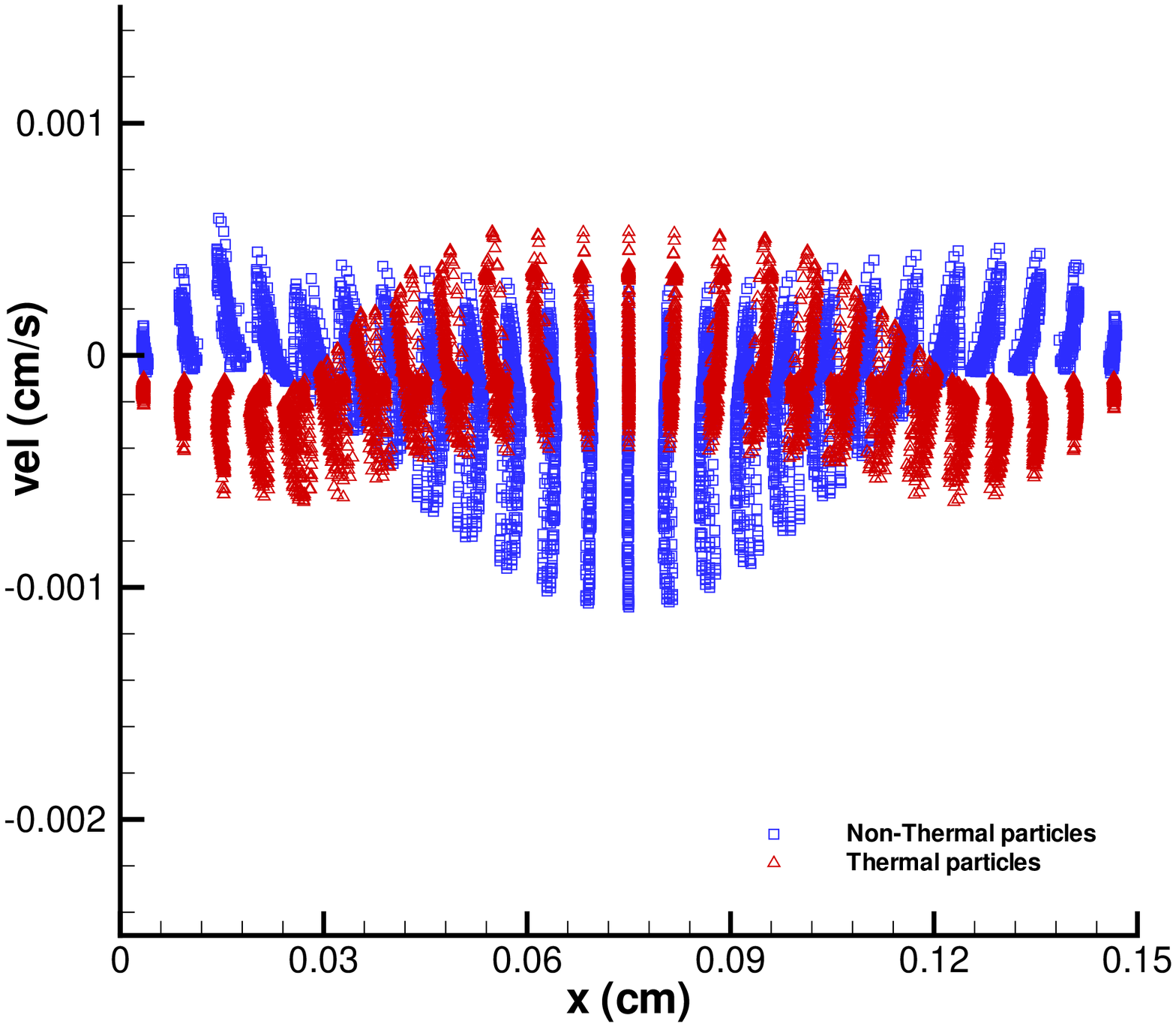}
\hspace{2mm}
\includegraphics[angle=  0,width=0.45\textwidth]{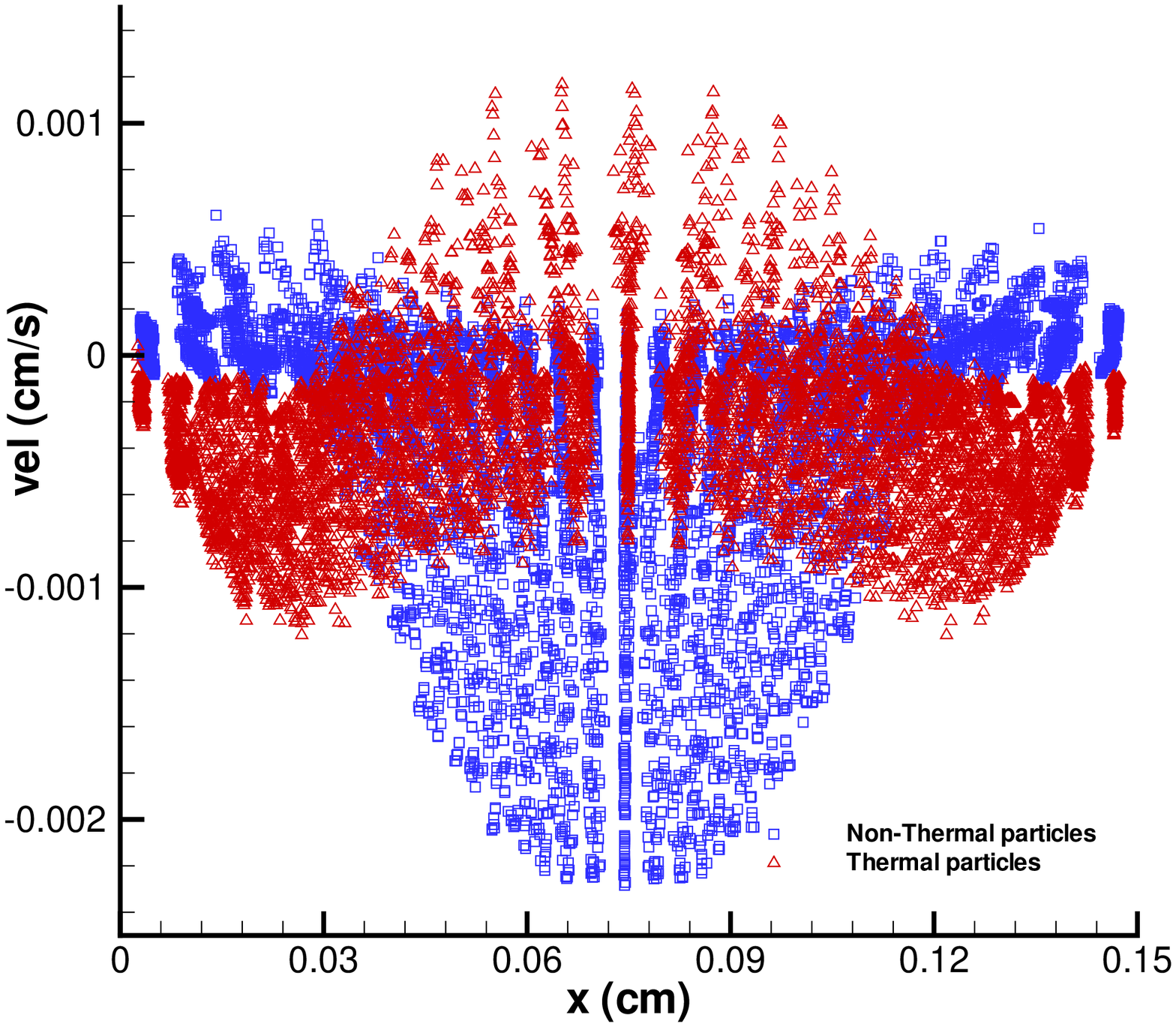}
\\
(a) \hspace{52mm} (b)
\\
\includegraphics[angle=  0,width=0.45\textwidth]{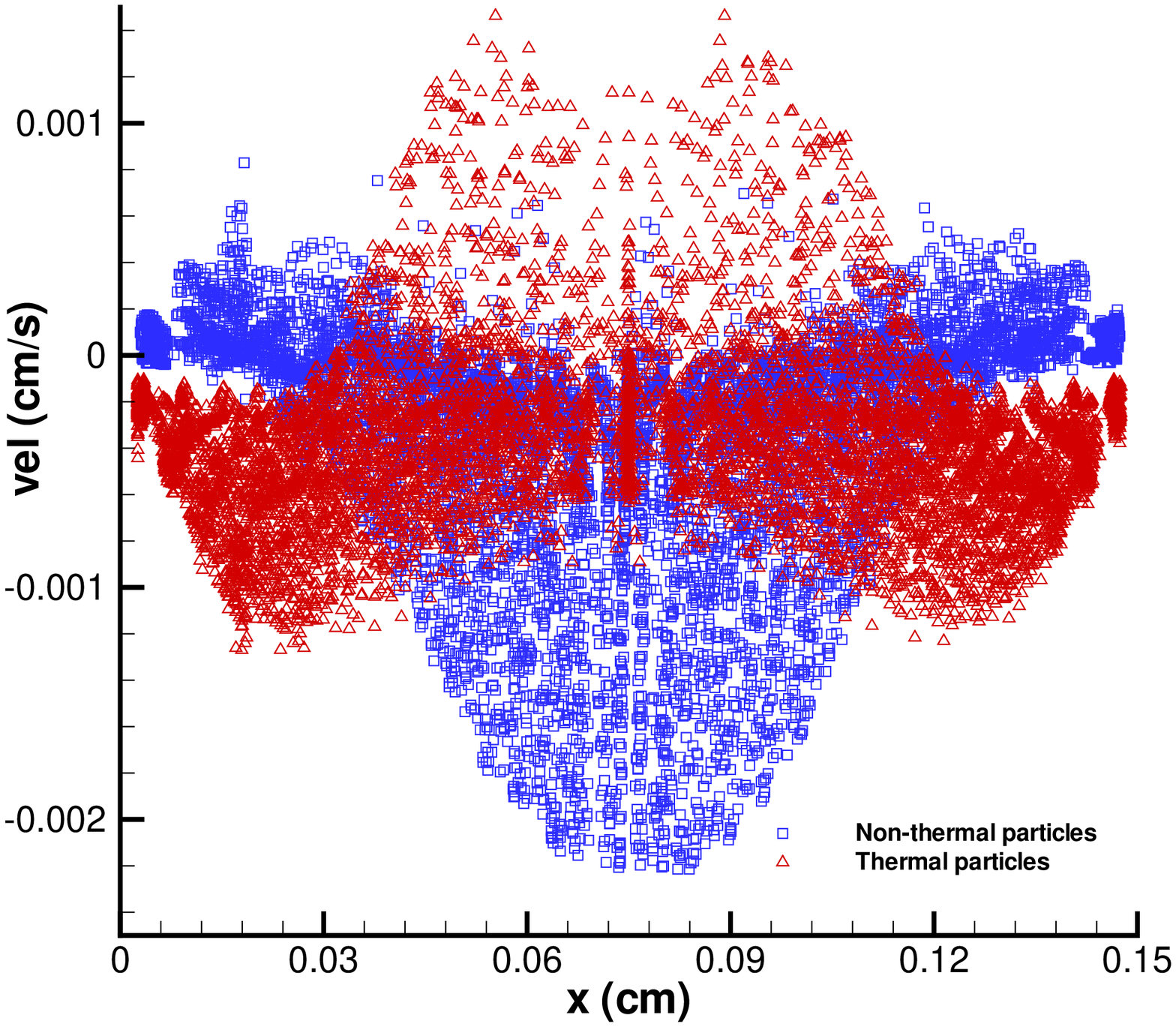}
\hspace{2mm}
\includegraphics[angle=  0,width=0.45\textwidth]{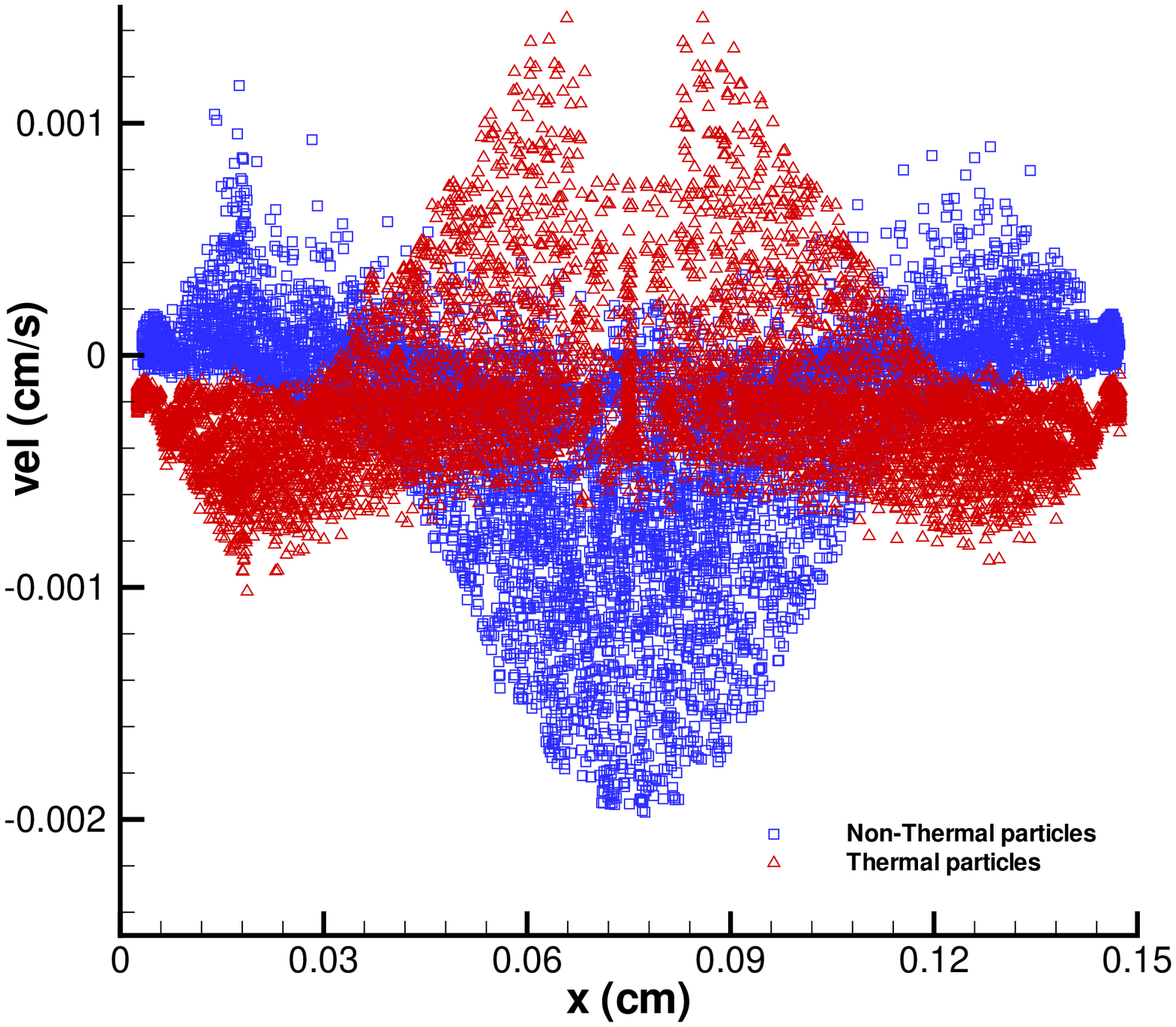}
\\
(c) \hspace{52mm} (d)
\caption{Particle deposition velocity along the x-direction with and without heat transfer at time 
(a) $t=2.5 s$, (b) $t=5.0 s$, (c) $t=7.5 s$, (d) $t=10.0 s$.}
\label{3dparticlevelocity}
\end{figure}

 Figure~\ref{3dparticleposition1} displays the three-dimensional particle distribution during the beginning $10.0 s$ when 
 heat transfer is not considered. As shown, unlike the two-dimensional case, the three-dimensional particles in the center 
 region settle more efficiently than others. The particles close to the walls move fairly slow as a result of the no-slip 
 boundary condition employed between the fluid and wall. 
 The particle matrix is rapidly hollowed out forming a downstream particle-constructed pestle. The particle-constructed 
 pestle falls down directly until impacts on the bottom and then the particles scatter in all directions. This process has 
 been reported in~\cite{zhang2015powderEffect,Robinson2014121} without considering heat transfer.

\begin{figure}[!ht]
\centering
\includegraphics[angle=  0,width=0.45\textwidth]{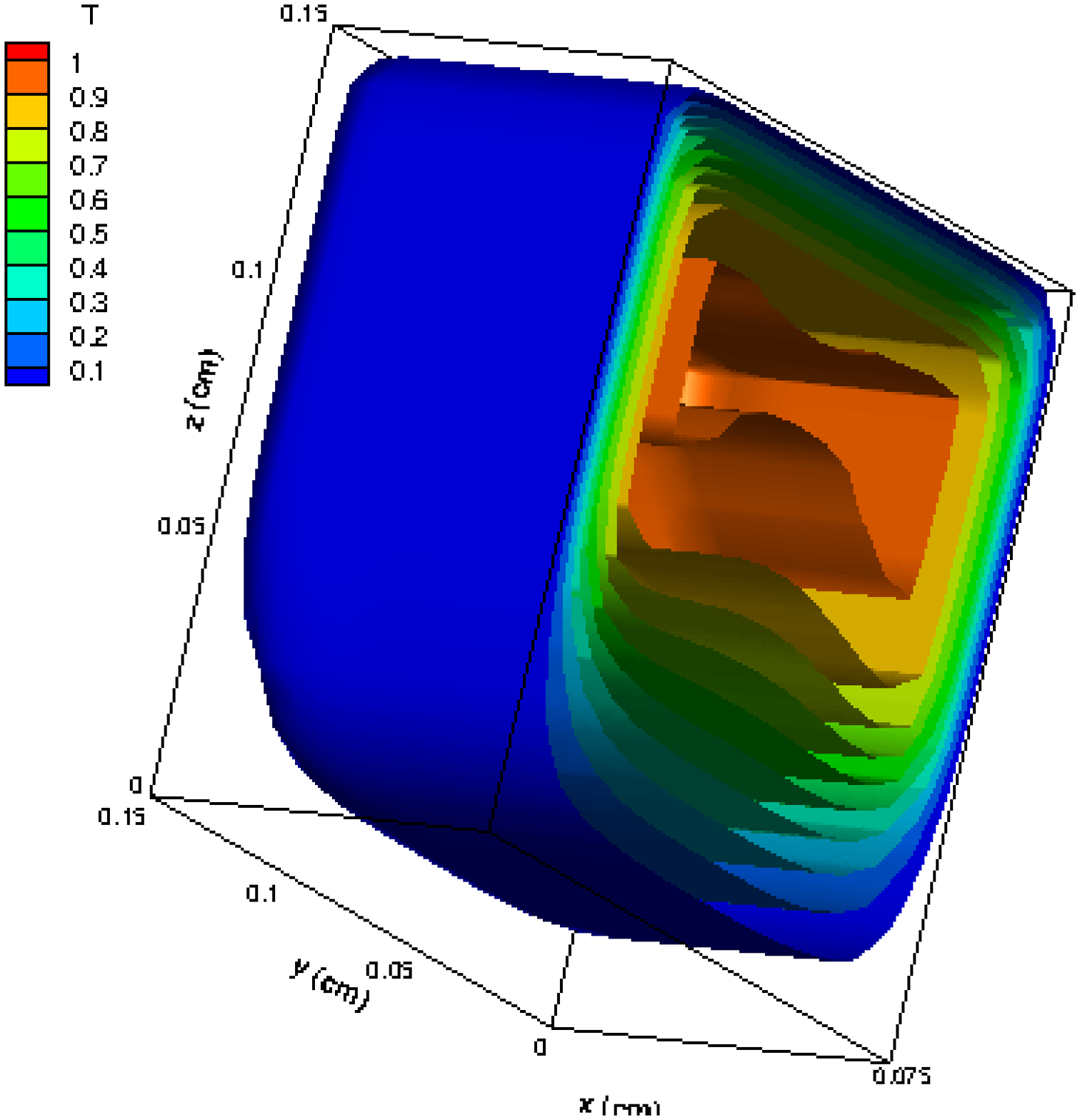}
\hspace{2mm}
\includegraphics[angle=  0,width=0.45\textwidth]{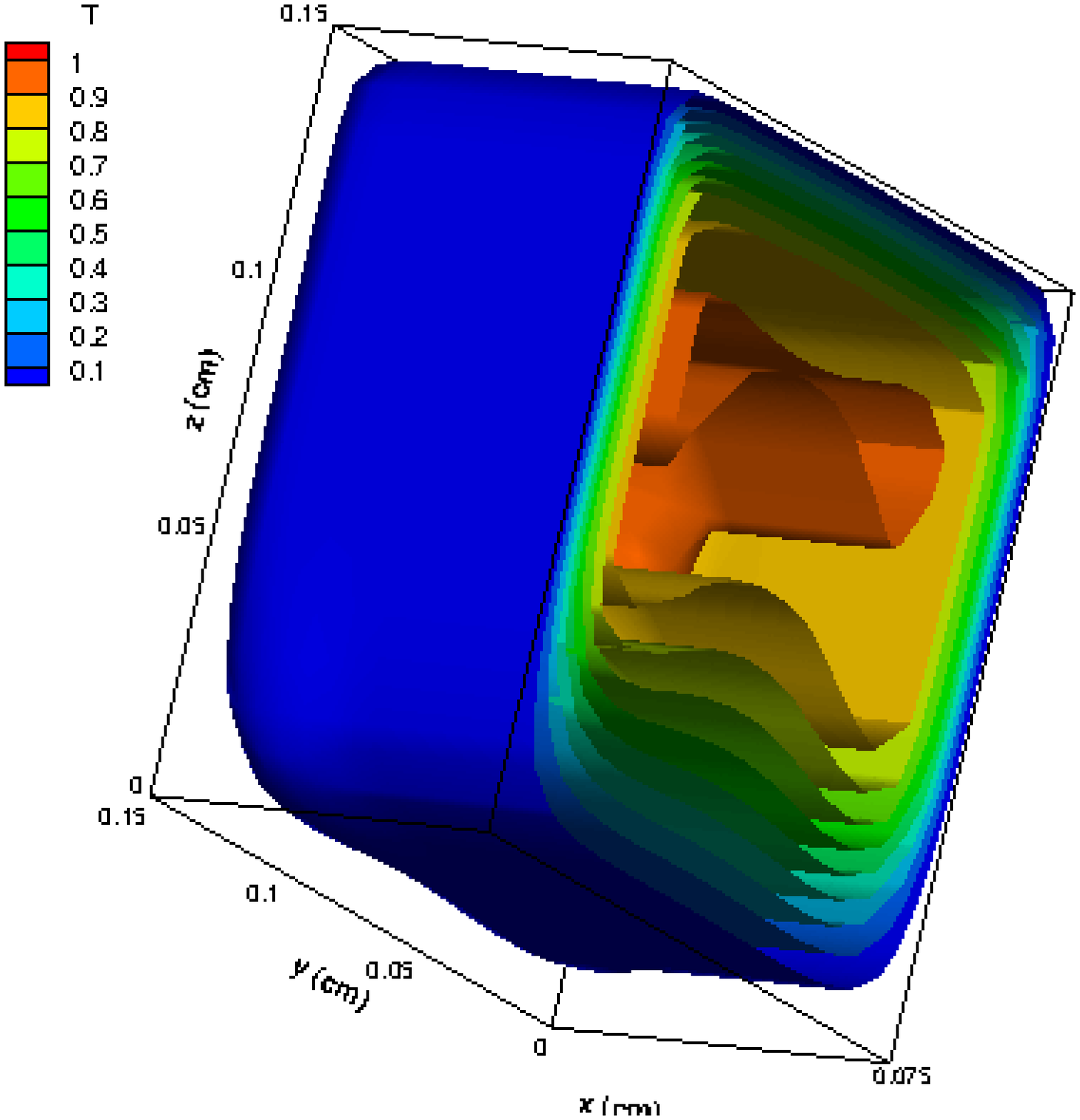}
\\
(a) \hspace{52mm} (b)
\\
\includegraphics[angle=  0,width=0.45\textwidth]{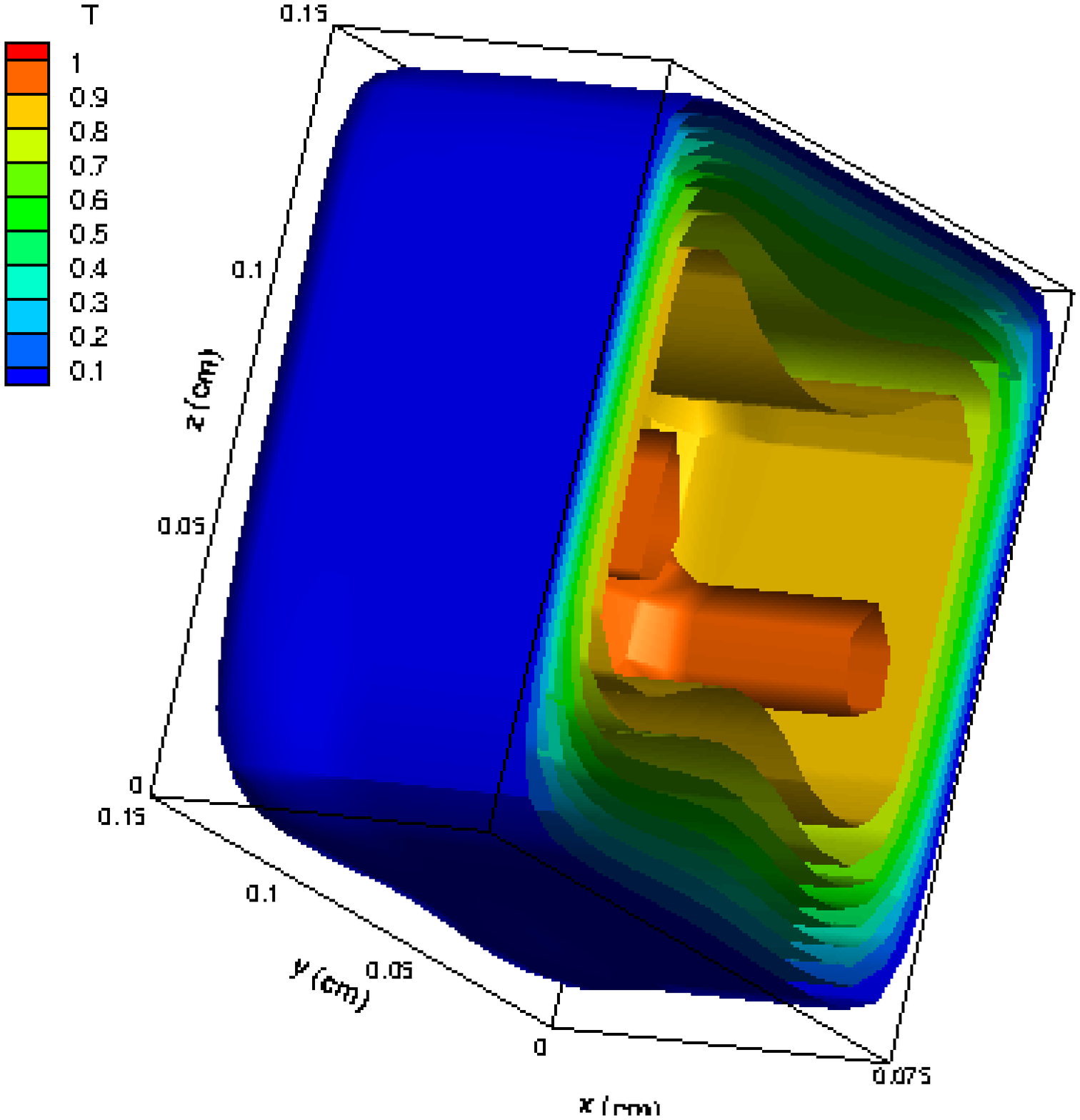}
\hspace{2mm}
\includegraphics[angle=  0,width=0.45\textwidth]{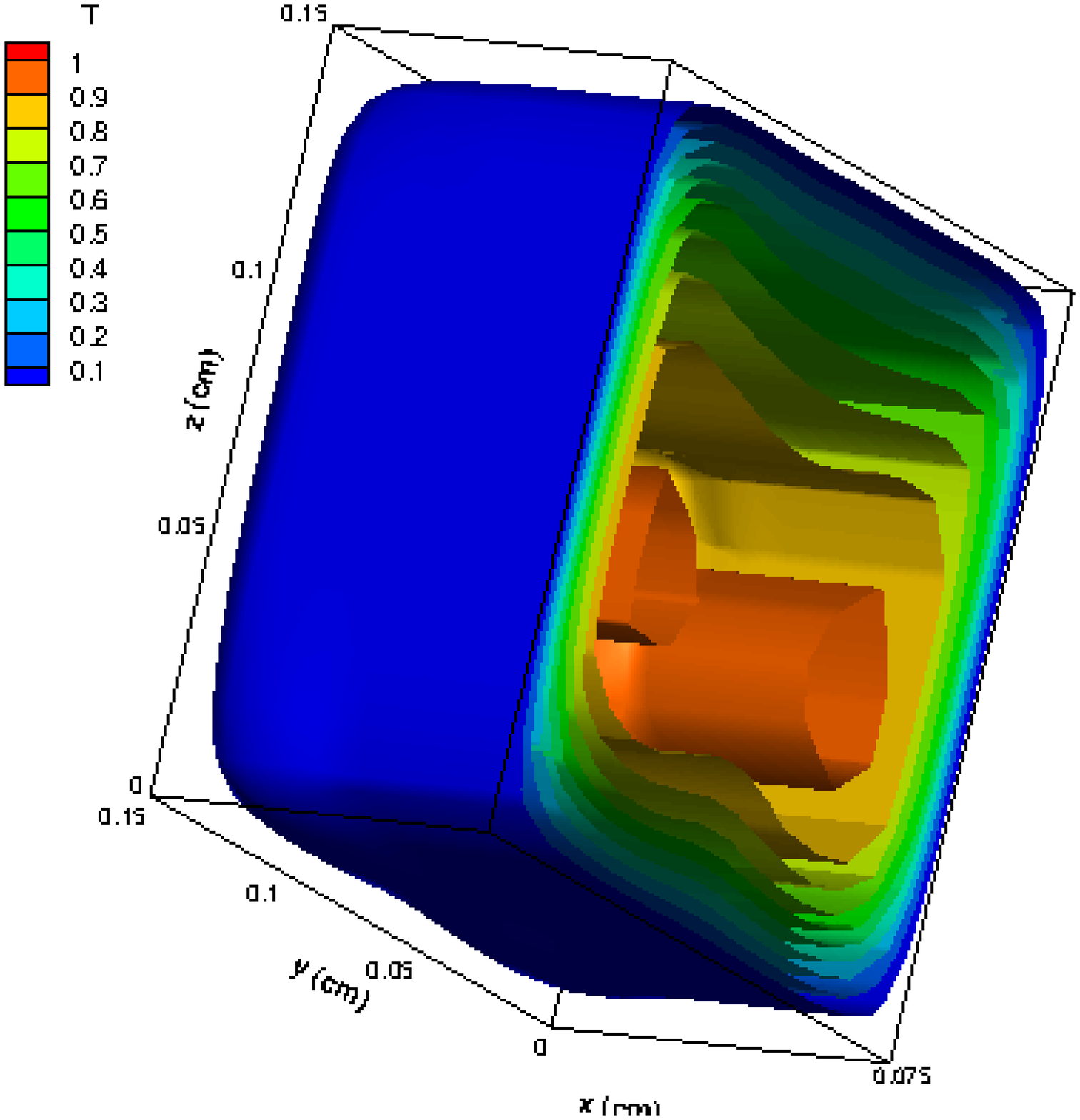}
\\
(c) \hspace{52mm} (d)
\caption{Isothermal surfaces in the cubic cavity at time 
(a) $t=2.5 s$, (b) $t=5.0 s$, (c) $t=7.5 s$, (d) $t=10.0 s$.}
\label{3dIsosurfaces}
\end{figure}
  
\begin{figure}[!ht]
\centering
\includegraphics[angle=  0,width=0.45\textwidth]{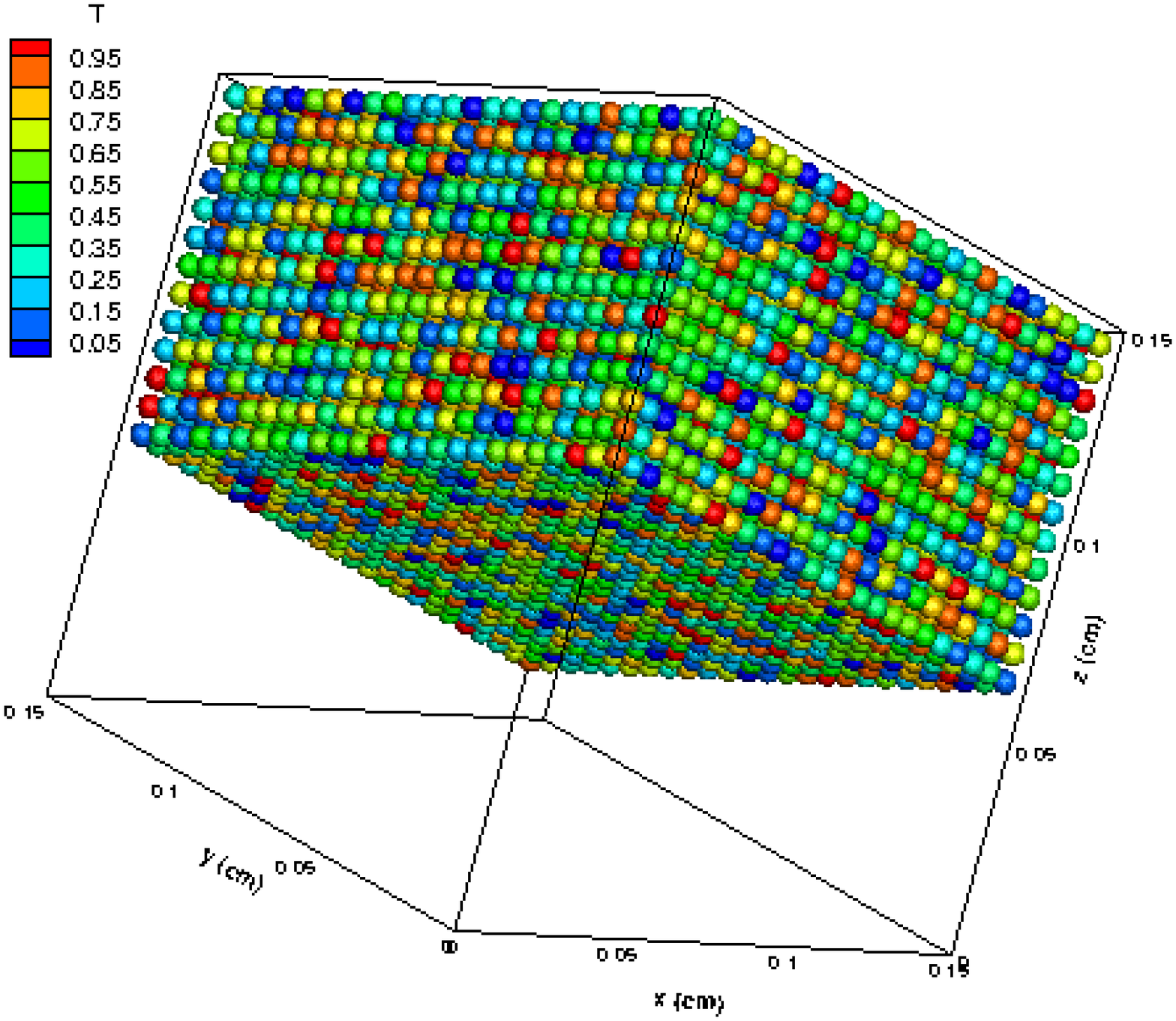}
\hspace{2mm}
\includegraphics[angle=  0,width=0.45\textwidth]{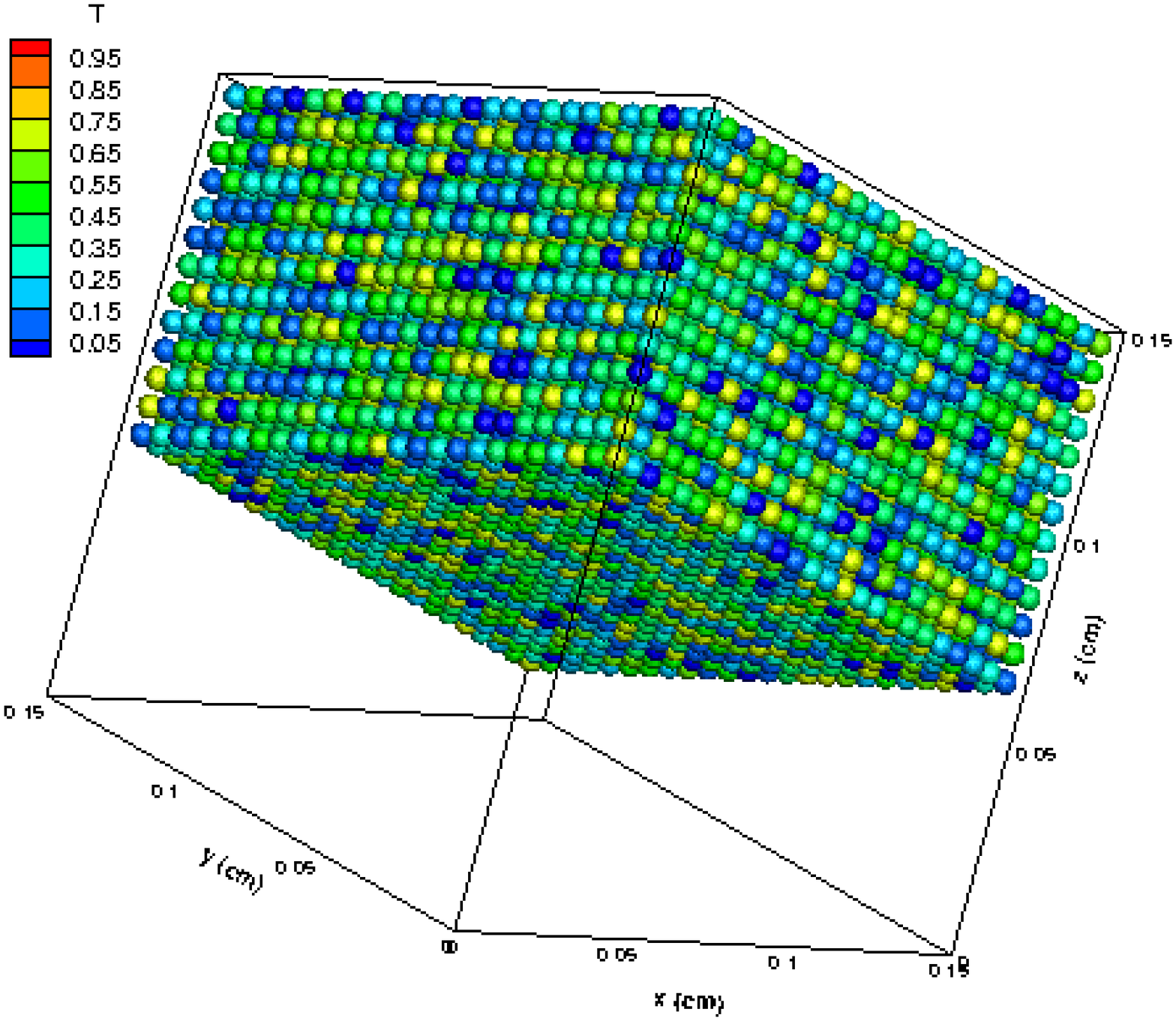}
\\
(a) \hspace{52mm} (b)
\\
\includegraphics[angle=  0,width=0.45\textwidth]{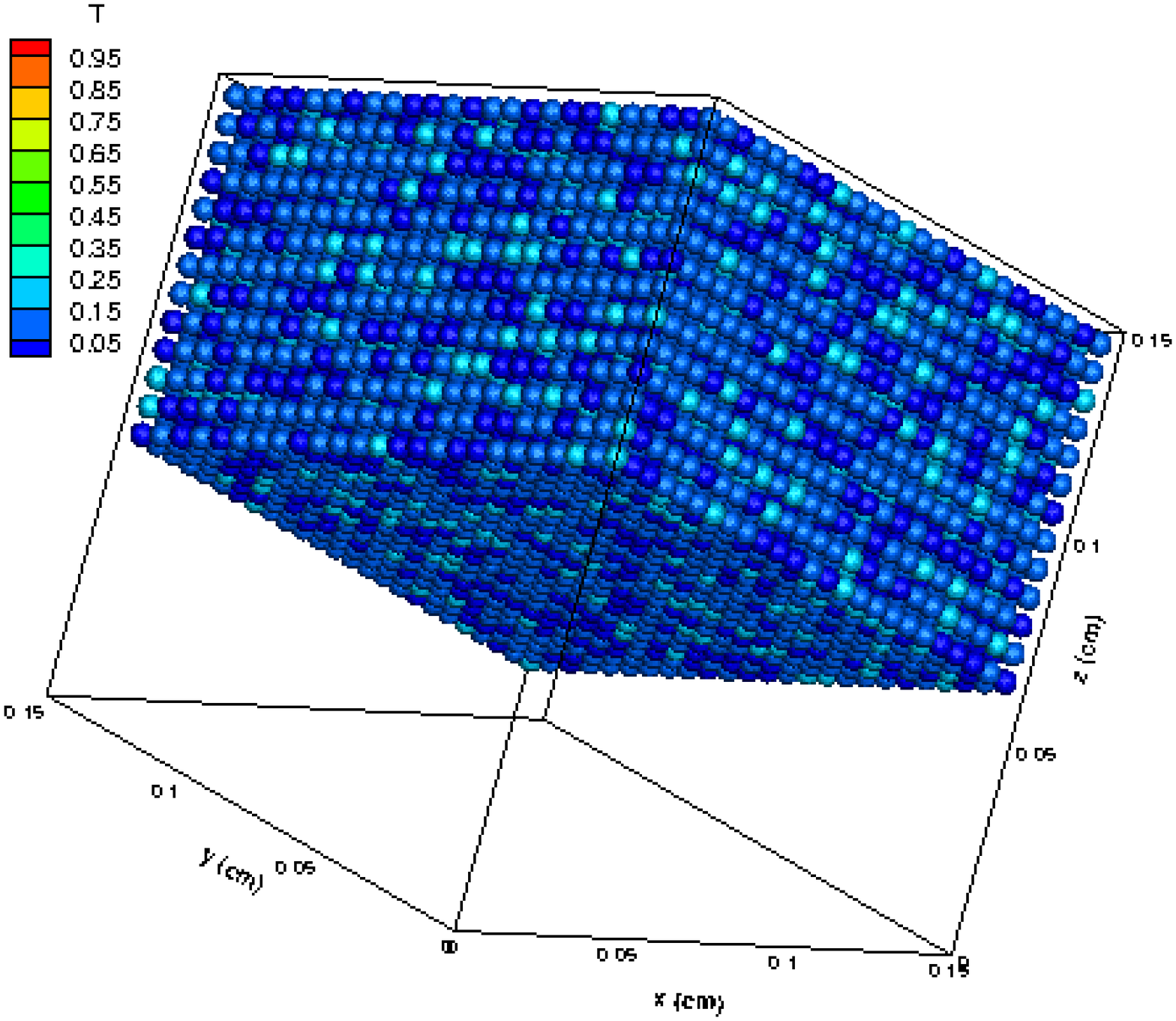}
\hspace{2mm}
\includegraphics[angle=  0,width=0.45\textwidth]{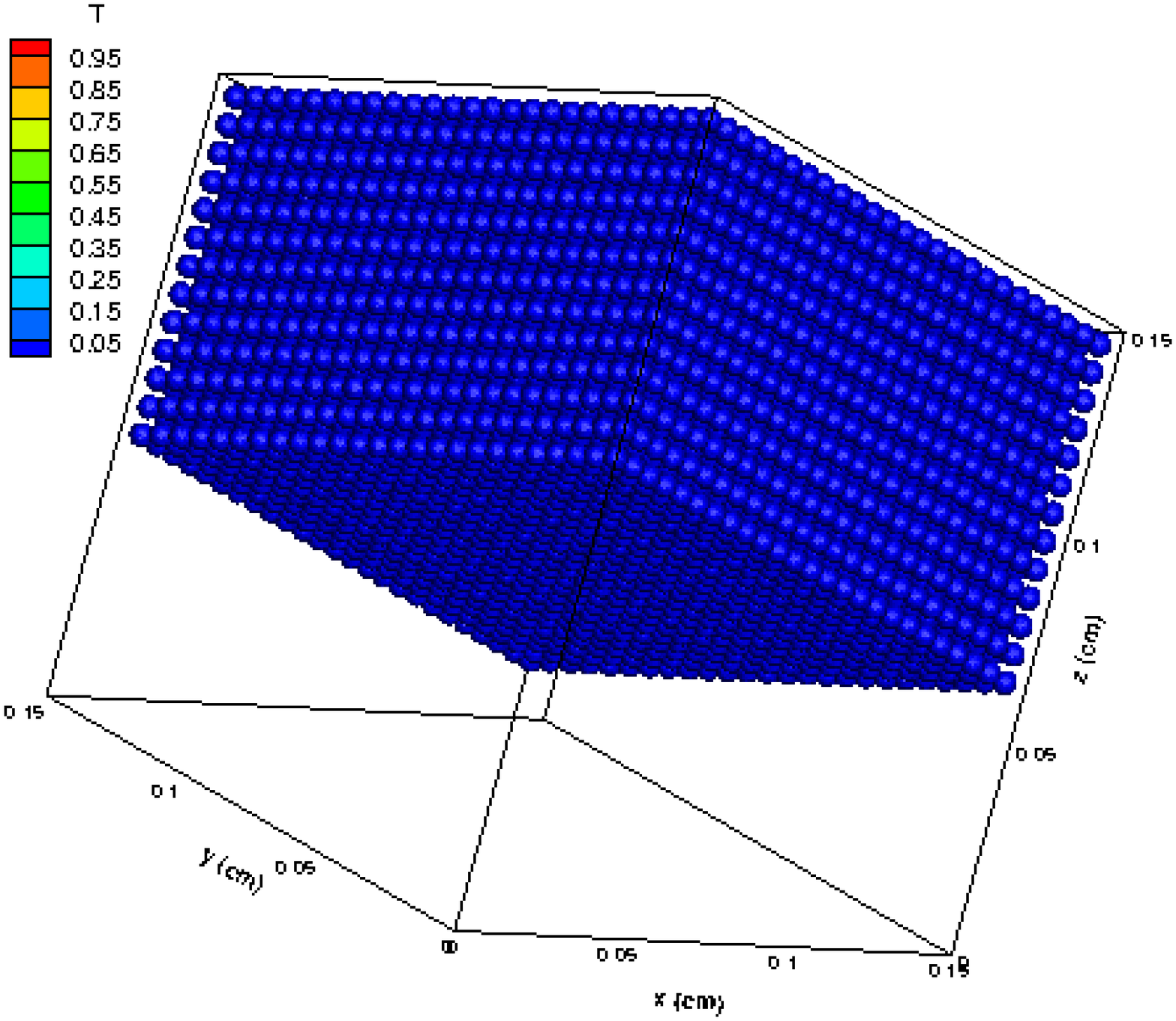}
\\
(c) \hspace{52mm} (d)
\caption{The $8125$ thermosensitive particles colored by temperature at time 
(a) $t=0.0 s$, (b) $t=0.01 s$, (c) $t=0.03 s$, (d) $t=0.12 s$.}
\label{3dparticlepositionrandom}
\end{figure}

\begin{figure}
 \centering
 \includegraphics[width=0.54\textwidth]{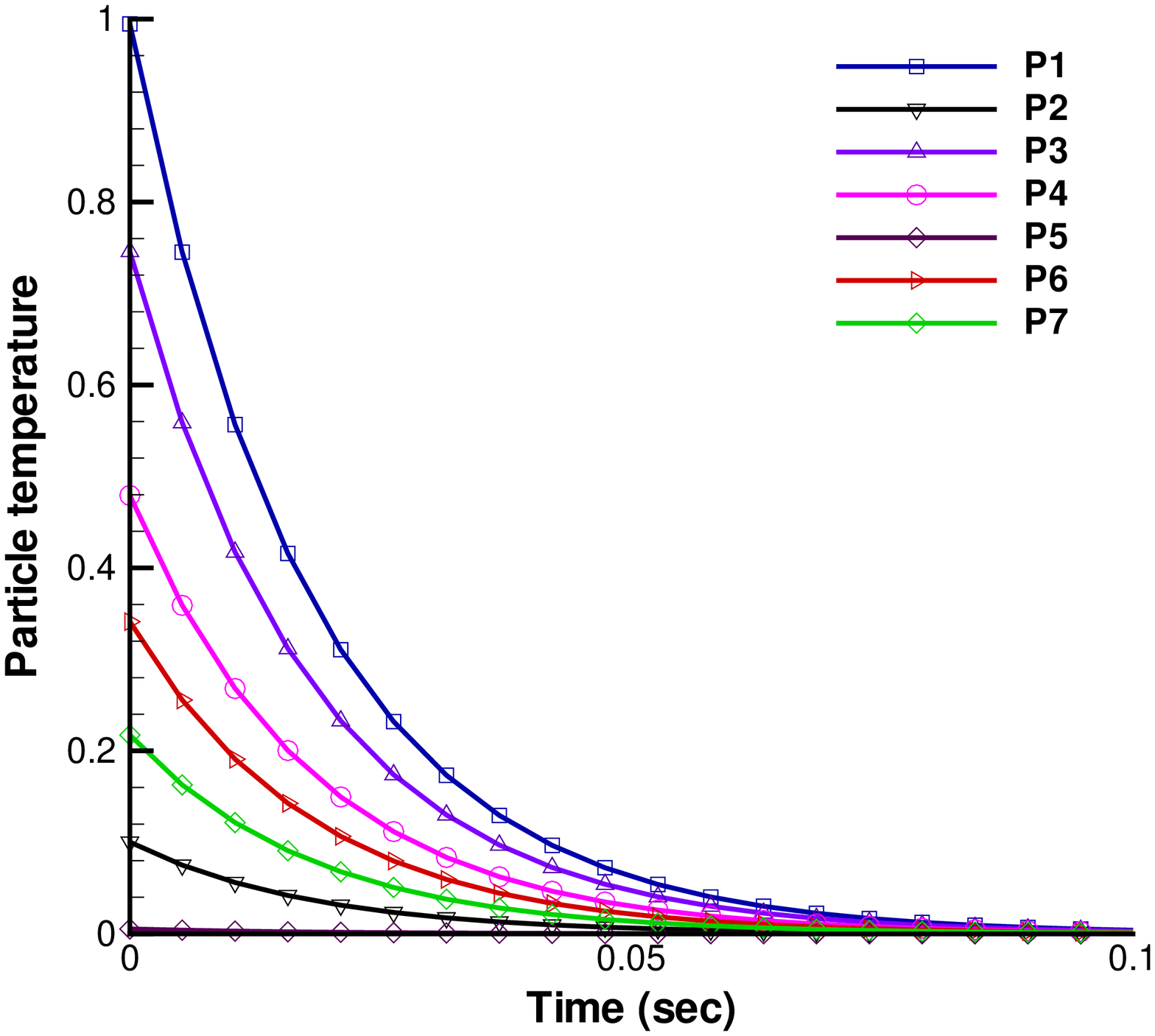}
 \vskip-0.2cm
 \caption{Temperature evolution histories of $7$ particles.} \label{guiyi}
 \end{figure}

Figure~\ref{3dparticleposition2} shows the particle distribution of the $8125$ particles with heat transfer at the same 
moments as Figure~\ref{3dparticleposition1}. It can be seen that the particle behaviors are obviously different to the former 
case when heat transfer is not considered. Instead of forming a particle-constructed pestle in the center region, the bottom 
surface of the isothermal particle aggregate is more fluctuant with bulges. This finding is in line with the two-dimensional 
one, the interface of the solid and lower fluid is more complex when heat transfer is considered. At $t=2.5 s$, the particle 
distributions exhibit completely contrary trends with and without heat transfer. It shows a hump in 
Figure~\ref{3dparticleposition1}(a) whereas a hollow 
in Figure~\ref{3dparticleposition2}(a). The discrepancy is due to the high thermal buoyancy in the interior, the particles in 
the corners have the priority to settle since the temperature in these regions is low. The subsequent sedimentation feature 
in Figure~\ref{3dparticleposition2} follows this trend where the bulging at the four corner region can be observed.

Figure~\ref{droptime} shows the changing history of the minimum particle height in the two cases. This is a good measurement 
of the overall sedimentation efficiency. It can be seen that the two profiles are quite comparable before $2.4s$.
Then, the non-thermal particles begin to accelerate whereas the sedimentation of the thermal particles is significantly 
delayed by the buoyancy. 
Detailed comparison between the particle deposition velocity along the x-direction with and 
without heat transfer is given in Figure~\ref{3dparticlevelocity}. The contrary distribution of the particle deposition velocity validates 
the afore-mentioned observations.

At last, the temperature distributions in the cubic cavity are displayed in Figure~\ref{3dIsosurfaces} in terms of isothermal 
surfaces. The cavity is dissected at the bisector of the x-direction which visualizes the temperature distribution feature 
inside. It is shown that the temperature at the interior of the particle aggregate is the highest. The thermal buoyancy at 
this region is also the highest and thus the solid particles possess the highest upwards velocities as shown in 
Figure~\ref{3dparticlevelocity}. The location of the heat core moves downwards with sedimentation of the solid particles. 
The temperature between the particle aggregate and surrounding walls changes gradually due to the low $Ra$ adopted. 

\subsection{Sedimentation of three-dimensional thermosensitive particles in fluid}\label{Sedimentationrandom3D}

In this subsection, three-dimensional thermosensitive particles are considered in the same model 
as Section~\ref{Sedimentation3D}. Different to all the above cases where the temperature of the solid
particles is constant. Here, the thermosensitive particles could lose or receive heat according to the surrounding 
temperature. The initial temperature of the $8125$ thermosensitive particles is set randomly between
$0$ and $1$ as shown in Figure~\ref{3dparticlepositionrandom} (a). The walls are set cold with $0$ temperature. 
The initial temperature of the stagnant fluid is $0.5$. All the other computational parameters are the 
same as Section~\ref{Sedimentation3D}. 
 
Since the walls are cold, it can be expected that the temperature of the whole system would
reach a final steady-state which is the same as the wall temperature. This process can be clearly observed in
Figure~\ref{3dparticlepositionrandom} which displays the particle temperature distribution at different time instants.
Along with that the cold walls keep sucking heat, the temperature of the whole system drops rapidly to the steady-state 
value. Here, we monitor the temperature evolution histories of $7$ solid particles randomly picked from the assembly. 
As shown in Figure~\ref{guiyi}, the different profiles again indicate the feature of the cooling process.

\section{Concluding remarks}\label{conclusion}

In this paper, the LBM-PIBM-DEM scheme was employed to solve thermal interaction problems between spherical particles and 
fluid. The LBM was adopted to solve the fluid flow and temperature fields, the PIBM was responsible for the no-slip velocity 
and temperature boundary conditions at the particle surface, and the kinematics and trajectory of the particles were evaluated 
by the DEM. 

Four case studies were implemented to certify the capability of the 
proposed coupling scheme. First, numerical simulation of natural convection in a two-dimensional square cavity with 
an isothermal concentric annulus was carried out for verification purpose. Then, sedimentation of two- 
and three-dimensional isothermal particles in fluid was numerically studied. The instantaneous 
temperature distribution in the cavity was captured. The effect of thermal buoyancy on the particle behaviors was discussed. 
Finally, sedimentation of three-dimensional thermosensitive particles in fluid was numerically investigated.
Our results revealed that the thermal buoyancy has a great effect on the particle behaviors both in two- and three-dimensional 
simulations. When heat transfer is considered, the interface between the solid particle aggregate and lower fluid is more 
unstable. The heat buoyancy could influence on the sedimentation efficiency of the solid particles very much. All the 
simulations demonstrate that the LBM-PIBM-DEM coupling scheme is a promising one for the solution of 
complex fluid-particle interaction problems with heat transfer. 

One critical assumption made in this study is that the heat conduction between solid particles or particle 
and wall is ignored. This assumption is reasonable when intense inter-particle collisions dominate the system, 
for example, the fore part of the sedimentation process where the collision process is instant. Nevertheless, 
this may be against those actual engineering processes where the heat conduction between solids cannot be 
neglected~\cite{zhou2009particle,hou2012computational1}. 
As shown in Section~\ref{Naturalconvectioncavityannulus}, the LBM-PIBM scheme is competent to pure natural convection 
problems. Further meshing on the solid particle can be performed to calculate heat conduction through particles 
like Feng et al.~\cite{feng2008discrete,feng2009discrete}. This work is interesting but will of course increase the 
calculation burden. An alternative approach is to consider fluid flow and heat transfer just at the particle scale as 
done in~\cite{zhou2009particle, zhou2010new,hou2012computational1,hou2012computational2}. Therefore, it is in the 
authors' opinion that the current LBM-PIBM-DEM scheme is more suitable for dynamic processes such as gas fluidization, 
hydrocyclone and pneumatic conveying. It can also be used to generate sub-particle information to support particle scale 
modeling.

\section*{Acknowledgments}

This work has been financially supported by the \textit{Ministerio de
Ciencia e Innovaci\'{o}n}, Spain (ENE2010-17801).
F. Xavier Trias would like to thank the financial support by the 
\textit{Ram\'{o}n y Cajal} postdoctoral contracts (RYC-2012- 11996)
by the \textit{Ministerio de Ciencia e Innovaci\'{o}n}. 
The authors are thankful to Dr. Yang Hu from Beijing Jiaotong University, China for his insightful suggestions
on the coupling simulation.

\section*{References}

\bibliography{./ThermalPIBM}

\end{document}